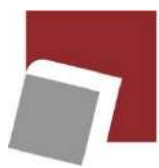

Institute for Advanced Studies in Basic Sciences
Gava Zang, Zanjan, Iran

# A Survey on Patch-based Synthesis: GPU Implementation and Optimization

Bachelor Thesis
**Hadi Abdi Khojasteh**

**Supervisor: Dr. Ebrahim Ansari**

August 2016

# Abstract

This thesis surveys the research in patch-based synthesis and algorithms for finding correspondences between small local regions of images. We additionally explore a large kind of applications of this new fast randomized matching technique.

One of the algorithms we have studied in particular is PatchMatch, can find similar regions or "patches" of an image one to two orders of magnitude faster than previous techniques. The algorithmic program is driven by applying mathematical properties of nearest neighbors in natural images. It is observed that neighboring correspondences tend to be similar or "coherent" and use this observation in algorithm in order to quickly converge to an approximate solution. The algorithm in the most general form can find k-nearest neighbor matching, using patches that translate, rotate, or scale, using arbitrary descriptors, and between two or more images. Speed-ups are obtained over various techniques in an exceedingly range of those areas. We have explored many applications of PatchMatch matching algorithm. In computer graphics, we have explored removing unwanted objects from images, seamlessly moving objects in images, changing image aspect ratios, and video summarization. In computer vision we have explored denoising images, object detection, detecting image forgeries, and detecting symmetries. We conclude by discussing the restrictions of our algorithmic program, GPU implementation and areas for future analysis.

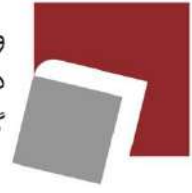

# بررسی روش های آمایش بر پایه تکه: پیاده سازی و بهینه سازی با واحد پردازش گرافیکی

پایان‌نامهٔ کارشناسی

هادی عبدی خجسته

استاد راهنما: دکتر ابراهیم انصاری

شهریور ۱۳۹۵

# چکیده

در این پایان نامه یک الگوریتم تطابق تصادفی سریع نوین برای یافتن تناظرات بین ناحیه‌های مختلف تصویر بررسی می‌شود. همچنین انواع مختلفی از اپلیکیشن‌های این شیوه تطابق تصادفی سریع مورد بررسی قرار می‌گیرد.

هسته اصلی الگوریتم که ما آن را PatchMatch می‌نامیم؛ می‌تواند ناحیه‌ها یا patch‌های مشابه از یک تصویر اول با تصویر دوم را بر اساس اندازه تشابه با سرعت بیشتری نسبت به الگوریتم‌های مشابه، پیدا کند. این الگوریتم از خواص آماری نزدیکترین همسایه در تصویر بهره می‌برد. مشاهده شده است که نزدیکترین همسایه به رابطه وابستگی تمایل دارد و از این مشاهده در الگوریتم برای همگرایی سریع به جواب تقریبی استفاده شده است. این الگوریتم در عمومی ترین شکل می‌تواند k–نزدیکترین همسایه مطابق را با استفاده از patch‌هایی که انتقال، چرخش یا مقیاس داده شده است، در بین یک یا چند تصویر پیدا کند. افزایش سرعت در اینجا با شیوه‌های جایگزین با تغییر بخش‌هایی از شیوه‌های قبلی مشابه امکان پذیر شده است که در این شیوه‌ها همگرایی به صورت تجربی و تئوری به اثبات رسیده است.

ما همچنین بسیاری از برنامه‌ها و کاربردهای الگوریتم PatchMatch را بررسی کرده‌ایم. حذف اشیاء ناخواسته از تصویر، جابجایی اشیاء در تصویر به صورت یکپارچه، تغییر نسبت تصویر و خلاصه سازی ویدیو در گرافیک کامپیوتری را مورد بررسی قرار داده‌ایم. این الگوریتم همچنین در بینایی کامپیوتر برای حذف نویز از تصویر، تشخیص اشیاء، تشخیص تصویر جعلی یا Forgery Detection، Symmetries Detection کاربر دارد.

در آخر این پایان نامه با بحث در مورد محدودیت‌های الگوریتم، پیاده‌سازی و بهینه‌سازی با واحد پردازش گرافیکی و زمینه‌های تحقیقاتی آینده به پایان می‌رسد.

أ

# فهرست

# فصل اول

# مقدمه

## ۱. ۱   نگاه کلی

از آنجایی که عکاسی دیجیتال به لوغ کامل رسیده است، محققان شیوه‌هایی را برای درک بیشتر از تصویر و ویدیو در سطح بالا توسعه داده اند. به طور مثال در الگوریتم‌های اخیر برای بازسازی تصویر[1]، ماشین به صورت خودکار بهترین تشابه را از تصویر اصلی اما با یک نسبت متفاوت محاسبه می‌کند [94,116]. الگوریتم‌های دیگر تکمیل تصویر[2] به کاربر اجازه می‌دهد که به آسانی ناحیه‌های ناخواسته از تصویر را حذف کند و باقیمانده ی نواحی تصویر به طرز قابل قبولی به صورت خودکار پر می‌شوند [29,60].

بیشتر این الگوریتم‌ها از این واقعیت که می‌توان تصویر را به بخش‌های کوچک مربع شکل با اندازه ثابت تقسیم کرد، بهره می‌برند (مثلا بخش‌های ۷ در ۷). تصویر می‌تواند با یک رویکرد بر پایه patch با مرتب کردن بخش‌های مختلف (مشابه آنچه انسان یک پازل را تکمیل می‌کند.) در این الگوریتم مجددا ایجاد شود. به صورت عمومی قبلا ایجاد تصاویر با patchها، مقدار قابل توجهی زمان و حافظه برای یافتن بخش‌های مشابه در تصاویر نیاز داشت.

در این پایان نامه الگوریتم PatchMatch [8] و الگوریتم‌های تعمیم یافته [10] از آن بررسی می‌شوند که تا حد زیادی به الگوریتم‌های جستجو برای بخش‌های مختلف تصویر سرعت بخشیده است. برای

---

1. image retargeting ؛ در این الگوریتم‌ها از روش هایی برای تغییر نسبت تصویر مثلا از ۳:٤ به ١٦:٩ استفاده می‌شود.
2. image completion

مسئله یافتن تطابق ساده، این الگوریتم ۲۰ تا ۱۰۰ برابر سرعت بیشتری نسبت به کارهای قبلی دارد و به اندازه قابل ملاحظه‌ای حافظه کمتری نیاز دارد.

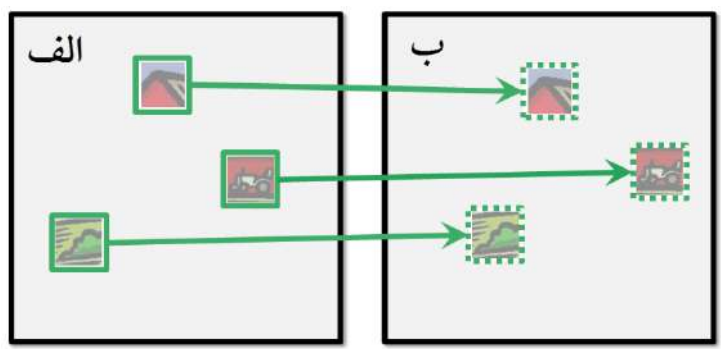

شکل ۱. ۱ – مسئله یافتن تطابق؛ به ازای هر patch در تصویر (الف)، نزدیکترین همسایه در تصویر (ب) را پیدا میکنیم.

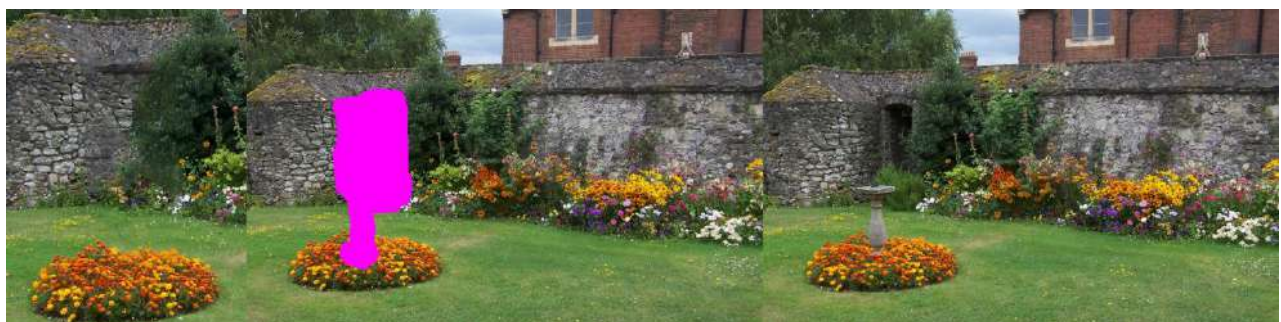

(الف) تصویر ورودی        (ب) ناحیه انتخاب شده        (ج) تصویر خروجی

شکل ۱. ۲ – تکمیل تصویر؛ تصویر (الف) در اختیار است. کاربر یک ناحیه را برای حذف انتخاب میکند (ب). سپس تصویر خروجی (ج) با استفاده از تطبیق الگوهای بافتی مشابه در الگوریتم ایجاد میشود. بدین ترتیب شی انتخاب شده حذف می‌شود.

هدف در این الگوریتم یافتن ناحیه نزدیکترین همسایه [3] در یک تصویر برای تمام ناحیه‌های دارای اشتراک در تصویر دیگر می‌باشد که در شکل ۱. ۱ نمایش داده شده است. دقت کنیم در اینجا مسئله برای تصاویری مطرح می‌شود که به شکل زیادی با هم اشتراک دارند و هر patch برای پیکسل‌های اطرافش

---

۳. ناحیه نزدیکترین همسایه یا nearest neighbors patch در اینجا به معنای ناحیه است که بیشترین شباهت را از لحاظ تفاوت رنگی یا میانگین تفاضلات رنگ‌ها برای آن ناحیه در یک تصویر در مقایسه با تصویر دیگر داشته باشد.

تعریف شده است [4]. بنابراین الگوریتم با توجه به فرض 'نتایج آماری تصاویر واقعی' همگرا می‌شود که در آن به طور خاص گفته شده است patchها در این تصاویر بسیار با هم همبستگی دارند.

این الگوریتم سرعت یافته به نوبه خود برنامه‌های جدیدی برای تغییر تصاویر، خلاصه سازی ویدیو و حوزه بینایی کامپیوتر ارائه کرده است که در اینجا بررسی خواهند شد.

## ۱.۲   چرا روش‌های Accelerate Patch-Based ؟

برای درک بهتر نیاز استفاده از الگوریتم‌های جستجوی سریع برای patchهای تصویر، مسئله تکمیل تصویر که در شکل ۱.۲ نمایش داده شده است را در نظر بگیرید. کاربر یک شی ناخواسته را برای حذف از عکس انتخاب می‌کند. با استفاده از الگوریتم .Wexler et al [119]، ابتدا این بخش با یک حدس اولیه درون یابی شده از مرزها مقداردهی می‌شود. تصویر به patchهای دارای اشتراکی تقسیم می‌شوند. سپس الگوریتم به صورت تکراری تناظرات بین patchهای داخل بخش انتخاب شده و خارج بخش را در نتیجه نیمه کامل تا به هر مرحله می‌یابد. در مرحله بعد ویژگی [5]های مطابق شده خارج از بخش انتخاب شده، در محل متناظر در این بخش چسابنده می‌شود و یک میانگین گیری یا رای گیری برای patchهای مشترک با آن در نظر گرفته می‌شود. حلقه ی داخلی در این الگوریتم برای مقیاس‌های مختلف از وضوح کم (درشت) تا وضوح زیاد (ریز) تا رسیدن به یک نتیجه با بیشترین وضوح ادامه می‌یابد؛ یعنی ابتدا اندازه patchها بزرگ در نظر گرفته شده و در مراحل بعدی این اندازه برای رسیدن به وضوح بیشتر و یافتن مقادیر مشابه بهتر، کوچکتر می‌شود.

در این فرآیند bottleneck یافتن بهترین تطابق برای patchهای داخل به خارج بخش انتخاب شده است. با استفاده از یک رویکرد مشابه بالا، می‌توان نسبت ابعاد یک تصویر را به صورت خودکار تغییر داد. همین رویکرد را می‌توان برای کاربردهای دیگر بیان شده نیز در پیش گرفت. برای این برنامه نشان خواهیم داد که استفاده از الگوریتم PatchMatch می‌تواند این فرآیند را با تعامل بیشتر کاربر و افزایش

---

۴. بدین ترتیب فرض می‌شود تقریبا در تصاویر ورودی بخش های خارج از یک patch ارتباط معنایی با patch خواهند داشت و به صورت کلی از هم مستقل نیستند.

۵. منظور از ویژگی یا feature، patch هایی از تصویر هستند که برای یکی از patch ها به عنوان مشابه ترین بخش در تصویر برگزیده شده اند.

کنترل سرعت بخشد. بدین ترتیب این الگوریتم می‌تواند افزایش سرعت بیشتر و کاهش حافظه اشغال شده را با تغییرات کیفی با افزایش تعامل کاربر به همراه داشته باشد.

علاوه بر این تحقیقات در حال رشد و بزرگی در زمینه آنالیزهای بر پایه patch برای تصاویر، ویدیوها، عکاسی محاسباتی، هندسه سه بعدی و بینایی کامپیوتر در جریان است. برخی از این کارها در فصل‌های آینده بررسی خواهند شد. توجه داشته باشید الگوریتم ارائه شده در هر کدام از این زمینه‌ها میتواند با جایگزینی به عنوان یک ابزار مورد استفاده قرار گیرد. حتی در جاهایی که الگوریتم نمی تواند به شکل جعبه سیاه مورد استفاده قرار گیرد، می‌تواند برای آن مسئله مشخص با مشاهدات مشابه‌ای که بیان شد، سازگار شود.

## ۳.۱  بررسی اجمالی

در اینجا توضیحات اجمالی در مورد بخش‌های باقیمانده در این بررسی ارائه خواهد شد.

### ۱.۳.۱  الگوریتم PatchMatch

ما ابتدا یک الگوریتم برای تطابق بر اساس مشاهدات آماری درباره نزدیکترین همسایه در تصاویر واقعی معرفی خواهیم کرد (فصل دوم). سپس PatchMatch را که یک الگوریتم تصادفی سریع برای یافتن یک تقریب خوب برای نزدیکترین همسایه در یک نمونه شامل ارتباط بالا بین بخش‌های مختلف است را معرفی خواهیم کرد (فصل سوم). برای سادگی ما ابتدا الگوریتم را به ساده ترین صورت ممکن یعنی یافتن یک نزدیکترین همسایه که تنها انتقال داده شده است، توضیح خواهیم داد. سپس الگوریتم کلی را بیان خواهیم کرد که می‌تواند k-نزدیکترین همسایه را با هر مقیاس یا چرخشی بیابد.

با اینکه الگوریتم اصلی بدون تغییر قابل استفاده است، اما تعدادی از پیاده سازی‌ها یا استراتژی‌ها می‌تواند به الگوریتم شتاب بخشد. ابتدا ما در مورد استراتژی‌های موازی سازی الگوریتم بر روی CPU و GPU خواهیم پرداخت. سپس در مورد بهینه سازی الگوریتم به طور مثال با جلوگیری از محاسبات اضافی، بحث خواهیم کرد. در ادامه در مورد استراتژی‌های جایگزین جستجو که فضای جستجو را به خصوص در زمانیکه تعداد زیادی تطابق نیاز باشد، کمک می‌کنند، بررسی خواهیم کرد. بر همین اساس در نهایت الگوریتم PatchWeb برای یافتن تطابقات بین هزاران یا میلیون‌ها تصویر کاربرد دارد را بررسی

۸

خواهیم کرد. PatchWeb وظیفه مقیاس پذیری را زمانیکه تعداد زیادی تصویر در اختیار داشته باشیم که به علت حجم زیاد قابل بارگذاری در حافظه نباشند، برعهده دارد. همچنین می‌تواند به صورت معماری خوشه‌ای توسعه یابد. در نهایت در مورد کاربردهای PatchMatch در زمینه ویرایش تصاویر، خلاصه سازی ویدیو در قالب "پرده‌های فیلم"[6] و برخی کاربردهای بینایی کامپیوتر توضیح داده شده است.

۱.۳.۲ برنامه‌های کاربردی ویرایش تصاویر

در فصل پنجم نگاهی خواهیم انداخت بر روش‌های ویرایش عکس‌ها با استفاده از کنترل‌های سطح بالایی که به کاربر داده می‌شود. ابتدا الگوریتم بازسازی تصویر که برای تغییر نسبت تصویر بدون تغییر در ظاهر یا کشش تصویر یا هرگونه اعوجاج ناشی از الگوریتم استفاده می‌شود، را بررسی می‌کنیم. سپس تکمیل تصویر را که به کاربر اجازه می‌دهد تا به راحتی بخش‌های ناخواسته از تصویر را حذف کند و الگوریتم با محاسبه خودکار، این بخش را با مقدار احتمالی محاسبه شده پر کند، توضیح خواهیم داد و در ادامه در مورد الگوریتم بُرزنی[7] که امکان حرکت دادن بخشی از تصویر به محل دیگر و محاسبه تغییرات تصویر به صورت خودکار را فراهم می‌کند، بحث خواهیم کرد.

در هر کدام از این موارد تعامل کاربر برای رسیدن به بهترین نتیجه ضروری است که به دو دلیل است. اول؛ الگوریتم در برخی موارد نیاز به تعامل با کاربر برای رسیدن به بهترین نتیجه دارد. دوم؛ اهداف مسئله همیشه نمی‌توانند از قبل توسط کاربر به روشنی بیان شوند و نیاز به یک فرآیند مکرر آزمون و خطا برای بهینه سازی نتیجه است. الگوریتم ارائه شده در اینجا می‌تواند تا حد زیادی برای استفاده در این کاربردها شتاب بخشیده و زمان اجرا برای این موارد را از دقیقه به ثانیه کاهش دهد[8].

---

[6]. پرده های فیلم یا Tapestries، طرح‌ها و تصویرهایی هستند که همچون کوبلن روی پارچه بافته می‌شوند. پرده‌های فیلم را با استفاده از نخ‌های ابریشمی رنگی روی ردیف‌های کتان و پشم که درون یک کادر (کارگاه گل‌دوزی) بسته شده‌اند، می‌بافند. از این روش برای بیان یک داستان به صورت کامل در یک تصویر همچون داستان رستم و سهراب شاهنامه استفاده می‌شود.

[7]. Image Reshuffling

[8]. به علت سرعت بسیار زیاد این الگوریتم و همچنین برای اولین بار تعاملی بودن برای رسیدن به بهترین نتیجه مورد نظر کاربر، این شیوه برای حذف اشیا ناخواسته کاربر در تصویر بر اساس الگوریتم ارائه شده، در نرم افزار Adobe Photoshop CS5 به عنوان قابلیت جدیدی با نام content aware اضافه شده است.

۹

### ۱.۳.۳ ویدیو در یک تصویر

در فصل شش، یک رویکرد جدید برای خلاصه سازی ویدیو در یک ساختار چند مقیاسی که هم در بعد مکانی و هم در طول مقیاس زمان ادامه یافته اند؛ بدین صورت که مرزی بین لحظات گسسته در زمان وجود ندارد و کاربر می‌تواند با بزرگنمایی بر روی تصویر جزئیات بیشتری از تغییرات زمان را مشاهده کند. این مفهوم در اینجا tapestries یا پرده‌های فیلم نامیده شده است چرا که شباهت زیادی به تصاویر اولیه به وجود آمده قدیمی قبل از تصاویر محرک برای بیان یک واقعه یا داستان همچون تصاویری که راویان برای بیان داستان‌های شاهنامه استفاده می‌کنند، دارند.

برای این خلاصه سازی تعدادی شاخص تعریف شده است که بهینه سازی‌های مختلفی هم برای آن در نظر گرفته می‌شود. برای رسیدن به یک خروجی با کیفیت بالا، این کار می‌تواند کاملا به صورت آفلاین انجام شود یا به تغییر برخی از پارامترهای بهینه سازی مسئله این مرحله به صورت آنلاین انجام شود. در هر دو روش آفلاین یا آنلاین کارآیی بستگی به الگوریتم تطابق دارد که در هر دو روش از PatchMatch استفاده شده است.

### ۱.۳.۴ برنامه‌های کاربردی بینایی کامپیوتر

در فصل هفت، برنامه‌های کاربردی متعددی در بینایی کامپیوتر که از قدرت الگوریتم تطابق عمومی استفاده می‌کنند، بررسی خواهند شد. برنامه حذف نویز متوسط غیر محلی برای حذف نویز از تصویر با یافتن و میانگین گیری الگوهای غیر محلی تکراری در این بخش بررسی شده است. همچنین ما شناسایی تصویر دیجیتالی جعلی را هم بررسی کرده‌ایم. در آخر تشخیص اشیا با استفاده از patchهایی که چرخیده شده اند و برای نورپردازی تغییر کرده اند، همانند خصوصیاتی در SIFT بررسی شده اند.

# فصل دوم
# بررسی آماری نزدیکترین همسایه

هدف یافتن ناحیه نزدیکترین همسایه (خصوصیات تصویر در یک ناحیه) در یک تصویر برای تمام ناحیه‌های دارای اشتراک در تصویر دیگر می‌باشد، که پیش تر توضیح داده شد. نزدیکترین همسایه در اینجا به ازای تفاوت هر دو patch که آن را با D نمایش می‌دهیم تعریف می‌شود که میزان تشابه دو patch را اندازه گیری می‌کند. به طور مثال عموما یک $L^2$ در نظر می‌گیریم که تفاوت را بین هر دو patch با در نظرگرفتن $L^2\ norm$ مربوط به تفاضل مقدار RGB متناظر با هر patch اندازه گیری می‌کند. یک تفاوت عمده این نوع یافتن نزدیکترین همسایه با دیگر موارد مشابه، این است که در اینجا ثابتی برای smoothness در نظر گرفته نمی شود و بهترین تطابق بدون در نظر گرفتن تطابق‌های همسایگان و به صورت مستقل محاسبه می‌شود.

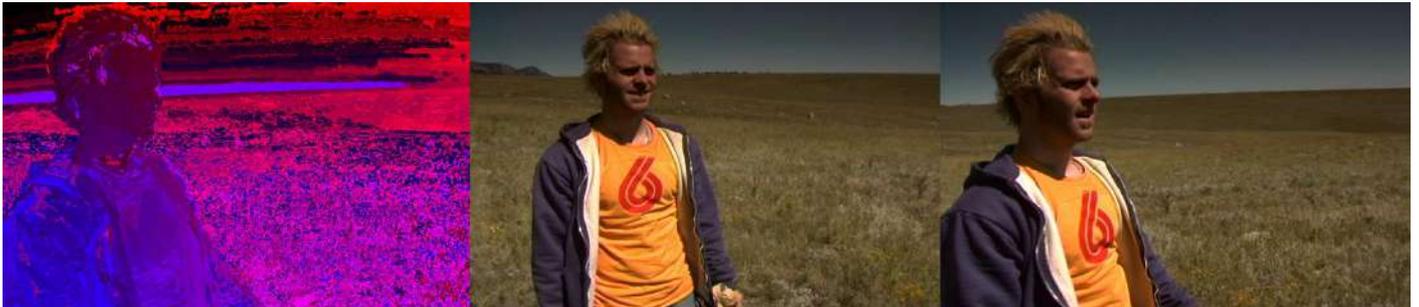

(الف) تصویر اول  (ب) تصویر دوم  (ج) محل‌های منطبق شده

شکل ۲. ۱ – Coherence؛ دو تصویر متفاوت اول و دوم انتخاب شده اند. در تصویر (ج) محل‌های واقعی نزدیکترین همسایه‌ها نمایش داده شده است. به ازای هر patch در تصویر اول، نزدیکترین همسایه را در تصویر دوم با مختصات نسبی x به رنگ قرمز و y به رنگ آبی نمایش میدهیم. ناحیه‌های وابسته همان رنگ را دارند. وابستگی‌های تاثیر گذار پررنگ تر هستند. چرا که منطبق کردن بسیاری از ناحیه‌ها به سادگی بین چند نزدیکترین همسایه ممکن؛ که همه‌ی آنها خطای تطابق برابر دارند، امکان پذیر است.

دو خصوصیت از تصاویر را در نظر بگیرید: وابستگی نزدیکترین همسایه [9] و توزیع peak مربوط به اینکه در چه موقعیتی بهترین نزدیکترین همسایه‌ها قرار گرفته اند. دو مرحله از الگوریتم یعنی انتشار و

---
9. coherence of nearest neighbors

جستجوی تصادفی که برای فرآیند جستجوی تکراری (با توجه به فرض‌های قبلی برای تصویر ورودی که در فصل ۳ بررسی خواهیم کرد.) استفاده می‌شوند، از این دو خصوصیت بهره می‌برند.

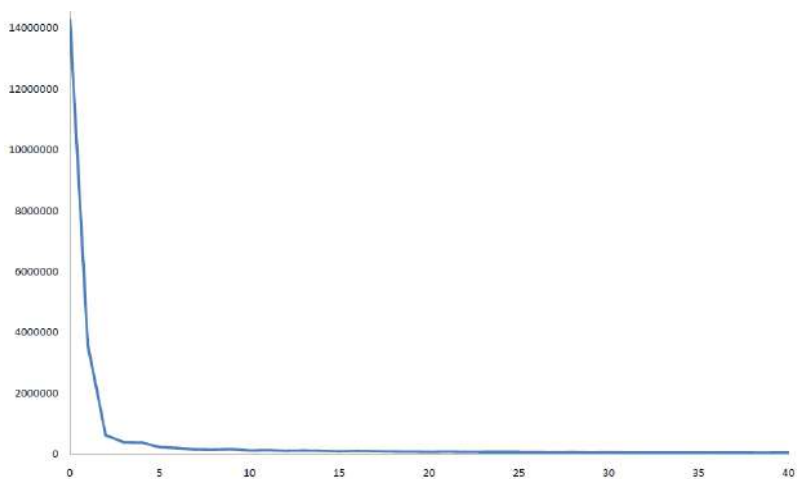

شکل ۲.۲ – نمودار هیستوگرام همبستگی بین نزدیکترین همسایگی patchهای همسایه. در محور افقی مقدار دوری نزدیکترین همسایه patch همسایه بر اساس فاصله دو بعدی اقلیدسی اندازه گیری شده است. در محور عمودی تعداد patchها با فاصله داده شده حساب شده است. قابل توجه است که تعداد زیادی از patchها فاصله‌ی صفر دارند که نشان دهنده‌ی وابستگی کامل است. در واقع یک دنباله کاهشی عموما با احتمال درست نمایی (likelihood) فاصله ۱، ۲ یا ۳ پیکسل داریم.

ابتدا در مورد همبستگی صحبت می‌کنیم. عموما با حرکت دادن یک patch به اندازه یک پیکسل در هر جهتی، انتظار داریم تا ظاهر patch به خصوص برای patchهای با اندازه‌ی بزرگ با سرعت کمی تغییر کند. بدین ترتیب انتظار داریم تا همبستگی زیادی در تطابق نزدیکترین همسایه داشته باشیم. هر دو تناظر در تصویرها همبستگی دارند اگر مختصات نسبی مشابهی داشته باشند؛ یعنی موقعیت در تصویر دوم منهای موقعیت در تصویر اول با هم برابر باشند. برای هر دو تصویری می‌توان این مقدار همبستگی را از تصویر اول به دوم برای محل‌های واقعی هر همبستگی محاسبه کرد و به صورت شکل ۱.۲ نمایش

داد. ناحیه‌های همبسته رنگ‌های مشابه دارند. این الگوریتم با انتشار تطابق‌های خوب به همسایه‌های مکانی پیکسل‌ها مزیت‌های این خصوصیت همبستگی را حفظ می‌کند.

با این حال، حتی اگر نزدیکترین همسایه برای patchها وابسته نباشد، به صورت مکانی به هم وابسته هستند و از توزیع peaked پیروی می‌کنند. این خصوصیت برای همه ورودی‌های ممکن درست نیست اما برای تصاویر واقعی – عکس‌هایی که توسط افراد گرفته شده اند – درست است.

می توان این توزیع peaked را به روش‌های مختلفی نمایش داد. اول می‌توان به این سوال پاسخ داد: برای هر patch همسایه به صورت عمودی یا افقی چقدر نزدیکترین همسایه برای آن‌ها دورتر است؟ این رابطه در شکل ۲.۲ برای زوج تصویرهای عریض دوتایی از یک مجموعه داده بزرگ از تصاویر واقعی نمایش داده شده است. قابل توجه است که تعداد زیادی از patchها از همسایه‌های عمودی و افقی خود فاصله ی مکانی صفر دارند که نشان دهنده ی وابستگی کامل است.بسیاری از patchهای باقیمانده فاصله مکانی بسیار کمی به اندازه ۱، ۲ یا ۳ پیکسل داشته اند و توزیع آن به صورت عمومی کاهش می‌یابد. مجموعه داده ی تصاویر مورد استفاده در شکل ۲.۳ نمایش داده شده است.

یک روش دیگر برای نمایش توزیع، با فرض اینکه مقدار فعلی را داریم، یک حد مطلوب از تناظرها برای یک patch است که فاصله D را دارد. حال سوال این است که اگر مکان مبدا در تصویر اول ثابت باشد، محل بهتر هدف برای تناظرات در تصویر دوم مرتبط با هدف کدام است؟ با تلفیق تمام تناظرات

۱۳

ممکن می‌توان یک هیستوگرام دو بعدی برای شدت در مقابل موقعیت نسبی x و y رسم کرد که شدت نشان‌دهنده احتمال مکان داده شده در تصویر دوم است که تناظر بهتری را ایجاد کرده است.

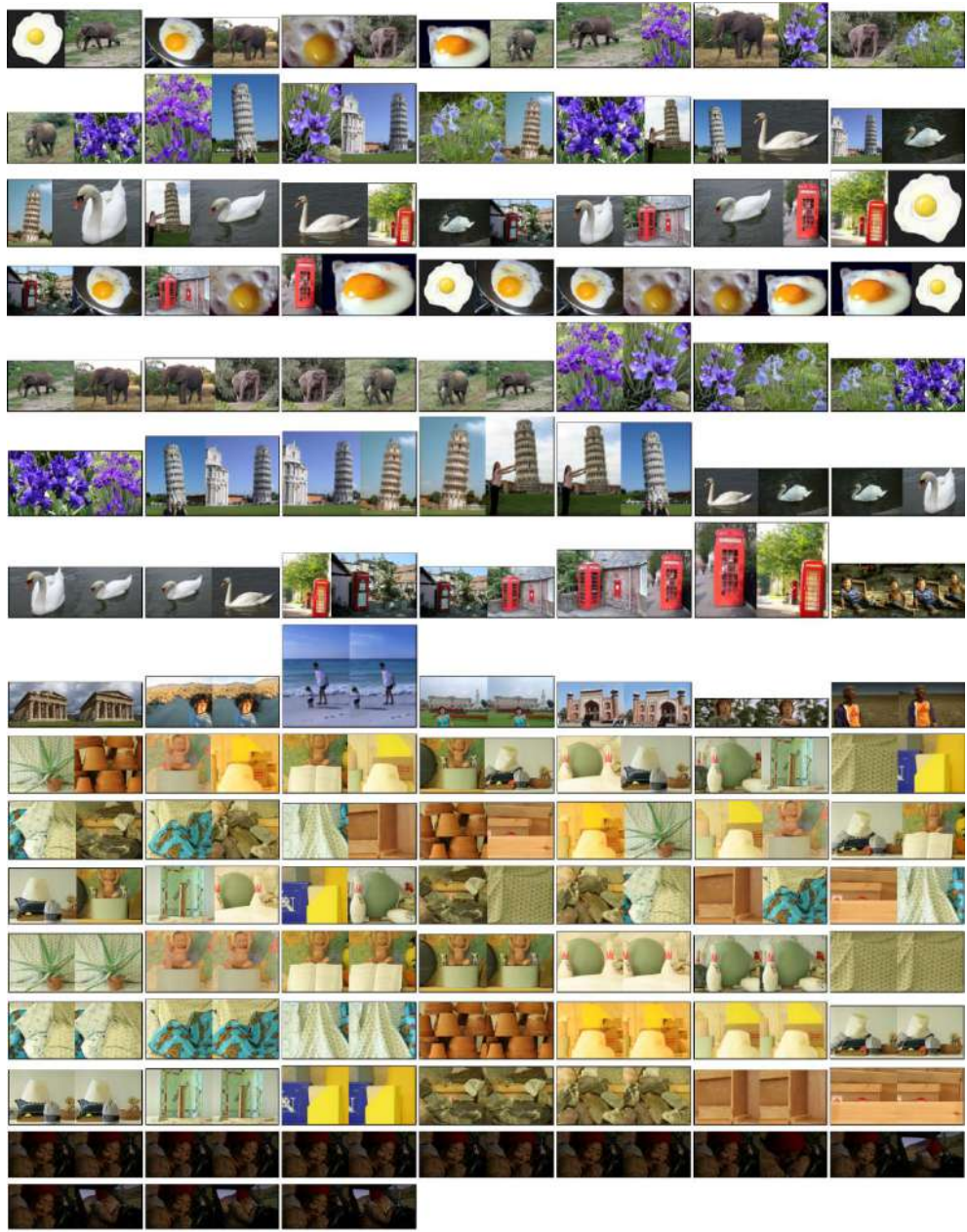

شکل ۲. ۳ – مجموعه داده از ۱۸۰ تصویر برای ایجاد هیستوگرام که از زوج تصاویر مشابه و غیر مشابه و فریم‌های زمانی مختلف از ویدیو تشکیل شده است. ۲۴ زوج تصویر مختلف از کلاس‌های مختلف، ۲۴ زوج تصویر از یک کلاس و ۶ زوج مشابه ورودی و خروجی برای برنامه‌های ویرایش تصاویر، ۲۱ زوج تصویر عریض، ۲۱ زوج تصویر بدون تطابق و ۱۲ زوج از ویدیو.

۱٤

روش دوم بیان شده در شکل ۲. ٤ نمایش داده شده است که بعد از تلفیق برای تمام مجموعه داده از ۱۰۸ تصویر رسم شده است.

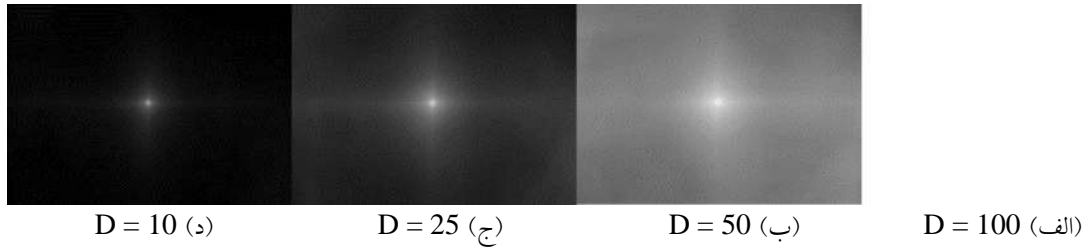

(الف) D = 100    (ب) D = 50    (ج) D = 25    (د) D = 10

شکل ۲. ٤ – هیستوگرام دوبعدی که توزیع peaked مکان قرار گیری تطابق‌های بهتر را نشان می‌دهد. مرکز مکان نقطه تطابق فعلی را نشان می‌دهد. برای تناظرهای با فاصله زیاد D (الف) تطابق‌ها به طور یکنواخت در سراسر تصویر توزیع شده اند. زمانیکه فاصله کم می‌شود (ب)، تطابق‌ها بهتر بیشتر peak دارند و همراه با تطابق‌های بهتر در نزدیکی نقطه فعلی هستند. این هیستوگرام با استفاده از یک مجموعه داده از ۱۰۸ زوج تصویر هم مشابه و هم غیر مشابه برای تطابق تشکیل شده است.

قابل توجه است که هیستوگرام دوبعدی از اینکه همسایه بهتر در کجا واقع شده است غیر یکنواخت است. اما در عوض از توزیع peaked پیروی می‌کند. زمانیکه فاصله D برای تناظرات بدست آمده زیاد باشد، مکان‌های بهتری که به دنبال آنها هستیم به صورت یکنواختی در طول تصویر توزیع شده اند. چنانچه D کم باشد، مکان‌های بهتر بیشتر در اطراف موقعیت فعلی خوشه بندی شده هستند، تا در نهایت در فاصله‌های خیلی کم، بهتر است که جستجو در اطراف نقطه فعلی انجام پذیرد.

مشاهده شده است که فرض بالا، برای تمام تصاویر ممکن برقرار نبوده است، اما تنها یک مجموعه بزرگ از تصاویر واقعی تمایل به پیروی از این فرض داشته اند. به طور مثال شکل ۲. ٥ یک هیستوگرام دو بعدی مشابه از اینکه در چه محل‌هایی بهترین همسایه‌ها قرار دارند را به صورت میانگین بر روی تصاویری با نویز گووسی، یکنواخت و نویز تصادفی نشان می‌دهد. توزیع این نویزها تقریبا به صورت یکنواخت است که هیچ نقطه اوجی ندارد و نشان می‌دهد که تطابق‌های بهتر در تصاویر تصادفی لزوما به

صورت مکانی در نزدیکی آن‌ها نیستند. از این رو الگوریتم ارائه شده می‌تواند با سرعت کمی بر روی ورودی همگرا شود، چراکه فرضیه قبلی گفته شده را دنبال نمی کند.

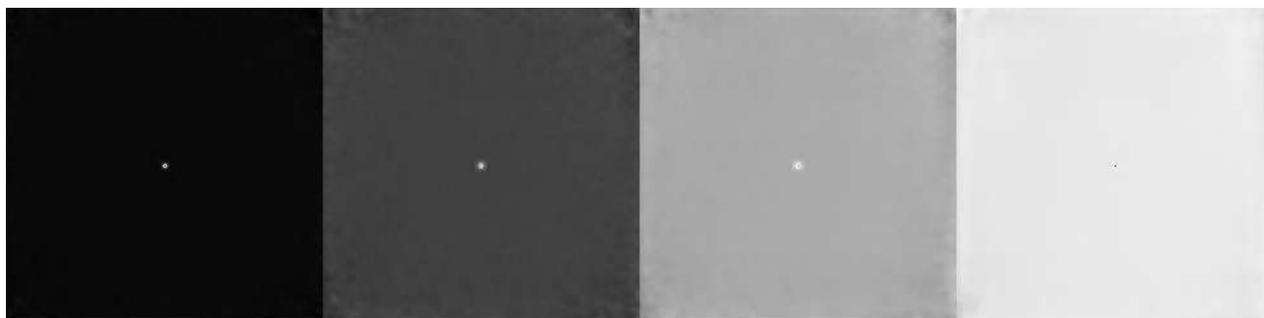

(الف) D = 100          (ب) D = 75          (ج) D = 50          (د) D = 25

شکل ۲. ۵ – هیستوگرام دوبعدی از اینکه در چه محل‌هایی بهترین تطابق‌ها قرار دارند. یافتن تطابق‌ها در تصاویر تصادفی ساخته شده با استفاده از گووسی، یکنواخت یا تابع تصادفی ایجاد نویز. بر خلاف تصاویر واقعی یک نقطه اوج وجود دارد و تطابق‌های بهتر به صورت بکنواخت در راستای تصویر توزیع شده اند. نقطه اوج در وسط به دلیل اندازه patchهای ۷ در ۷ است که باعث وابستگی کمی می‌شود

در نهایت برای کار آینده می‌توان بسیاری از خصوصیات آماری تصاویر واقعی را مورد مطالعه قرار داد. برای مثال، هیستوگرام دوبعدی شکل ۲. ۴ روی مجموعه داده ی بزرگی میانگین گرفته شده است، اما زمانیکه ما تنها یک زوج از تصاویر جداگانه را بررسی می‌کنیم، تغییرات بزرگی را می‌یابیم. زوج تصاویر با لبه‌های افقی قوی باعث به وجود آمدن انحراف عمودی در توزیع می‌شوند و تصاویر با الگوهای تکراری باعث الگوهای تکراری در توزیع می‌شوند که در شکل ۲. ۶ آمده است. در حوزه‌های دیگر نظیر ویدیو

یا هندسه سه بعدی، حتی خصوصیات نامحسوس کوچک نیز ممکن است روی این توزیع‌ها تاثیر گذار باشند.

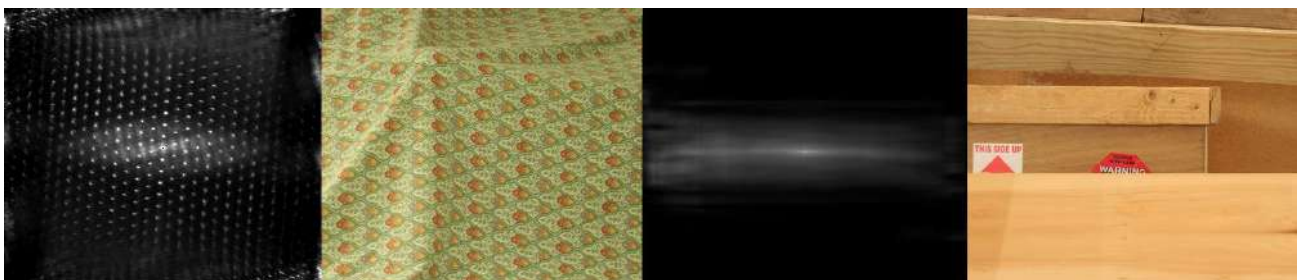

(الف) تصویر لبه‌ها (ب) هیستوگرام لبه‌ها (ج) تصویر الگوها (د) هیستوگرام الگوها

شکل ۲. ۵ – هیستوگرام دو بعدی از اینکه برای یک زوج تصویر مشخص، تطابق‌های بهتر در کجا قرار گرفته اند. برای اولین زوج تصویر، لبه‌ها (الف)، یک لبه افقی قوی دارند. در (ب) هیستوگرام دو بعدی نشان می‌دهد که در کجا تطابق‌های بهتر نسبت به تطابق فعلی پیدا شده اند. لازم به ذکر است انحراف افقی نشان می‌دهد که این برای نمونه‌هایی با چرخش در امتداد لبه برای یافتن تطابق‌های بهتر کاراً خواهد بود. زوج تصویر (ج) یک الگوی تکراری دارد. در هیستوگرام دو بعدی برای تطابق‌های بهتر (د) نشان می‌دهد که تطابق‌های بهتر با نمونه برداری در امتداد فاصله بندی شبکه‌ای یافت می‌شوند.

# فصل سوم
# الگوریتم PatchMatch

در این فصل هسته‌ی اصلی الگوریتم تطابق را بررسی خواهیم کرد که بر اساس مسئله یافتن Patchهای نزدیکترین همسایه ۲۰ تا ۱۰۰ برابر سرعت بیشتری نسبت به رویکردهای قبلی دارد. الگوریتم یک الگوریتم تقریبی تصادفی است که همیشه یک نزدیکترین همسایه دقیق باز گرداند اما یک نزدیکترین همسایه خوب باز می‌گرداند. این الگوریتم بر اساس جستجو در تصویر توسعه داده شده است اما الگوریتم تعمیم یافته است و می‌تواند حداقل به سیگنال‌های یک تا سه بعدی به کار برده شود.

برای سادگی ابتدا یک حالت خاص از الگوریتم که تنها یک نزدیکترین همسایه را با استفاده از فاصله $L^2$ بین patchهایی که تنها انتقال داده شده اند می‌یابد، ارائه می‌شود. سپس الگوریتم کامل تعمیم یافته که می‌تواند با هر چرخش یا مقیاسی تطابق را انجام دهد، k-نزدیکترین همسایه را پیدا کند و یک توصیف دلخواه را منطبق کند، ارائه خواهد شد [1].

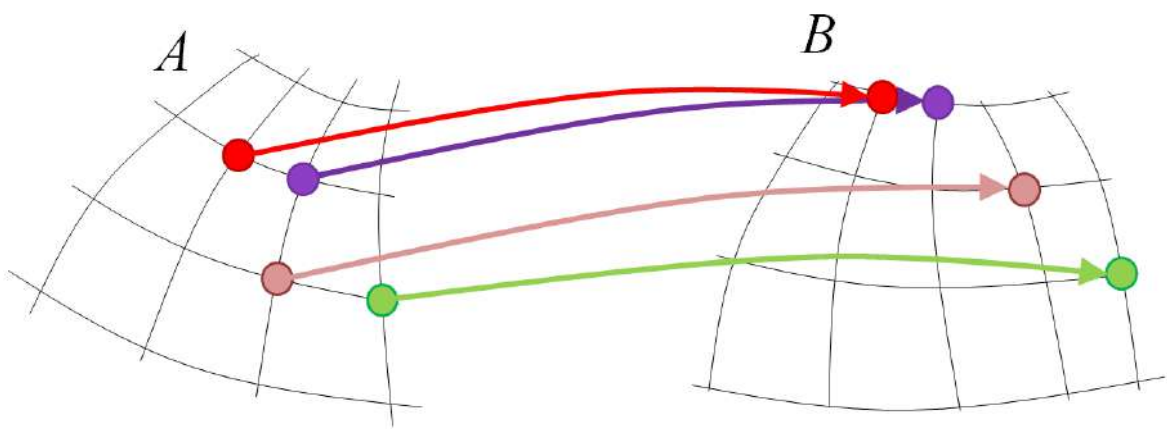

شکل ۳. ۱ – ایده سطح بالا پشت الگوریتم. توصیف گرها بر روی دو منیفولد محاسبه شده اند و به صورت دایره‌های رنگی نمایش داده شده اند. هر توصیف گر به صورت مستقل مشابه ترین را در منیفولد دیگر منطبق کرده است. وابستگی در اینجا با مشابه شدن توصیف گر قرمز به توصیف گر بنفش مشخص شده است. چراکه نزدیکترین همسایه‌ی آن‌ها به صورت مکانی نزدیک به هم هستند.

## ۳.۱ ایده سطح بالا

ایده سطح بالای پشت این الگوریتم در شکل ۳.۱ نمایش داده شده است. ما دو منیفولد [10] با توصیف گرهایی برای نقاط شبکه‌ای داریم که با دایره‌های رنگی مشخص شده اند. هدف یافتن مشابه ترین توصیف گر در تصویر دوم برای توصیف گرهای تصویر اول است. ما این کار را با بهره گیری از خصوصیت مکان محلی که در فصل قبل دیدیم، انجام خواهیم داد: زمانیکه ما یک تطابق خوب داشته باشیم، می‌توانیم این را به نقاط همسایه روی شبکه انتشار دهیم و اگر ما یک تطابق معقول داشته باشیم می‌توانیم این کار را با جستجوی تصادفی در اطراف مکان هدف بهبود ببخشیم. مرحله اول یا انتشار استفاده از این خصوصیت است که بسیاری از تطابق‌ها با هم وابستگی دارند یا یک مختصات تطابق نسبی یکسان دارند که در فصل قبلی بررسی شدند (شکل ۲.۱ و شکل ۲.۲). مرحله دوم یا جستجوی تصادفی یافتن تناظرات نسبی برای تناظر نقطه هدف فعلی با توجه به توزیع جستجو peaked مشابه توزیع‌های اندازه گیری شده در شکل ۲.۴ است.

در این فصل، الگوریتم را برای تصاویر دوبعدی که یک شبکه منظم بررسی خواهیم کرد. با این حال الگوریتم می‌تواند بر روی کانتورهای یک بعدی یا هندسه سه بعدی اعمال شود. بنابراین دو مرحله انتشار و جستجوی تصادفی می‌تواند برای هر فضای اقلیدسی یا نزدیک به آن تعمیم داده شود.

## ۳.۲ مقدمه

شیوه‌های اخیر برای درک بیشتر از تصویر و ویدیو در سطح بالا توسعه داده اند. مانند بازسازی تصویر که الگوریتم بهترین تشابه را از تصویر اصلی برای تغییر نسبت تصویر به کار میگیرد. یا الگوریتم‌های تکمیل تصویر که ناحیه‌های ناخواسته از عکس را حذف میکند. بسیاری از این شیوه‌های

---

۱۰. منیفولد یک فضای توپولوژیک است که به طور موضعی، اقلیدسی است. بدین معنی که حول هر نقطه، همسایگی موجود است به طوری که از نظر توپولوژیک مانند یک گوی واحد باز در فضای اقلیدسی می‌باشد، ولی از نظر ساختار کلّی می‌تواند از یک فضای اقلیدسی پیچیده‌تر باشد.

قدرتمند بر پایه patch هستند؛ یعنی تصویر را به بخش‌های کوچک مربعی با اندازه ثابت به نام patch تقسیم می‌کنند.

برای فهمیدن الگوریتم تطابق اجزای متداول الگوریتم‌های بر پایه patch ؛ یعنی شیوه نمونه گیری غیر پارامتری که یک جستجوی تکراری برای همه patchها در یک ناحیه برای یافتن ناحیه مشابه در تصویر دیگر است را در نظر می‌گیریم. به بیان دیگر با داشتن دو تصویر یا ناحیه، برای تمام patchهای ناحیه اول نزدیک‌ترین همسایه را در ناحیه دوم بر اساس فاصله $L^P$ می‌یابیم. این نگاشت را Nearest-Neighbor Field (NNF) یا نزدیک‌ترین همسایه درست می‌نامیم که به صورت شماتیک در شکل ۳. ۲ (د) نشان داده شده است. جستجو با رویکرد جامع (فراگیر) بسیار پر هزینه خواهد بود؛ یعنی $O(mM^2)$ به ترتیب برای ناحیه‌های تصویر و patchهایی با اندازه M و m تعداد پیکسل. حتی استفاده از روش‌های بهبود دهنده نظیر نزدیک‌ترین همسایه تقریبی [78] و کاهش ابعاد هنوز هم این بخش bottleneck شیوه نمونه گیری غیر پارامتری خواهد بود و از رسیدن به یک سرعت تعاملی جلوگیری می‌کند. علاوه بر این، این روش‌های بر پایه درخت حافظه به اندازه $O(M)$ یا بیشتر با ثابت‌های به نسبت بزرگی را نیاز دارند که برنامه‌ها را برای استفاده از تصاویر با وضوح بالا محدود می‌کنند.

برای محاسبه‌ی نزدیک‌ترین همسایه درست تقریبی الگوریتم ارائه شده در اینجا متکی به سه مشاهده کلیدی زیر خواهد بود:

**ابعاد فضا**: با اینکه ابعاد فضای patch بزرگ است (m بُعد)، خلوت است ( $O(M)$ تا patch ). بسیاری از شیوه‌های قبلی با صدمه زدن به ابعاد فضای patch با استفاده از ساختمان داده‌های درختی (به طور مثال kd-tree که جستو در آن با زمان $O(mM \log M)$ انجام می‌گیرد.) و شیوه‌های کاهش ابعاد (به طور مثال PCA) باعث افزایش سرعت جستجوی نزدیک‌ترین همسایه می‌شدند. در مقابل الگوریتم ارائه شده در اینجا کل فضای دوبعدی برای patchهای ممکن را بررسی کرده و سرعت بیشتر داشته و حافظه بهینه تری را نیاز دارد.

**ساختار واقعی تصاویر**: جستجوی مستقل معمول برای هر پیکسل بسیاری از ساختار واقعی در تصاویر را نادیده می‌گیرد. در الگوریتم patch-sampling synthesis ، خروجی شامل تکه‌های بزرگ

پیوسته داده‌ها از ورودی است (که در [3] دیده شده است). بنابراین می‌توان کارآیی را با انجام جستجو برای همسایه‌های هر پیکسل با رویکرد پیوسته بودن بهبود داد.

**قانون اعداد بزرگ:** در نهایت در حالیکه هر یک از انتخاب‌های تصادفی از patchهای مقدار دهی شده [11]، بسیار بعید است که حدس خوبی باشند، بخش کوچکی اما با اهمیت از یک بخش بزرگ از مقداردهی‌های تصادفی احتمال دارد که حدس خوبی باشند. اگر این بخش خیلی بزرگ‌تر شود، شانس اینکه هیچ patch ی مقدار دهی درستی نشده باشد، بسیار ناچیز می‌شود.

بر اساس این سه مشاهده یک الگوریتم تصادفی برای محاسبه نزدیکترین همسایه‌های درست تقریبی با استفاده از به روز رسانی‌ها افزایشی ارائه شده است (بخش ۳. ٤). الگوریتم با یک حدس اولیه آغاز می شود که بر اساس اطلاعات قبلی یا به سادگی با یک مقداردهی تصادفی مقداردهی می شود. فرآیند تکراری از دو بخش تشکیل شده است: انتشار که با استفاده از وابستگی باعث انتشار راه حل‌های خوب به پیکسل‌های مجاور می‌شوند؛ و جستجوی تصادفی که بردار انحراف‌های فعلی با استفاده از مقیاس‌های مختلف از انحراف‌های تصادفی جایگشت می‌یابد. پیاده سازی این الگوریتم بر روی CPU نشان داد که ۲۰ تا ۱۰۰ برابر در مقایسه با kd-tree با PCA افزایش سرعت داشته است. یک پیاده سازی GPU پیشنهاد داده شده است که تقریبا ۷ برابر سرعت بیشتری نسبت به نسخه پیاده سازی CPU برای تصاویر مشابه داشته است. الگوریتم ارائه شده در اینجا برخلاف الگوریتم‌های قبلی که یک ساختمان داده کمکی برای افزایش سرعت جستجو ایجاد می‌کردند، تنها کمی حافظه بیشتر فراتر از تصویر ورودی نیاز دارد. با تنظیمات معمول برای پارامترهای الگوریتم، زمان اجرا $O(mM \log M)$ و حافظه ی مورد نیاز $O(M)$ خواهد بود.

الگوریتم تعمیم سافته تطابق در بخش‌های ۳. ٥ و ۳. ۷ ارائه شده است که در حالیکه بسیاری از امکانات ارائه شده در کارهای قبلی یافتن نمی شوند، هنوز هم سادگی و سرعت خود را حفظ کرده اند. به طور مثال در بینایی ماشین، سختی انجام یک جستجوی ٤ بعدی در میان انتقال، چرخش و مقیاس‌ها قبلا با استفاده از خصوصیت‌های خلوت که به برخی از این تبدیلات نامتغیر هستند، ایده گرفته شده

---

اند. الگوریتم ارائه شده در اینجا، به طور موثری تناظرات متراکم را با وجود افزایش ابعاد می‌یابد. بنابراین یک روش جایگزین برای شیوه‌های نقاط مورد علاقه خلوت خواهد بود.

## ۳.۳  کارهای مرتبط

نمونه گیری بر پایه patch یک ابزار محبوب برای تحلیل و بررسی تصویر و ویدیو شده است. برنامه‌هایی برای تحلیل بافت، تکمیل برای تصویر و ویدیو، بازسازی و خلاصه سازی، بازترکیب و ویرایش تصویر، image stitching [12] ، حذف نویز و غیره. ما این برنامه‌ها را با جزئیات در فصل ۵ بررسی خواهیم کرد. در اینجا روش‌های جستجوی آن‌ها را به علاوه کارهای قبلی جستسجوی نزدیکترین همسایه بررسی خواهیم کرد.

شیوه‌های بر پایه Patch تا حد زیادی می‌توانند به شیوه‌های حریصانه و بهینه سازی تقسیم شوند. بهینه سازی می‌تواند برخی از تناقضات در طول فرآیند حریصانه را اصلاح کند. شیوه بهینه سازی نتایج با کیفیت بالایی تولید کرده است در حالیکه به سادگی با الگوریتم قابل استفاده است.

کیفیت شیوه بهینه سازی در قسمت پرهزینه تکرارهای جستجو زیادی می‌آید که به صورت واضحی bottleneck پیچیدگی برای همه‌ی این شیوه‌ها است. علاوه بر این در برنامه‌هایی نظیر تحلیل بافت، معمولا بافت مرجع یک تصویر کوچک است. در برنامه‌های دیگر نظیر تکمیل، بازسازی و بُرزنی، تصویر ورودی به صورت معمول بزرگ تر است. بنابراین هنوز هم مسئله جستجو بحرانی است.

انواع مختلفی از افزایش سرعت برای جستجوها پیشنهاد شده است. به طور کلی شامل ساختن داده‌های درختی مانند TSVQ [117] ، kd-tree‌ها [52, 61, 79, 119] و VP-tree‌ها که هر کدام از آن‌ها هم از جستجوهای دقیق و هم جستجوهای تقریبی (ANN) پشتیبانی می‌کنند. در برنامه‌های تحلیل گر جستجوهای تقریبی غالبا با عطف با روش‌های کاهش ابعاد مثل PCA [52, 61, 67] استفاده می‌شوند. به این خاطر که شیوه‌های ANN در ابعاد کم از نظر زمان و حافظه بسیار کارا هستند. در مقاله [3] یک روش انتشار محلی که وابستگی‌های محلی در فضای جستجو برای یک patch به مکان همسایه‌های نقطه مورد جستجو در بافت مرجع کاهش می‌دهد، معرفی می‌کند. مرحله انتشار در الگوریتم ارائه شده در اینجا نیز از فرض وابستگی یکسانی الهام گرفته است. روش k-وابستگی [109] ایده

---

۱۲. فرآیند ترکیب چند عکس در قالب تصاویر با یک بخش هم پوشانی برای تولید یک پانورامای قطعه قطعه شده یا یک تصویر با وضوح بالا را گویند.

انتشار را با مرحله پیش محاسبه در k نزدیکترین همسایه‌ی هر patch که ذخیره سازی شده است، ترکیب کرده است که جستجوهای بعدی از مزیت این مجموعه از قبل محاسبه شده استفاده می‌کنند. با اینکه این کار فاز جستجو را سرعت می‌بخشد، اما هنوز k-وابستگی نیاز به جستجوی کل نزدیکترین همسایه‌ها برای همه‌ی پیکسل‌های ورودی دارد که این در زمینه تحلیل بافت نشان داده شده است. این روش فرض می‌کند مقداردهی اولیه آنقدر نزدیک به مقدار واقعی هست که برای جستجوی نزدیکترین همسایه به تعداد کمتر کافی باشد. این برای یک ورودی کوچک از بافت‌ها درست است، اما برای تصاویر بزرگ پیچیده، نیاز است تا از جواب‌های محلی جلوگیری کنیم. در این مطلب سعی می‌کنیم سرعت و حافظه مورد استفاده الگوریتم ارائه شده را با kd-tree با کاهش ابعاد مقایسه کنیم. همان طور نشان خواهیم داد که به اندازه حداقل یک مرتبه بزرگی از بهترین ترکیب‌های رقیب (یعنی ANN+PCA) پرسرعت است و به طور قابل توجهی از حافظه کمتری استفاده می‌کند. همچنین این الگوریتم بسیار تعمیم یافته تر از kd-treeها است. چراکه این الگوریتم می‌تواند با معیارهای فاصله دلخواه اعمال شده و به سادگی برای تعامل‌های محلی مثل ایجاد محدودیت‌ها توسط کاربر اصلاح شود.

تعدادی از تحقیقات از این الگوریتم استفاده کرده اند. کاربردهایی در زمینه بینایی ماشین [10] و خلاصه سازی ویدیو [9] که در ادامه آمده است. محققان دیگری از این الگوریتم برای یافتن تکرارها در تصاویر [23]، تحلیل تصاویر ترکیبی [90]، افزایش تصاویر با استفاده از تناظرات [47]، شکل‌گیری بین

تصاویر مختلف [96]، تشخیص حالات افراد با مقایسه تصاویر [53]، کاهش نویز ویدیو [70]، تخمین حرکت [16] و بافت با اعداد [85] از این الگوریتم استفاده کرده اند.

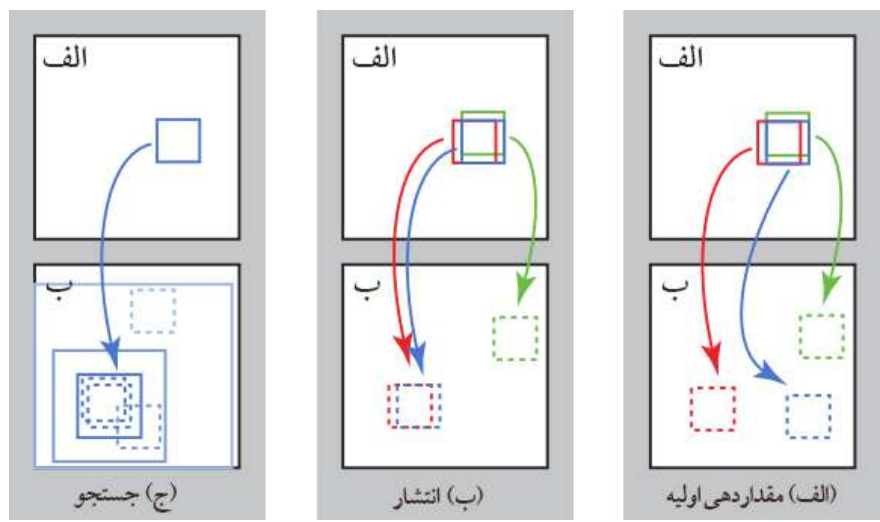

شکل ۳. ۲ – فازهای الگوریتم تصادفی نزدیکترین همسایه: (الف) patchها در ابتدا یک مقداردهی تصادفی دارند. (ب) patch آبی رنگ، همسایه‌های بالا/ سبز و چپ / قرمز را برای بهبود دادن نگاشت بررسی می‌کند و تطابق‌های خوب را انتشار می‌دهد. (ج) patch به صورت تصادفی برای بهبود در یک بخش محلی متحدالمرکز جستجو می‌شود.

## ۳. ٤   یافتن تطابق در انتقال

هسـته اصـلی یک الگوریتم برای محاسبه تناظرات بین patchهاست. نزدیکترین همسـایه درسـت (NNF) در اینجا به صـورت یک تابع $f: A \mapsto \mathbb{R}^2$ بر روی تمامی مختصـات patchها [۱۳] در تصـویر اول برای یک تابع فاصله دو جفت ptach (که با D نشان میدهیم) تعریف می‌شود. با داشتن مختصات a در تصـویر A و تناظر نزدیکترین همسـایه b در تصـویر B، $f(a)$ ناحیه b را تشکیل می‌دهد. مقدار

---

۱۳. منظور محل قرارگیری هر patch در تصویر است.

۲٤

خروجی تابع f را نزدیکترین همسایه گوییم که این مقادیر خروجی در آرایه‌ای به اندازه ابعاد A ذخیره خواهد شد.

در این بخش الگوریتم تصادفی برای محاسبه نزدیکترین همسایه‌های درست تقریبی ارائه خواهد شد. همانطور که قبلا اشاره کردیم، ایده اصلی الگوریتم جستجو در فضای ممکن مختصات، جستجو در همسایه‌های یک ناحیه است. همچنین این نکته حائز اهمیت است که در یک تصویر بزرگ ممکن است حدس‌های اولیه زیادی صحیح باشند.

الگوریتم سه بخش اصلی دارد که در تصویر ۳.۲ نمایش داده شده است. در ابتدا آرایه نزدیکترین همسایه‌ها با فرض‌های قبلی یا به صورت تصادفی مقداردهی می‌شود. سپس یک فرآیند به روزرسانی تکراری اعمال می‌شود که نزدیکترین همسایه‌های خوب به پیکسل‌های مجاور انتشار می‌یابند. در ادامه این کار با جستجوهای تصادفی در همسایه‌های بهترین نزدیکترین همسایه برای یافتن یک تناظر بهتر ادامه می‌یابد. بخش ۳.۴.۱ و ۳.۴.۲ این گام‌ها را با جزئیات بیشتر توضیح می‌دهند.

### ۳.۴.۱   مقداردهی اولیه

نزدیکترین همسایه می‌تواند با یک مقدار تصادفی یا یک فرض قبلی مقداردهی اولیه شود. برای مقداردهی با مقدار تصادفی، از تعدادی نمونه تصادفی در تصویر دوم به صورت یکنواخت استفاده می‌کنیم. در برخی کاربردها از یک فرآیند تغییر اندازه تدریجی از اندازه بزرگ تا کوچک استفاده خواهیم کرد. در این حالت، می‌توان از حدس اولیه قبلی به مقادیر سطح بالا رسید. با این حال، اگر این حدس اولیه انتخاب شود، ممکن است گاهی اوقات الگوریتم به بهینه محلی همگرا شود. برای حفظ کیفیت این مقدار اما با حفظ قابلیت فرار از مقادیر بهینه محلی، تکرارهای اولیه الگوریتم با مقداردهی‌های تصادفی انجام شده و سپس با مقداردهی‌های اولیه سطح بالا تنها در ناحیه همان patch ادغام می‌شود و سپس باقیمانده تکرارها انجام می‌شود. از آنجایی که مرحله تکراری بسیار حساس به نتایج تکرارهای قبلی است، بنابراین این کار برای فرار از بهینه‌های محلی انجام می‌شود.

### ۳.۴.۲   تکرار

بعد از مقداردهی اولیه، یک فرآیند تکراری برای بهبود NNF انجام می‌شود. هر تکرار به صورت زیر است: هر نزدیکترین همسایه به صورت چپ به راست و از بالا به پایین مورد بررسی قرار می‌گیرند، و

۲۵

بعد از هر مرحله انتشار یک مرحله جستجوی تصادفی انجام می شود. اینکار در سطح هر patch انجام می‌شود. اگر $P_j$ و $S_j$ به ترتیب انتشار و جستجوی تصادفی در j امین patch باشند، بنابراین به این ترتیب به پیش می‌رویم: $P_1.S_1.P_2.S_2.\ldots.P_n.S_n$ .

**انتشار :** با فرض اینکه مختصات patchها، به جز انتقال‌های نسبی برای یک پیکسل به راست یا پایین، به احتمال زیاد به هم شبیه هستند، هدف بهتر کردن $f(x.y)$ با استفاده از نزدیکترین همسایه $f(x-1.y)$ و $f(x.y-1)$ است. به عنوان مثال، اگر یک نگاشت خوب در $(x-1.y)$ داشته باشیم، سعی داریم تا از یک نگاشت به راست برای نگاشت $(x.y)$ استفاده کنیم. با قراردادن $z = (x.y)$ ، یک جواب احتمالی برای $f(z)$ ، $f(z-\Delta_p)+\Delta_p$ است که در آن $\Delta_p$ مقادیر $(1.0)$ و $(0.1)$ را اختیار می‌کند.

نتیجه آنچه انجام شد نشان می‌دهد که اگر $(x.y)$ یک نگاشت صحیح داشته باشد و در یک ناحیه R وابسته با شد، بنابراین تمام Rهای پایین و سمت راست $(x.y)$ با مقادیر صحیح نگاشت تکمیل شده اند. علاوه بر این در تکرارهای فرد، اطلاعات به بالا و سمت چپ با بررسی patchها به صورت برعکس و استفاده از جواب‌های احتمالی پایین و به سمت راست، انتشار می‌یابند. انتشار به سرعت همگرا می‌شود، اما اگر به تنهایی مورد استفاده قرار گیرد، با یک بهینه محلی خاتمه می‌یابد. بنابراین بخش دوم به جستجوهای تصادفی اختصاص دارد.

**جستجو تصادفی :** یک توالی از جواب‌های احتمالی از توزیع نمایی نمونه برداری می‌شود و در صورتیکه هر کدام از این نمونه‌ها فاصله D کمتری داشت ، نزدیکترین همسایه فعلی به روز رسانی می شود. اگر $v_0$ نزدیکترین همسایه فعلی برای $f(z)$ با شد، هدف تلاش برای بهبود $f(z)$ با آزمایش توالی‌های مختلف جواب‌های احتمالی نگاشت شده در یک فاصله کاهش نمایی از $v_0$ است:

$$u_i = v_0 + w\,\alpha^i R_i \qquad (3.1)$$

که در آن $R_i$ یک عدد تصادفی یکنواخت در $[-1.1] \times [-1.1]$ ، $w$ شعاع بیشینه جستجو ، و $\alpha$ یک نسبت ثابت بین اندازه پنجره‌های جستجو است. Patchها برای $i = 0.1.2\ldots$ تا زمانیکه $w\,\alpha^i$ به کمتر از یک پیکسل کاهش یابد، بررسی می شوند. در اینجا $w$ بیشترین ابعاد تصویر است و $\alpha = 1/2$ است.

**معیارهای توقف :** از آنجایی که می‌توان معیارهای مختلفی برای توقف بر اساس کاربرد برای توقف را به کار برد، اما در عمل تکرار بر اساس تعداد مشخصی از تکرار بهترین نتیجه را دا شت. تمام نتایج

۲٦

آورده شده در اینجا بر اساس ٤ یا ٥ تکرار الگوریتم هستند، پس از آنکه NNF تقریبا همیشه همگرا شده است. همگرایی در شکل ٣.٣ نمایش داده شده است.

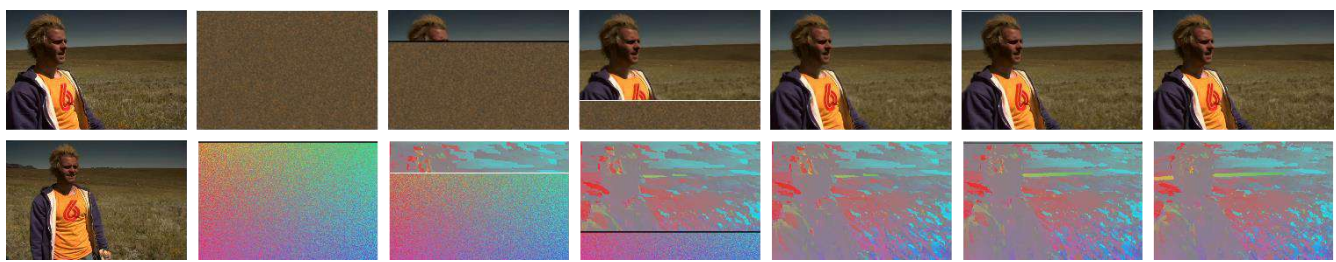

(و) تکرار پنجم   (ی) تکرار دوم   (هـ) تکرار اول   (د) $\frac{3}{4}$ تکرار اول   (ج) $\frac{1}{4}$ تکرار اول   (ب) مقدار تصادفی   (الف) تصویر اصلی

شکل ٣.٣ – همگرایی؛ (الف) تصویر بالا تنها با استفاده از patchهایی از تصویر پایین بازسازی شده است. (ب) بالا: بازسازی با استفاده از رای دهی patchها [14]، پایین: یک مقداردهی تصادفی، که زاویه با رنگ و مقدار با اشباع رنگ نمایش داده شده است. (ج و د) نمایش بخشی از مسیر برای انجام تکرار اول، مقادیر انحراف در بالای نوار افقی پویش انتشار یافته اند. (هـ) پایان تکرار اول. (ی) تکرار دوم. (و) بعد از انجام پنج تکرار، تقریبا تمام patchها تغییر نکرده اند.

**کارآیی** : کارآیی این روش ساده به روش‌های مختلفی قابل بهبود است. در فاز انتشار و جستجوی تصادفی، زمانیکه در حال بهبود $f(z)$ با یک جواب احتمالی $u$ هستیم، یک توقف زودهنگام زمانیکه حاصل جمع جزئی برای فاصله بیش از فاصله فعلی باشد، قابل انجام است. همچنین در مرحله انتشار، زمانیکه از patchهای مربعی به طول ضلع $p$ و نُرم $L_p$ استفاده می‌کنیم، تغییر در فاصله می‌تواند با توجه به مقدار زائد مربوط به جمع بر روی ناحیه‌های دارای همپوشانی، به صورت افزایشی با زمان $O(p)$ به جای $O(p^2)$ محاسبه شود. با این حال اینکار میزان حافظه بیشتری برای فاصله‌های فعلی به اندازه $D(f(x,y))$ را متحمل خواهد شد.

## ٣. ٥ یافتن تطابق در چرخش و مقیاس‌های مختلف

در برخی برنامه‌ها، نظیر تشخیص اشیا در بینایی کامپیوتر، مطلوب است تا patchها در سراسر طیف وسیعی از چرخش‌ها یا مقیاس‌ها تطابق یابند. برای این کار بدون از دست دادن کلیت مسئله، یک

---

١٤. رای گیری با نگاه کردن به رنگ نزدیکترین همسایه و میانگین گیری بر روی تمام patch هایی که هم پوشانی دارد، انجام می‌گیرد.

patch از تصویر اول بدون تغییر اندازه یا چرخش با یک patch از تصویر دوم که بر اساس مرکز، چرخش یا تغییر اندازه یافته است، تطابق می‌یابد.

برای جستجو در بازه چرخش $\theta \in [\theta_1, \theta_2]$ و بازه مقیاس $s \in [s_1, s_2]$، بازه‌ی جستجو الگوریتم اصلی را از $(x, y)$ به $(x, y, \theta, s)$ تغییر می‌دهیم. بنابراین تعریف نزدیکترین همسایه صحیح به شکل نگاشت $f: \mathbb{R}^2 \mapsto \mathbb{R}^4$ تغییر خواهد کرد. در اینجا $f$ با نمونه برداری یکنواخت در بازه مکان‌ها، جهت‌ها و مقیاس‌های مختلف مقداردهی اولیه خواهد شد. در فاز انتشار patchهای همسایه تنها با یک انتقال ساده ارتباط نخواهند داشت. بنابراین باید همسایگان با ماتریس ژاکوبی تبدیل شوند. اگر $T(f(x))$ یک تبدیل کامل به شکل $(x, y, \theta, s)$ باشد، جواب‌های احتمالی به شکل $f(x - \Delta_P) + T'(f(x - \Delta_P))\Delta_P$ خواهند بود. در فاز جستجوی تصادفی هم از یک پنجره با سایز کاهشی در اطراف ناحیه استفاده خواهیم کرد.

همگرایی این رویکرد در شکل ۳.۴ نمایش داده شده است. با توجه به جستجو در ٤ بعد به جای یک بعد، ترکیب انتشار و جستجوی تصادفی به طور موفقیت آمیزی از فضای جستجو نمونه برداری می‌کند و به طور کارایی تطابق‌های خوبی در میان patchها انتشار می‌دهد. در مقابل، با درخت kd، جستجو در

سـراسـر مقیاس‌ها و چرخش‌ها نا بدیهی اسـت. همه ی چرخش‌ها و مقیاس پذیری‌ها یا باید به درخت اضافه شوند یا درخواستی دیگر، که منجر به هزینه‌های زمان و حافظه می‌شود.

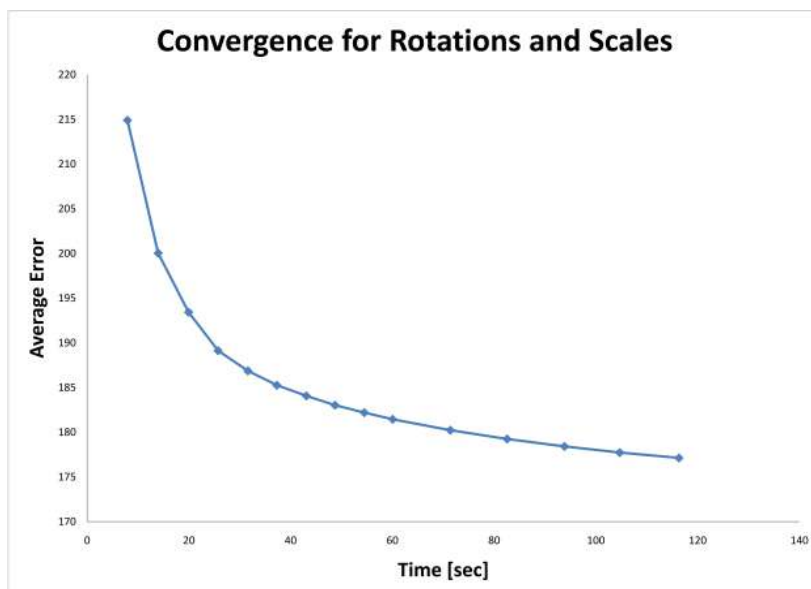

شکل ۳. ٤ – جستجو در میان چرخش‌ها و مقیاس‌ها. این میانگینی در میان مجموعه داده‌ای تصاویر ۰٫۳ مگاپیکسلی است. هم زمان در تکرار و هم تکرارهای مورد نیاز در همگرایی نسبت به الگوریتم انتقال تنها بالاتر هستند. زمان در تکرار اصولا بخاطر فیلترینگ دوسویه استفاده شده در محاسبه فاصله تکه بیشتر است: اگر فیلترینگ نزدیکترین همسایه استفاده شود هر تکرار چندین برابر سریع تر می‌شود اما فاصله تکه با توجه به نمونه برداری اندکی بالاتر می‌رود.

یادآوری می‌کنیم که زمان در تکرار در شـکل ۳. ٤ نسـبت به الگوریتم انتقال تنها بالا تر اسـت. این ا صولا بخاطر فیلتر دو سویه‌ای ا ست که برای ج ستجوی الگو در تکه چرخانده شـده/تغییر مقیاس داده شده ، استفاده شده است. متناوبا، فیلترینگ نزدیکترین همسایه می‌تواند استفاده شود که درآن هر تکرار چندین برابر سریعتر می شود اما می‌توان نمونه برداری را در موقعیت بهترین نزدیکترین همسایه تعریف کرد. این شـاید یا شـاید نه مسـئله‌ای وابسـته به برنامه کاربردی باشـد. یک جایگزین سـوم برای نمونه برداری تصویر هدف و سپس نمونه برداری با کمک فیلتر نزدیک ترین همسایه است. این سرعت بالا و

نیز برخی افزایش دقت‌ها را به دنبال دارد. ما برای پژوهش‌های آتی پیشنهاد می‌کنیم که می‌توان الگوریتم‌های GPU را با توجه به فیلترینگ دوسویه پیاده سازی شده در سخت افزار کارآمد تر کرد.

### ۳.۶ تطابق توصیف گره‌های دلخواه

الگوریتم PatchMatch اساسا با استفاده از جمع مجذور اختلاف فاصله تکه پیاده سازی شده است اما هیچ نیازمندی صریحی برای تابع فاصله بیان نمی کند. تنها فرض صریح این است که تکه‌هایی با مجاورت مکانی نزدیک باید با احتمال بیشتری بهترین نزدیک ترین همسایه را داشته باشند به طوری که PatchMatch بتواند در انتشار بهترین نزدیک ترین همسایه‌ها و یافتن یکی از آنها موثر باشد. به نظر می‌رسد این برای انواع مختلفی از توابع فاصله و توصیف گره‌ها درست باشد. در واقع، الگوریتم می‌تواند حتی به سرعت استفاده از توصیف گره‌های پشتیبانی کننده از محدوده وسیعی را پوشش دهد که که تکه‌های تصویر کوچک کار می‌کند زیرا تمایل به تنوع نسبی در تصویر دارد. درکل، "تابع فاصله" می‌تواند هر نوع الگوریتمی باشد که یه ترتیب کلی را فراهم می‌آورد که آن نیازی به تبعیت از ویژگی‌های تابع فاصله متریک ریاضی معمولی ندارد. برای مثال، تفارن می‌تواند شکسته شود: "فاصله" از تکه مرجع الف تا تکه هدف الف می‌تواند تغییر کند اگر تصاوری تغییر کنند یا می‌توان از معادله مثلثی تبعیت نکرد[۱۵]. بنابراین، تطابق پذیری ما بسیار انعطاف پذیر است. در فصل ۷، مثال‌های زیادی در بینایی کامپیوتری مطرح می‌کنیم، همچون patch‌هایی که تغییرات روشنایی را جبران کرده و تطبیق توصیف گره‌های SIFT را انجام می‌دهند.

### ۳.۷ تطبیق k نزدیکترین همسایه

برای برنامه‌های خاصی مانند حذف نویز با میانگین گیری تکه‌های مشابه، ما امیدواریم در هر نقطه $(x, y)$ بیش از یک نزدیکترین هم‌سایه را بیابیم. این می‌تواند با جمع آوری k نزدیکترین هم‌سایه برای هر تکه انجام پذیرد. بنابراین NNF، $f_k$ یک نگاشت چند مقداری، با k مقدار است. NNF می‌تواند به صورت آرایه‌ای با کانال‌های A و k ذخیره گردد. تغییرات ممکن زیادی در PatchMatch برای محاسبه k-NN وجود دارند. ما کارایی برخی از آنها را با یک رویکرد استاندارد مقایسه کردیم: کاهش بعد با

---

۱۵. تطابع می‌تواند در میان تصاویر کاملا متفاوت انجام شود – نرخ همگرایی تنها به اندازه نواحی منطبق بستگی دارد.

۳۰

PCA، با تبعیت از ساختار درخت kd [78] با افکندن همه‌ی تکه‌های تصویر B بر مبنای PCA، سپس جستجوی ε نزدیکترین همسایه در درخت kd برای هر تکه از تصویر A بر همان مبنا.

از آنجایی که هریک از این الگوریتم‌ها می‌توانند برای دقت بالاتر و سرعت بالاتر نیز بهبود یابند، ما هر یک را با برخی تنظیمات ارزیابی کردیم. برای PatchMatch، ما به سادگی تکرارهای اضافی را محاسبه کردیم و برای درخت kd پارامترهای PCA و ε را تنظیم کردیم. کارایی نسبی این الگوریتم‌ها در شکل ۳. ۵ رسم شده است. ما همچنین آن را با FLANN، بسته‌ای که شامل درخت kd، درخت معنایی k، یک الگوریتم ترکیبی و تعداد زیادی پارامتر که می‌توانند در کارایی موثر باشند می‌شود مقایسه کردیم.

**الگوریتم پشته‌ای:** در ساده‌ترین نوع، ما k نزدیک‌ترین همسایه با موقعیت هر تکه جمع‌آوری کردیم. در طی انتشار، نزدیک‌ترین همسایه‌ها در موقعیت فعلی را به طور مستمر با هر k نزدیک‌ترین همسایه با سمت چپی یا بالایی (یا پایینی یا سمت راستی در هر تکرار) مقایسه می‌کنیم. کاندیداهای جدید $f_i(x - \Delta_p) + \Delta_p$ هستند که $\Delta_p$ مقدار (۰و۱) و (۱و۰) می‌گیرد و $i = 1,2,...,k$ است. اگر هر کاندید نسبت به بدترین کاندید ذخیره شده در x نزدیک‌تر است، آن بدترین کاندید با کاندیدی از patch همسایه جایگزین می‌شود. این کار می‌تواند به طور موثری با یک پشته بیشینه انجام شود که پشته فاصله تکه D را ذخیره می‌کند. فاز جستجوی تصادفی نیز به طرز مشابهی کار می‌کند: n نمونه پیرامون هر k نزدیک‌ترین همسایه برداشته شد که در کل nk نمونه است. بدترین عنصر از پشته خارج می‌شود اگر فاصله کاندیدا بهتر باشد. در هنگام آزمون کاندیداها، ما یک جدول Hash نیز برای شناسایی سریع کاندیداهای قبلی لیست k، و برای اجتناب از اضافه کردن‌های تکراری ایجاد می‌کنیم. k نزدیک‌ترین همسایه در پشته به صورت آرایه مستطیلی f ذخیره می‌شود. برای کاهش استفاده از حافظه، جدول Hash را برای زمانی نیازی به بازخوانی هر patch از حافظه داریم بازسازی می‌کنیم.

**الگوریتم پشته‌ای با k نمونه‌گیری خلوت:** k همسایه اغلب انحرافات خوشه‌ای دارند زیرا تصاویر طبیعی وابستگی‌های زیادی در خود دارند. بنابراین، ممکن است تعجب برانگیز باشد که آیا جستجوی جامع در میان همه‌ی k داوطلب در فاز جستجوی تصادفی و انتشار واقعا ضروری است، یا آیا می‌توانیم k داوطلب را با پراکندگی بیشتری نمونه‌گیری کنیم. برای انتشار، ما تنها می‌توانیم انحراف همسایه را با کوچکترین فاصله ("بهترین p" در شکل ۳. ۵) مقایسه کنیم یا یک عنصر تصادفی به عنوان داوطلب برای انتشار (به صورت "p تصادفی" نشان داده می‌شود) انتخاب می‌کنیم. همچنین، در جستجوی

۳۱

تصادفی، می‌توان به طور تصادفی پیرامون انحراف با کوچکترین فاصله (به صورت "بهترین RS" نشان داده می‌شود" نمونه گیری کرد یا پیرامون یک انحراف دلخواه تصادفی (با "RS تصادفی" نشان داده می‌شود) نمونه گیری کرد. نهایتا، می‌توان انتشار یا جستجوی تصادفی را روی n انحراف انجام داد که m به طور تصادفی به صورت یکنواخت بین ۱ و k انتخاب می‌شود (با "P متغیر" و "RS متغیر" نشان داده می شود). گرچه آنها تعداد عملیات در هر تکرار را کاهش می‌دهند، همه ی این استراتژی‌ها برای نسبت به الگوریتم هیپ طبیعی همگرایی آهسته تری دارند که در شکل ۳. ۵ نشان داده شده است.

**استفاده از الگوریتم ۱-NN برای یافتن k-NN:** مشاهده دیگر آن است که کاندیداهای مختلفی در دوره الگوریتم PatchMatch 1-NN اصلی ملاحظه شده اند. بنابراین، در استراتژی دیگر، ما همه ی انحرافات داوطلبان را به کمک الگوریتم برای هر پیکسل به دست می‌آوریم، و سپس لیست را برای یافتن k بالایی (با "لیست ۱-NN" نشان داده می‌شود) مرتب می‌کنیم. به طور مشابه، می‌توانیم الگوریتم ۱-NN را k بار اجرا کنیم، هر اجرا در صورتی محدود می‌شود که انحرافات برابر با هر یک از انحرافات انتخابی قبلی (با "k دفعه اجرای ۱-NN" نشان داده می‌شود) باشند. همانند استراتژی پشته خلوت، محاسبه تکرار سریع انجام می‌شود اما نسبت به الگوریتم پشته اصلی کند تر همگرا می‌شود.

**تغییر k طی تکرارهای متعدد:** می‌تواند تکرار ۱ را با تعدادی از نزدیک ترین همسایگان $k_0$ آغاز نمود و پس از آنکه نیمی از تکرارها انجام شدند، آن را به تعداد نهایی مطلوبی از نزدیکترین همسایه‌های k، یا با حذف بدترین عنصر از پشته یا با افزودن عناصر تصادفی یکنواخت مورد نیاز، افزایش یا کاهش داد. افزایش تعداد کمی از نزدیکترین هم‌سایگان $k_0 = k/2$ (با "k افزایش" نشان می‌دهیم)، و کاهش تعداد زیادی از نزدیک ترین همسایگان $k_0 = 2k$ (با "k کاهش" نشان می‌دهیم) به باعث شد تا دوباره این الگوریتم‌ها نسبت به پشته ساده کندتر عمل کند.

برخی از این الگوریتم‌ها، تکرارهای تنها را نسبت به الگوریتم پشته پایه توصیف شده در بالا سریعتر انجام می‌دهند، اما همگرایی آهسته تر انجام می‌شود زیرا آنها در یک تکرار اطلاعات کمتری را منتشر

می‌کنند. درکل، الگوریتم پشته اصلی انتخاب خوبی برای محدوده وسیعی از منحنی سرعت/کیفیت هست.

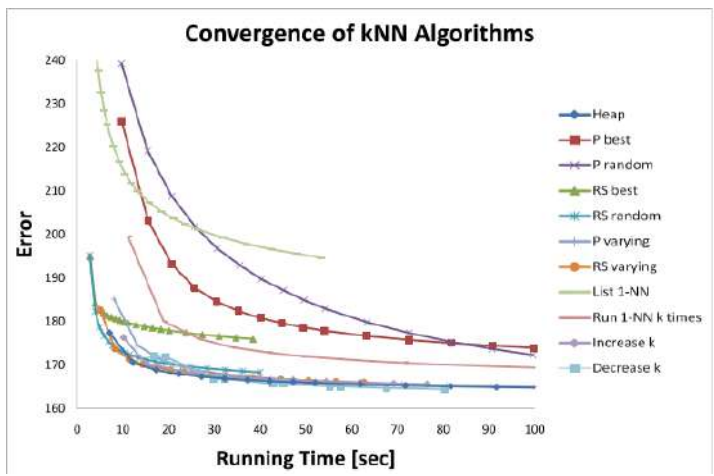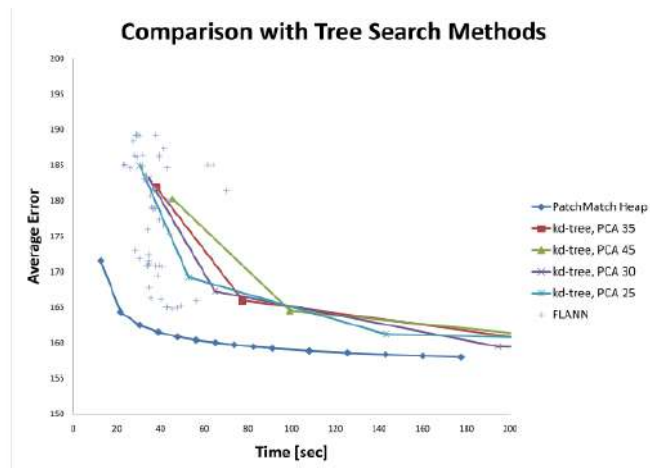

شکل ۳. ۵ – چپ: عملکرد انواع k-PatchMatch، با $k = 16$ نزدیکترین همسایه، در همه‌ی تصاویر در شکل ۷. ۲ میانگین گیری شده، به ۰٫۲ مگاپیکسل تغییر یافته و با آنها تطبیق یافت. خطا متوسط فاصله تکه $L_2$ در سراسر k تطبیق ، برای تکه‌های ۷ در ۷ است. نقاط روی هر منحنی پیشرفتی را در هر تکرار نشان می‌دهد. راست: مقایسه با درخت kd و FLANN، در ۰٫٤ مگاپیکسل، در همان مجموعه داده این میانگین گیری شدند. مجددا تطبیق برای $k = 16$ نزدیک ترین همسایه است. الگوریتم PatchMatch از روشهای درختی کارا تر است.

الگوریتم پشته پایه در محدوده وسیعی از k و اندازه‌های تصویر مختلف کاراتر است؛ برای مثال: الگوریتم برای $k = 16$ و تصاویر ورودی ۰٫۱ تا ۰/۱ مگاپیکسل، چندین برابر سریعتر از درخت kd باشد. در مقایسه با پیاده سازی درخت kd ی Mount و Arya [78] و FLANN [79]، این الگوریتم در تغییر همه‌ی پارامترهای ممکن، در هنگام تنظیم تنها تعداد تکرارهای الگوریتم پشته کنترل شد. FLANN الگوریتم‌های متعددی را پیشنهاد می‌کند، بنابراین ما محدوده بزرگی از پارامترها را که در شکل ۵-۳ با علامت + نشان داده شده است نمونه برداری کردیم. FLANN می‌تواند به طور خودکار پارامترها را بهینه سازی کند اما دریافتیم که عملکرد حاصل همواره به نمونه گیری نقطه‌ای ما وابسته است. در هردو مورد، این تنظیم پارامتر گسترده در عملکرد رویکرد الگوریتم پشته‌ای که ارائه کردیم –نه فراتر از آن– تاثیر میگذارد. بنابراین، پیشنهاد می‌کنیم که الگوریتم پشته‌ای k-PatchMatch کلی انتخاب

۳۳

بهتری برای طبقه وسیعی از مسائل خواسته شده در تناظر تکه تصویر است. با بهینه سازی بیشتر الگوریتم ما، فاصله جواب‌های یافته شده کمتر شده و عملکردی بیشتر خواهد شد [4].

## ۳. ۸   تحلیل همگرایی

در این بخش، ما همگرایی الگوریتم NN-۱ خود را برای تطابق میان انتقالات، برای نمونه‌های ترکیبی و واقعی تحلیل می‌کنیم. انتقال الگوریتمی است که تحلیل آن نسبت به کلی سازی‌های توصیف شده در بخش‌های ۳-۵ تا ۳-۷ توصیف شده اند.

### ۳٬۸٫۱   تحلیل یک نمونه ترکیبی

الگوریتم PatchMatch با حد گیری به NNF دقیق همگرا می شود. در اینجا یک تحلیل تئوری برای این همگرایی پیشنهاد می‌کنیم. تحلیل ما در اینجا تنها به الگوریتم تنها نزدیکترین همسایه تک انتقالی اعمال می‌شود. نشان می‌دهیم که با احتمال بالایی به سرعت در اندکی ازتکرارهای اولیه همگرا می‌شود. به علاوه، نشان می‌دهیم که در موارد اینچنینی که تنها تطابق‌های تقریبی مورد نیاز است، الگوریتم حتی سریع‌تر عمگرا می شود. . بنابراین الگوریتم ما بهتر ا ست به عنوان یک الگوریتم تخمینی با محدود کردن محاسبات به تعداد کمی از تکرارها استفاده شود. ما با تحلیل همگرایی به حوزه دقیق نزدیک ترین همسایه شروع کردیم و سپس این تحلیل را به مورد سودمندتری از همگرایی به راه حل تقریبی گسترش دادیم. فرض کنید الف و ب اندازه یکسانی (M پیکسل) دارند و مقداردهی اولیه تصادفی انجام شده ا ست. گرچه احتمالات هر موقعیت که در فرض اولیه با بهترین انحراف تخصیص یافتند بسیار کوچک (۱/M) هستند، احتمال حداقل یک انحراف که به درستی تخصیصی یافته باشد بسیار خوب ($1 - \left(1 - \frac{1}{M}\right)^M$) یا برای مقادیر بزرگ M تقریبا $1 - 1/e$ هستند. چون جستجوی تصادفی در نواحی محلی کوچک بسیار متراکم است می‌توانیم اختصاص "درستی" برای هر تخصیص در همسایگی کوچکی از پیکسل‌های به اندازه C در پیرامون انحراف صحیح داشته باشیم. این انحرافات در تقریبا یک تکرار از جستجوی تـ صادفی  صحیح خواهند بود. احتمالاتی که در آنها حداقل یک انحراف تخـ صیص

یافته یک همسایگی اینچنینی عالی هستند: $\left(1 - \left(1 - \frac{C}{M}\right)^M\right)$ یا برای مقادیر بزرگ M،
$1 - \exp(-C)$.

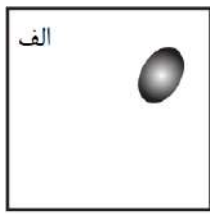

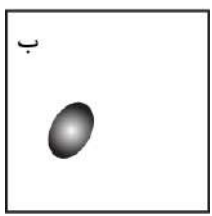

اکنون یک مورد آزمون ترکیبی چالش برانگیز برای الگوریتم خود در نظر می‌گیریم: یک ناحیه مجزای R با سایز m پیکسل در دو موقعیت مختلف در دو تصویر الف و ب واقع شده اند (در شکل نشان داده شده). این تصویر نمونه پیچیده‌ای است زیرا هیچ اطلاعاتی در مورد جایی که انحرافات نواحی مجزا ممکن است یافت شوند ارائه نمی شود. تکه‌ها در پس زمینه‌ای یکنواخت می‌توانند تعداد زیادی از تکه‌های مشابه دیگر را تطبیق دهند، که با تخمین‌های تصادفی در یک تکرار با احتمال بالایی یافت شده اند، بنابراین همگرایی را تنها برای ناحیه R بررسی می‌کنیم. اگر تنها یک انحراف در ناحیه مجزای R در همسایگی C از انحراف درست باشد آنگاه فرض می‌کنیم که پس از تعداد اندکی از تکرارها، با توجه به تراکم جستجوی تصادفی در نواحی محلی کوچک (که قبلا بیان شد)، همه ی R با انتشار (برای سادگی فرض کنید این لحظه‌ای است) صحیح خواهد بود. اکنون فرض کنید R هنوز همگرایی ندارد. جستجوهای تصادفی انجام شده توسط الگوریتم ما در بیشینه مقیاس w را در نظر بگیرید. تکرارهای جستجوی تصادفی در مقیاس w به طور مستقل تصویر ب نمونه برداری می‌کنند و احتمال p که هرکدام از این نمونه‌ها در همسایگی C از انحراف صحیح قرار می‌گیرند برابر است با:

$$p = 1 - (1 - C/M)^m \qquad (2.3)$$

قبل از انجام هر تکراری، احتمال همگرایی $p$ است. احتمال اینکه در تکرارهای 0, 1, ..., t-1 همگرایی انجام نشود و در تکرار t همگرا شود برابر $p(1-p)^t$ است. بنابراین احتمالات، توزیع هندسی هستند و زمان مورد انتظار $t = 1/p - 1$ است. برای سادگی، اندازه نسبی را $\gamma = m/M$ بگیرید، سپس همچنان که تفکیک پذیری M بزرگ می‌شود حد بگیرید:

$$t = [1 - (1 - C/M)^{\gamma m}]^{-1} - 1 \qquad (3.3)$$

$$\lim_{M \to \infty} t = [1 - \exp(-C\gamma)]^{-1} - 1 \qquad (4.3)$$

با بسط تیلور برای $\gamma$، می‌شود. که تعداد تکرارهای مورد انتظار ما در همگرایی برای تفکیک تصویر بزرگ و اندازه کوچک m نسبت به تفکیک پذیری تصویر M ثابت باقی می‌ماند. شبیه سازی‌ها برای تصاویر با تفکیک پذیری M از ۰٫۱ تا ۲٫۰ مگاپیکسل انجام شده است که این مدل را تایید می‌کند. برای

۳۵

مثال، برای یک ناحیه m = 202، الگوریتم با احتمال بالایی پس از ۵ تکرار برای یک تصویر M = 20002 همگرا می‌شود.

آزمون بالا دشوار است اما برای آنها تطابق دقیق نیست. بدترین نمونه برای تطابق دقیق زمانی است که تصویر ب از الگویی با تکرار بالا با انحرافاتی شبیه به ویژگیهای مجزا در الف باشد. ممکن است انحراف توسط یکی از عوامل منحرف کننده به "دام" بیفتد و اندازه موثر ناحیه همسایگی C ممکن است به ۱ کاهش یابد (مثلا، تنها تطابق دقیق می‌تواند راه حل را از عامل منحرف کننده در طی جستجوی تصادفی بیرون بکشد). هرچند در عمل، برای تحلیل تصویر و برنامه‌های ترکیبی همچون آنهایی که در این پایان نامه نشان دادیم، یافتن یک تطابق تقریبی (با توجه به شباهت تکه) موجب هیچ اختلاف قابل توجهی نخواهد شد. شانس یافتن یک تطابق تقریبی موفق در زمانی که عوامل منحرف کننده مشابه حضور دارند واقعا بالا است زیرا هر عامل منحرف کننده خودش یک تطابق تقریبی است. اگر فرض کنیم Q عامل منحرف کننده در تصویر ب وجود دارند که با تطابق دقیق در برخی مرزها تشابه دارند، در اینجا هر تقریب همان ناحیه همسایگی C را دارد، سپس با پیروی از تحلیل بالا تعداد تکرارهای مورد انتظار برای همگرایی به $M/(QC_m) - 0.5$ کاهش می‌یابد.

### ۳.۸.۲   تحلیل برای تصاویر دنیای واقعی

در اینجا ما تقریب‌های ایجاد شده توسط الگوریتم خود در تصاویر دنیای واقعی را تحلیل می‌کنیم. ما عملکرد الگوریتم انتقال را تنها برای یافتن نزدیک ترین همسایه بررسی می‌کنیم. برای رسم عملکرد کلی الگوریتم کلی شکل ۳.٤ و ۳.٥ را ببینید. برای ارزیابی اینکه الگوریتم چطور به درجه‌های مختلفی از

تشابه بصری در میان تصاویر ورودی و خروجی آدرس دهی می‌کند، تحلیل خطا بر روی مجموعه‌های داده‌ای متشکل از زوج تصاویر یک محدوده از تشابهات بصری اعمال شده است.

| اندازه (مگاپیکسل) | زمان (ثانیه) | | حافظه (مگابایت) | |
|---|---|---|---|---|
| | PatchMatch | kd-tree | PatchMatch | kd-tree |
| ۰٫۱ | ۰٫۶۸ | ۱۵٫۲ | ۱٫۷ | ۳۳٫۹ |
| ۰٫۲ | ۱٫۵۴ | ۳۷٫۲ | ۳٫۴ | ۶۸٫۹ |
| ۰٫۳۵ | ۲٫۶۵ | ۸۷٫۷ | ۵٫۶ | ۱۱۸٫۳ |

جدول ۳. ۱ - مقایسه حافظه و زمان اجرا برای ورودی نشان دهده شده در شکل ۳٫۳ . الگوریتم را با روشی که برای جستجوی مبتنی بر patch انجام میشود، مقایسه می‌کنیم: درخت kd، تطابق نزدیک ترین همسایه را تقریب می‌زند. الگوریتم از $n = 5$ تکرار استفاده می‌کند. پارامترها برای درخت kd برای تعیین خطای متوسط الگوریتم بهبود یافته اند.

اینها شــامل ورودی‌ها و خروجی‌هایی از عملیات ویرایش تصویر (یا موارد خیلی مشــابه) می‌شــدند، جفت‌های استریو[16] و فریم‌های تصویر متوالی (یا چیزی مشابه)، تصاویر از یک طبقه بندی در مجموعه داده Caltech-256[17] (یا کمی مشابه) و زوج تصویرهای غیر وابسته می‌شود. برخی از اینها در تفکیک پذیری‌های مختلف (۰٫۱ تا ۰٫۳۵ مگاپیکســـل) و اندازه patchهای (۴ در۴ و ۱۴ در ۱۴) مختلف تحلیل شــدند. الگوریتم ارائه شــده و درخت kd ANN+PCA هردو روی هر جفت اجرا شــدند و مقایســه شدند (با NN دقیق محاسبه شدند). به یاد داشته باشید که چون زمان پیش محاسبه برای برنامه‌ها اهمیت دارد، از تصویر کردن PCA برای کاهش چند بعدی بودن داده‌های ورودی، برخلاف kumar و همکاران [63]، که بردارهای ویژه را برای تصـــویر کردن PCA مختلف در هز گره از درخت kd محســابه کردند،

---

ا ستفاده شده ا ست. چون هر الگوریتم پارامترهای قابل بهبودی دارد، این پارامترها نیز برای د ستیابی به محدوده‌ای از خطاهای تقریبی تغییر داده شده است.

خطا برای هر مجموعه داده‌ای به طور میانگین محا سبه شده ا ست و صدک ۹۵ ام تکه‌ها بین فا صله تکه RMS الگوریتم و مقدار واقعی فاصله تکه RMS اصلی اختلاف دارند. برای ۵ تکرار از الگوریم، دریافتیم متوسط خطاها بین ۰٫۲ و ۰٫۵ سطوح خاکستری تصاویر مشابه و بین ۰٫۶ و ۱٫۵ سطوح خاکستری در تصاویر غیر مشابه (از ۲۵۶ سطح خاکستری) است. در صدک ۹۵ ام، خطاها از ۰٫۵ تا ۲٫۵ سطوح خاکستری تصاویر مشابه و بین ۰٫۹ تا ۶٫۰ سطوح خاکستری برای تصاویر غیر مشابه است.

الگوریتم ارائه شده اساسا نسبت به درخت kd سریع تر است و با توجه به محدوده وسیعی از پارامترها از حافظه ی کمتری استفاده می‌کند. برای اندازه patch ۷ در ۷ استفاده شده برای اغلب نتایج، دریافتیم که الگوریتم ۲۰ برابر تا ۱۰۰ برابر سریع تر است و حدود ۲۰ برابر حافظه کمتری نسبت به درخت kd بدون توجه به تفکیک پذیری ا ستفاده می‌کند. جدول ۳٫۱ مقایسه ی حافظه و زمان میانگین استفاده شده در الگوریتم را در برابر درخت‌های kd ANN برای نوعی از ورودی نشان می‌دهد: جفت‌ها در شکل ۳٫۳ نشان داده شده است. بقیه مجموعه داده‌ها نتایج مشابهی می‌دهند. برای مقایسه دقیق زمان اجرا، پارامترهای درخت kd ANN برای دستیابی به متوسط خطای تقریبی الگوریتم خود پس از ۵ تکرار تنظیم شده است.

خطاها و تسریع‌ها به دست آمده تابعی از اندازه patch و تفکیک پذیری تصویر هستند. برای patchهای کوچکتر، تسریع‌های کوچکتری به دست آوردیم (۷ برابر تا ۳۵ برابر برای تکه‌های ۴در۴)، و الگوریتم مقادیر خطای بالاتری دارد. به طور عکس، تکه‌های بزرگ تسریع‌های بالاتر (۳۰۰ برابر یا بیشتر برای تکه‌های ۱۴ در ۱۴) و مقادیر خطای پایین تر دارند. تسریع‌ها در تفکیک پذیریهای کوچکتر، پایین تر هستند اما در تفکیک پذیری‌های بالاتر سطوح کمتری داریم.

پژوهش‌ها [63] نشان می‌دهد که درخت‌های vp نسبت به درخت‌های kd برای جستجوهای دقیق نزدیکترین همسایه کاراتر هستند. این مورد در مجموعه داده‌های استفاده شده نیز مشخص بود. هرچند، تطابق دقیق در برنامه پیاده سازی شده خیلی آهسته صورت می‌گیرد. در هنگام انجام تطابق تقریبی با

PCA، دریافتیم که درخت‌های vp نسبت به درخت‌های kd برای مقادیر خطای معادل کند تر هستند، بنابراین آنها را از مقایسات خود حذف کردیم.

## ۳. ۹   تشریح مطالب

ماهیت الگوریتم ما برخی تشابهات صوری LBP را در بر دارد و الگوریتم‌های Graph Cuts اغلب برای حل حوزه‌های تصادفی مارکو در یک شبکه تصویر استفاده شده اند [107]. هرچند، اختلافات اساسی وجود دارند: الگوریتم ارائه شده در اینجا، برای بهینه سازی تابع انرژی بدون هیچ عبارت همسایگی طراحی شده است. MRFها اغلب از عبارت همسایگی برای تنظیم بهینه سازی‌ها بدون هیچ نویز یا از دست رفتن داده‌ها، بر اساس وابستگی موجود در مدل کلی استفاده می‌کنند. متقابلا، الگوریتم هیچ مدل کلی صریحی ندارد اما از وابستگی در داده‌ها برای هرس کردن جستجو برای راه حل‌های احتمالی یک مسئله جستجوی موازی ساده استفاده می‌کند. چون الگوریتم این نواحی وابسته را در تکرارهای آغازین می‌یابد، تطابق‌های ما در وابستگی‌ها دچار خطا می‌شوند. بنابراین، وابستگی‌ها صریحا انجام نمی شوند، رویکرد برای برنامه‌های ترکیبی عملی متعددی همانطور که در فصل‌های ۵ و ۶ نشان داده خواهد شد کفایت می‌کند. به علاوه، الگوریتم از محاسبات پر هزینه سازگاری patchهای به هم پیوسته و الگوریتم‌های استنتاجی/بهینه سازی اجتناب می‌کند.

به طور خلاصه، این فصل الگوریتم تطابق اصلی را در محتوای تطابق توصیفگرهای با نمونه برداری متراکم تر نشان می‌دهد. ما همگرایی عملی و تئوری را تحلیل کردیم و دریافتیم که الگوریتم حداقل نسبت به پژوهش‌های قبلی، در میان محدوده وسیعی از پارامترها سریع تر است. در ادامه جزئیات پیاده سازی برای معماری‌های موازی و مجموعه‌های بزرگی از تصاویر را تشریح خواهیم کرد.

# فصل چهارم
# استراتژیهای موازی سازی و جستجو

در این فصـــل، ما جزئیاتی در مورد پیاده ســـازی الگوریتم خود در معماری موازی تشـــریح می‌کنیم همانند GPU و CPU چند هسته‌ای و "PatchWeb" که الگوریتمی مشابه PatchMatch است که به مســـائل جســـتجو و مقیاس پذیری در مجموعه تصاویر بزرگ اشـــاره دارد. ما همچنین دو عملیات جســـتجوی جدید معرفی می‌کنیم: enrichment و binning، که در فرایند جســـتجو، به خصوص مجموعه داده‌های تصویر بزرگ کمک می‌کنند.

## ۴. ۱   موازی سازی

انواع موازی از الگوریتم تطابق روی GPU و CPU پیاده سازی شده است. عمل جستجوی تصادفی اندکی قابل موازی سازی است به طوری که چالش اصلی موازی سازی عمل انتشار است. در اینجا الگوریتم‌های موازی مختلفی را تشریح می‌کنیم.

یک الگوریتم GPU کاملا موازی در زبان CUDA پیاده سازی شده است. برای اینکه عملیات وابسته تر نگه داشته شود، بین تکرارهای جستجوی تصادفی و انتشار جاگزین شده است که هر گام همه‌ی حوزه انحراف را به صورت موازی آدرس دهی کند. گرچه انتشار ذاتا یک عمل سریال است، ما این الگوی Jump Flood Rong و Ten [91] را برای انجام انتشار در سراسر تکرارهای متعدد تطابق دادیم. . از آنجایی که نسخه CPU با اطلاعات انتشاری همه انواع خط اسکن سازگار است، دریافتیم که در عمل انتشارهای طولانی نیاز نیست و بیشینه فاصله پرش ۸ کفایت می‌کند. همچنین تنها از ۴ همسایه در هر فاصله پرش به جای ۸ همسایه توصیه شده توسط Rong و Tan استفاده می‌کنیم. با دقت تخمین مشابه، الگوریتم GPU، ۷ برابر سریعتر از الگوریتم CPU تک هسته‌ای است.

دو نسخه مختلف از الگوریتم موازی را روی CPU پیاده سازی شده است. نخست، الگوریتم Jump Flood تشریح شده در بالا روی CPU پیاده سازی شده است. خطاهای آن با الگوریتم CPU اصلی

قابل قیاس هستند و ازاینرو تقریبا ۲٫۵ برابر کندتر از الگوریتم CPU تک هسته‌ای است، زمان اجرا با تعداد هسته‌ها رابطه خطی دارد.

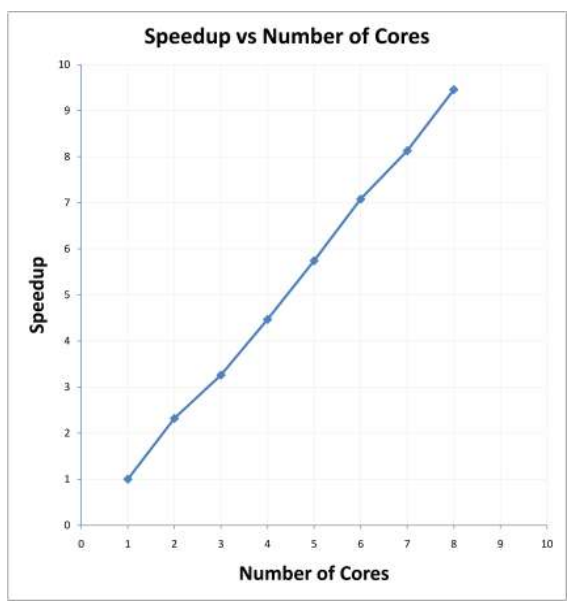

شکل ٤. ١ – تسریع خطی تقریبی الگوریتم موازی با افزایش تعداد هسته‌ها. انحرافات خطی تاثیرات حافظه نهان در نظر گرفته شده است.

از آن برای رویکرد دومی تلاش کردیم و الگوریتم CPU معمولی را با تقسیم NNF به قسمت افقی، و اداره هر قسمت با هسته‌ای مجزا موازی سازی کردیم. چون قسمت‌ها به صورت موازی گردانده می‌شوند، اطلاعات می‌توانند به طور عمودی در تمام طول تکه در یک تکرار مشخص منتشر شوند. برای تضمین اینکه اطلاعات شانس انتشار در تصویر را داشته باشند ما به طور همزمان از یک بخش مهم پس

از هر تکرار استفاده کردیم. برای اجتناب از نقض منابع با توجه به انتشار بین تکه‌های مجاور، نزدیکترین همسایه‌ها در ردیف آخر تکه پس از همگام سازی ذخیره شده است [18].

برای الگوریتم تکه کردن، یک تسریع خطی تقریبی، روی ماشین تست 8 هسته‌ای، همانطور که در شکل 1.4 نشان داده شد، مشاهده کردیم.

باور داریم برای پژوهش‌های آینده، پیاده سازی‌های GPU کاراتری می‌تواند با یافتن جایگزین‌هایی برای انتشار Jump Flood ایجاد شود؛ بنابراین ما بهبودهایی ارائه می‌کنیم. در الگوریتم سریال اصلی، هر قطر از تصویر به طور سریال تنها به قطرهای قبلی وابسته است، بنابراین، پیمایش سریال قطرها، پردازش موازی هرکدام، برای انتشار ممکن است . (چون انتشار تنها نیاز به کار با فاصله محدود شده دارد، ما می‌توانیم هر قطر n ام را در یک زمان پردازش کنیم، برای مثال $n = 8$). متناوبا، این می‌تواند روی تمامی ستون‌ها یا سطرهای تصویر، برای توازن بار بهتر انجام شود اما با استفاده از الگوهای انتشار (برای مثال، در هنگام پردازش ستون‌ها از چپ به راست و انتشار اطلاعات درست، که با کمک همسایه‌های چپ، بالا-چپ و پایین چپ، متشر می‌شود.

## 4.2    عملیات جستو: **Enrichment** و **Binning**

ما در اینجا دو عملیات جستجو جدید معرفی می‌کنیم که می‌تواند به صورت کلی استفاده شود، که در هنگام جستجوی مجموعه‌های تصویر بزرگ اهمیت دارند. این بخاطر آن است که آنها در انجام اکتشاف کاملا تصادفی انجام شده توسط عمل "جستجوی تصادفی" کمک می‌کنند.

### 4.2.1   **Enrichment** [19]

اولین عمل جستجو Enrichment نامیده می‌شود. گام انتشار PatchMatch تطابق‌های خوبی در میان ابعاد مکانی تصویر منتشر می‌کند. هرچند، در موارد مکانی، می‌توانیم تطابق‌های انتشاری در میان فضای خود patchها ملاحظه کنیم: برای مثال، در هنگام تطابق یک تصویر الف با خودش – همانند

---

18. متناوبا، برای اجتناب از این مسئله همگام سازی، روی یک ماشین n هسته ای، می‌توان ورودی را به 2n تکه عمودی تقسیم کرد وتکه های زوج را به دنبال تکه های فرد به صورت موازی پردازش کرد.

19. غنی سازی

٤٢

حذف نویز غیر محلی (بخش ۳. ۷)- بسیاری از k نزدیکترین همسایگان patch مرجع، patch، و k-1 باقی مانده patchهایی در لیست k-NN خود خواهند داشت.

ما enrichment را به صورت انتشار تطابق‌های خوب از یک patch برای k-NN آن یا برعکس تعریف می‌کنیم. این عمل را enrichment می‌نامیم زیرا حوزه نزدیک ترین همسایه را می‌گیرد و با در نظر گرفتن مجموعه "غنی تری" از تطابق‌های داوطلبین خوب بالقوه به جای انتشار یا جستجوی تصادفی تنها بهبود می‌بخشد. از یک دیدگاه تئوری گراف، ما می‌توانیم انتشار شار معمولی همچون حرکت تطابق‌های خوب یک شبکه مستطیلی که گره‌ها مراکز (پیکسل‌ها) patch هستند در حالی که enrichment، تطابق‌های خوب را در میان گرافی که هر گره آن به k-NN آن متصل است حرکت می‌دهد. ما دو نوع enrichment، برای حالت مکانی تطابق تکه‌هایی در تصویر الف با سایر تکه‌های درون همان تصویر معرفی می‌کنیم.

**Enrichment پیشرو** از ترکیبی از تابع $f$ با خودش برای شناسایی کاندیداهایی برای بهبود حوزه نزدیک ترین همسایه استفاده می‌کند. حالت متعارف enrichment پیشرو $f^2$ است. اگر $f$ یک NNF با k همسایه با شد، ما $f^2$ نزدیکترین هم‌سایه را با جستجو در همه‌ی نزدیک ترین همسایگان نزدیک ترین همسایه می‌سازیم: $k^2$ تا از اینها وجود دارند. کاندیداها در $f$ و $f^2$ با هم مقایسه شدند و بهترین k کلی به عنوان نزدیکترین هم سایه $f'$ بهبود یافته استفاده شده است. اگر $min()$، k تطابق بالاتر را نشان می‌دهد، سپس داریم: $f' = min(f, f^2)$. انواع دیگر enrichment پیشرو می‌توانند همچون $f^3$ یا $f^4$ و غیره در نظر گرفته شوند. همچنین، برای بهبود زمان اجرای مربوط از $O(k^2)$ به $O(k)$، می‌توانیم به طور تصادفی k عنصر نمونه از $f^2$ برداریم یا تنها $\sqrt{k}$ عنصر بالاتر از $f$ را قبل از محاسبه $f^2$ برداریم. برای تعدیل مقادیر k، همچون k = 16 الگوریتم $f$ متعارف نسبت به این جایگزین‌ها سریع‌تر همگرا می‌شود.

به طور مشابه، **enrichment معکوس** ، نزدیک ترین همسایه پسرو را برای ایجاد کاندیداهایی برای بهبود NNF پیمایش می‌کند. الگوریتم متعارف در اینجا $f^{-1}$ است. پس بهتر است معکوس چند مقداری $f^{-1}$ را محاسبه کنیم. به یاد داشته باشید که ممکن است $f^{-1}$ مقادیر صفر داشته باشد اگر هیچ تکه‌ای به تکه الف اشاره نکند یا به مقداری بیشتر از k اشاره کند اگر تکه‌های متعددی به الف اشاره کنند. ما $f^{-1}$ را با استفاده از لیستی از طول‌های متفاوت هر موقعیت ذخیره می‌کنیم. مجددا، برای بهبود NNF فعلی، بهترین همسایگان k فعلی و همه‌ی همسایگان $f^{-1}$ را با تولید NNF بهبود یافته $f''$

٤٣

رتبه‌بندی می‌کنیم: $f'' = min(f, f^{-1})$. همچنین توان‌های معکوس دیگر ($f^2$، $f^3$ و غیره) قابل بررسی هستند. اما با بررسی‌های بیشتر مشخص شده است که $f^{-1}$ متعارف سریع‌تر است. به خاطر داشته باشید که در اغلب موارد، تابع فاصله متقارن است، بنابراین فاصله‌های مسیر نیازی به محاسبه شدن برای $f^{-1}$ ندارند. نهایتا، می‌توانیم enrichment معکوس و پیشرو را به هم الحاق کنیم و در یافتیم که $f^{-1}$ دنبال شده توسط $f^2$ در کلی سریع‌ترین است. عملکرد این الگوریتم‌ها در شکل ٤.٢ محاسبه شده است [4].

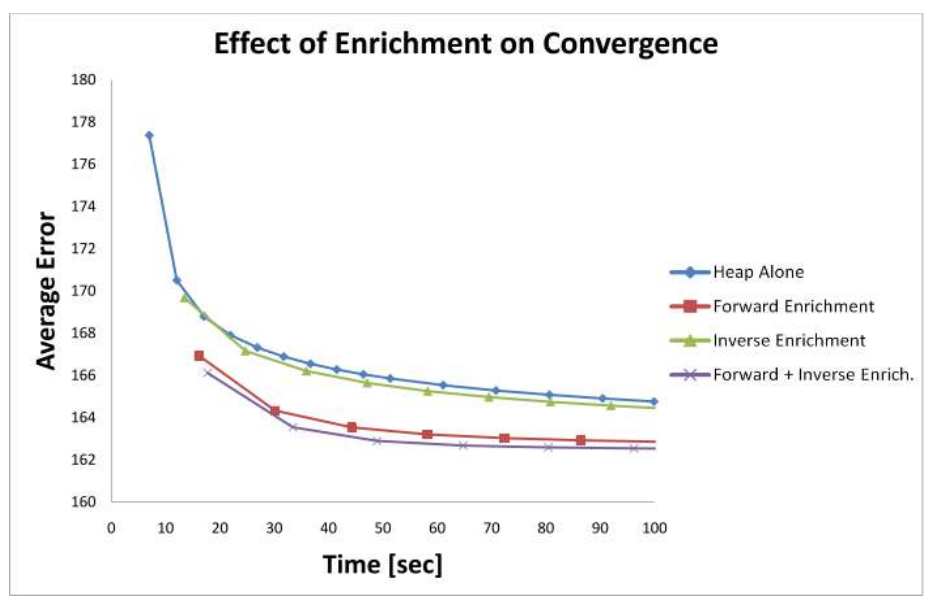

شکل ٤.٢ – مقایسه الگوریتم پشته با و بدون enrichment. الگوریتم پشته کند ترین است و الگوریتم سریع تر از هردویenrichment پیشرو و معکوس استفاده می‌کند. نمودار میانگین مجموعه داده‌ای از ٣٠ تصویر ٠٫٢ مگاپیکسلی و پارامتر $k = 16$ همسایه را نشان می‌دهد.

در این حالت از تطابق تصاویر الف و ب، enrichment معکوس می‌تواند گاهی انجام شود. Enrichment پیشرو می‌تواند با محاسبه‌ی نگاشت‌های تزدیک‌ترین هم‌سایه در هر دو جهت به کار بسته شود؛ این بررسی می‌تواند برای پژوهش‌های آتی در این حوزه ادامه داشته باشد.

### ٤.٢.٢ Binning [20]

دومین عمل جستجو را Binning می‌نامیم. این عمل می‌تواند به عنوان تحلیلی برای درج و صف بندی patchs در درخت kd با ابعاد کم استفاده شود. هدف ما شناسایی دقیق نحوه نمایش patch نیست اما در عوض اندکی از اجزای گرادیان و رنگ اولیه را بدست می‌آوریم.

در این عملیات، اولا برای هر تکه مشخص می‌کنیم که در کدام "بلوک" قرار می‌گیرد. این به عنوان یک پیش محاسبه انجام می شود. ما با کاهش ابعاد PCA نمایش patch برای هر تعداد خیلی کمی از $k$ بعد، بلوک را شناسایی می‌کنیم. سپس هر یک از این $k$ بعد را با تقسیم کردن به $p$ قسمت با اندازه دلخواه مشخص می‌کنیم. بنابراین در کل $p^k$ بلوک وجود دارند که یک شبکه چند بعدی یکنواخت نا سازگار شکل می‌دهد. ما معمولا p و k کوچکی را انتخاب می‌کنیم پس تعداد کل بلوک‌ها در مرتبه‌ای از ده هزار قرار می‌گیرد همچون p = 9 و k = 4. همچنین PCA باید با دقت تخمین زده شود، به طوری که patchها می‌توانند در هنگام محاسبه PCA به طور پراکنده نمونه برداری شوند. برای هر بلوک، ما برداری از همه ی مختصات patchهایی که در آن بلوک قرار گرفته اند لیست می‌کنیم [21].

پس از این پیش محاسبات، الگوریتم PatchMatch را با عملیات جستجوی بلوک اضافه شده اجرا می‌کنیم. این عملیات نزدیک ترین همسایه داوطلب جدیدی با فاصله ی بالقوه کمتری را ارائه می‌دهد. این کار با انتخاب تکه‌ای تصادفی از بلوکی که تکه درخواستی فعلی در آن قرار گرفته انجام می‌شود.

در این محتوای مجموعه‌های تصویر بزرگ، binning به همگرایی با شانس بیشتری در دستیابی تصادفی به هدف خوب کمک می‌کند. جستجوی تصادفی به تنهایی از تمام مجموعه به طور تصادفی نمونه برداری خواهد کرد و شانس پایینی در دستیابی به "هدف خوب" خواهد داشت، درحالیکه Binning تکه‌های با تشایه بالا را هدف قرار می‌دهد. برای ۱۰۰ تصویر ورودی در شکل ٤.٤، Binning از الگوریتمی با یک فاکتور ۲-٤ برای مقادیر خطای برابر کمک می‌گیرد. بخش بعدی الگوریتمی کامل برای تطابق مجموعه‌های تصویر بزرگ را تشریح می‌کند که Binning و Enrichment و نیز مدیریت حافظه و الگوریتم‌های مقیاس پذیر را برای تعداد زیادی از تصاویر به کار می‌گیرد.

## ٤.٣ PatchWeb: تطابق در مجموعه‌های تصویر بزرگ

---

[20]. دور ریزی

[21]. در برخی برنامه ها، کاهش مصرف حافظه اهمیت زیادی دارد پس هر تکه n ام که در بلوک قرار می‌گیرد می‌تواند ذخیره شود

این بخش PatchWeb، الگوریتمی که PatchMatch را برای محدودیت‌های مکانی جستجو در مجموعه‌های تصویر بزرگ تطبیق می‌دهد، معرفی می‌کند.

### ۴٫۳٫۱  مقدمه

اخیرا، مجموعه‌های وسیعی از تصاویر دیجیتالی به صورت آنلاین منتشر می‌شوند و کاربران مجموعه‌های تصویر قابل تغییر اندازه می‌سازند. نهایتا، تعداد پروژه‌های پژوهشی متمرکز روی استفاده از مجموعه‌های تصویر به عنوان یک ورودی در حال افزایش است. نتایج، حذف اشیاء ناخواسته در تصاویر [50]، افزایش کیفیت تصاویر کم کیفیت [31]، افزایش تفکیک پذیری تصویر [46]، و سایر پژوهش‌های گرافیکی دیگر را نشان می‌دهند.

بسیاری از این روش‌ها باهم در یافتن تناظر میان تصویر درخواستی و مجموعه عظیم تصاویر موجود اشتراک دارند. در پژوهش‌های اخیر، برای دستیابی به زمان‌های اجرای عملی، ویژگی‌های محاسبه شده متراکم شده در هر پیکسل صرف نظر شدند. در عوض، الگوریتم‌ها تمایل به کاهش فضای جستجو، با کاهش میزان داده‌های ورودی به میزانی کمتر دارند. این با ویژگی‌های خلوت [82, 40]، ویژگی‌های جهانی که دیدگاهی کلی را توصیف می‌کنند [50]، یا سایر تکنیک‌های کاهش ابعاد انجام شده است. هرچند، این ویژگی‌های خلوت می‌توانند عملکرد الگوریتم را کاهش دهند.

نتایج پیچیده متعددی از میزان زیادی از داده‌ها به این الگوریتم‌ها وارد می‌شوند. اول، در برخی موارد همچون حذف اشیاء از تصاویر [50]، میلیون‌ها تصویر باید برای دستیابی به نتایج کیفیت بالا استفاده شوند بنابراین هر تکنیک جستجو باید قادر به مقیاس پذیری در میلیون‌ها تصویر باشد. دوم، برای الگوریتم‌هایی همچون افزایش وضوح تصویر [43]، بالاترین کیفیت ممکن است با در نظر گرفتن همه‌ی تناظرهای متراکم ممکن حاصل شود. برای یک تصویر درخواستی یک مگاپیکسلی و یک میلیون پایگاه داده تصویر، یک الگوریتم Brute force برای یافتن بهترین تناظر در هر پیکسل باید ۱۰۱۸ تناظر را تست کند، بنابراین زمان پردازش بالایی نیاز دارد. بخاطر این، پژوهش قبلی به طور هوشمندی، مجموعه‌ای از تطابق‌های ممکن را برای کاهش کیفیت نتایج محدود کرد. نهایتا، بسیاری از روش‌های موجود، ویژگی‌های مقیاس پذیری ضعیفی دارند که به موجب آن یافتن تناظرهای میان تصویر

درخواستی و مجموعه‌ای از n تـ صویر، $O(n)$ زمان می‌گیرد. هرچه مجموعه داده‌ها سریعتر رشد کند، این مقیاس پذیری خطی هم نیازمند ویژگی‌های مطلوبی همچون تعامل پذیری کاربر می‌شود.

ما الگوریتمی را تشریح می‌کنیم که برای غلبه بر تعدادی از این کاستی‌های موجود روش‌های جستجو غلبه می‌کند. الگوریتم تطابق‌های نزدیک ترین همسایه تقریبی در میان یک تصویر درخواستی و مجموعه تـ صویر بزرگ را می‌یابد. در این الگوریتم نخـ ست با توجه به دانش ما در تلاش برای حل مـ سئله تطابق متراکم: برای یافتن بهترین تناظر(ها) در هر پیکسـ ل، که به طور مسـ تقل تمام مجموعه داده را تطبیق می‌دهد تلاش می‌کنیم. این برای تعداد زیادی از تصـ اویر مقیاس پذیر اسـ ت یعنی می‌تواند روی یک کامپیوتر یا یک کلاستر با منابع حافظه و زمان محدود اجرا شود.

الگوریتم جستجو از الگوریتم PatchMatch برای یافتن تناظر میان $n = 2$ تصویر ورودی الهام گرفته است. این الگوریتم را "PatchWeb" می‌نامیم زیرا یک ساختار گرافی یا "وب" درمیان تصاویر مجموعه می‌سازد. پس از فرایندهایی که برای تصاویر با اندازه n از مجموعه به صورت خطی زمان می‌برند، عملیات تقاضا می‌تواند به طرز کارایی با بازگرداندن نزدیکترین تطابق همسایه در میان تصویر درخواستی و مجموعه تصویر بزرگ اجرا شود. الگوریتم ما در شکل ٤. ٣ نشان داده شده است.

به طور خاص، الگوریتم مسئله یافتن هر patch از تصویر درخواستی یا ناحیه A با یک مجموعه تصویر دلخواه $B_1, B_2, ..., B_n$ را حل می‌کند که نزدیک ترین مسیر در تمام مجموعه تصویر، تحت یک معیار فاصله patch همچون معیار $L_p$ را می‌دهد.

الگوریتم ما با مرتب سـ ازی هر تصـ ویر $B_i$ در مجموعه یک نگاشـ ت (NNF) حوزه نزدیک ترین همـ سایه هر patch در $B_i$ برای شبیه ترین patch در بقیه مجموعه تـ صویر کار می‌کند. ما از تکنیک‌های مشابهی همچون الگوریتم PatchMatch استفاده می‌کنیم: با شروع از یک تخمین ضعیف برای حوزه، ما

حوزه را با ارائه تناظرهای کاندید بهتر بهبود می‌دهیم. این کاندادهای بهتر با رشته‌ای از عملیات پیشنهاد شده اند. تناظرهای میان مجموعه تصویر $B_i$ تا رسیدن به معیار همگرایی بهبود داده شدند.

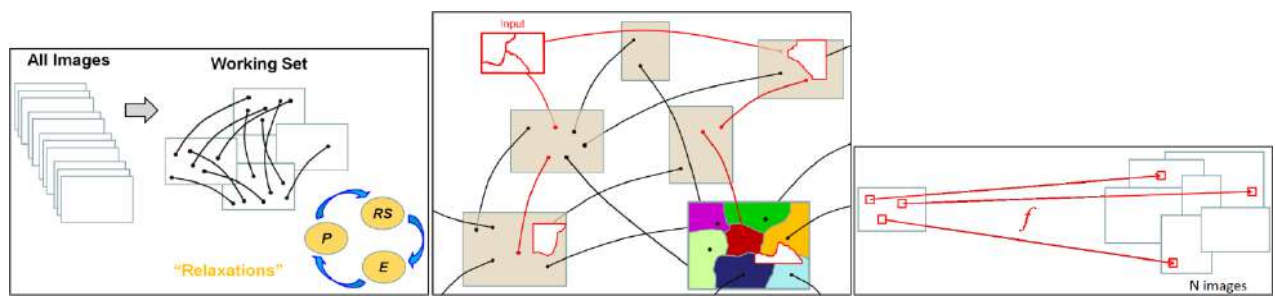

(الف) مجموعه تصویر پرس و جو   (ب) پیش پردازش ساختارهای داده‌های وب   (ج) چگونگی ایجاد ساختار

شکل ٤. ٣ – الگوریتم PatchWeb؛ (الف) عدف ما یافتن سریع تطابق‌های نزدیک ترین همسایه در میان تصویر درخواستی و یک مجموعه تصویر است. (ب) برای اینکار ما روی مجموعه تصویر پیش محاسبه انجام دادیم و گرافی ساختیم که تطابق‌های نزدیک‌ترین همسایه در میان مجموعه را می‌دهد. این گرف وب است: نزدیک ترین همسایه که به هر patch یک همسایه در هر کجای وب را می‌دهد، ذخیره می‌کند. (ج) الگوریتمی برای ساخت وب. در سراسر مجموعه تصویر، ما به طور تصادفی یک مجموعه کاری انتخاب می‌کنیم و تسهیلاتی برای ایجاد پیوندهای نزدیک‌ترین همسایه در میان مجموعه کاری با استفاده از عملگرهایی همچون نمونه گیری تصادفی، انتشار و enrichment انجام می‌دهیم.

در اغلب موارد، مجموعه تصویر B نمی تواند حافظه را پر کند. بنابراین، ما یک مجموعه کاری تعریف می‌کنیم، این، تصاویری است که می‌توانند خوانده شوند، تصاویری که از مجموعه‌ای از تصاویر در مجموعه تصاویر نمونه برداری شده اند. در یک کلاستر، هر پردازنده مجموعه کاری خودش را خواهد داشت. هر پردازنده تناظرهای بهبود یافته را در حد مطلوب ارائه می‌کند.

در اینجا الگوریتم کلی را relax می‌نامیم، زیرا با دادن مجموعه‌ای از تصاویر، تطابق‌های نزدیک‌ترین همسایه در میان آنها را به آرامی بهبود می‌بخشد. الگوریتم درخواستی به سادگی با افزودن تصویر درخواستی الف به مجموعه کاری و اجرای relax با NNFها برای مجموعه تصویر موجود در حالت فقط خواندنی کار می‌کند، به طوری که وب موجود پیوندی با تصویر درخواستی نخواهد داشت.

چون مسئله تطابق متراکم دشوار به نظر رسیده بود و هیچ پژوهش قبلی شناخته شده در این حوزه وجود نداشت، الگوریتم PatchWeb را با تعدادی الگوریتم رقیب همچون PatchMatch روی همه ی

٤٨

جفت تصاویر مقایسه کردیم و ویژگی‌های مقیاس پذیری بهتری از خود نشان می‌دهد. ادعا نمی کنیم که الگوریتم در استفاده از حافظه و زمان بهینه است اما این اولین روشی است که مسئله تطابق متراکم را به صورت مقیاس پذیر حل می‌کند.

این بخش شامل الگوریتمی برای یافتن تناظرهای متراکم درمیان میلیون‌ها تصویر و ویژگی‌های مفیدی همچون تقاضای کارا و مقیاس پذیری برای کلاستر و نیز کار در زمان محدود و محدودیت‌های حافظه می‌شود.

۲.۳.۴ پژوهش‌های مرتبط

ما در ابتدا ساختارهای داده‌ای توزیع شده و توصیف گره‌های سطح پایین که معمولا در تحلیل مجموعه‌های تصاویر استفاده شده اند بررسی می‌کنیم. سپس پژوهش قبلی که روی ابعاد برنامه‌های کاربردی بالقوه متنوع همچون تشخیص شی/منظره، رندر گیری مبتنی بر تصویر، افزایش تفکیک پذیری تصویر و مقایسه و اینکه این‌ها چه ارتباطی با الگوریتم پیشنهادی ما دارند بحث خواهیم کرد.

**ساختارهای داده توزیع شده:** مسئله تطابق متراکم به میزان حافظه خیلی زیادی نیاز خواهد داشت ؛ برای موردی با یک میلیون تصویر یک مگاپیکسلی، قبلا یک میلیارد patch را در بر گرفتن که در این میان تناظرها باید یافت شوند که انکار به محاسبات بیشتری نیاز دارد. بنابراین، فضای توزیع شده و محاسبات با هم تناسب دارند. دو ساختار داده توزیع شده، درخت‌های kd و LSH توزیع شده هستند.

درخت kd توزیع شده به صورت جنگلی از درخت‌های kd کار می‌کند: هر گره محاسباتی یک بخش فرعی از درخت را اداره می‌کند به طوری که همه ی گره‌ها با هم یک درخت kd بزرگتر را نشان می‌دهند. مشکلات مقیاس پذیری و مهندسی برای استفاده از درخت kd هنوز وجود دارند زیرا اگر توصیف گره‌ها خیلی فشرده نباشند درخت ممکن است در حافظه جای نگیرد و بخش‌ها باید در دیسک به صورت سریال ذخیره شوند. در حالی که در اصل یک درخت توزیع شده با انتخاب سیستم‌های درست ممکن است با الگوریتم وارد رقابت شود، الگوریتم ما به سادگی پیاده سازی می‌شود و مقیاس‌ها به سادگی از یک کامپیوتر بدون نیاز به تغییرات معماری به کلاستر تغییر می‌یابند. LSH توزیع شده در یک شبکه نظیر به نظیر بدون ساختار، بلوک‌های Hash را به جفت‌هایی در یک شبکه اشتراکی Chord-style تخصیص می‌دهد. یک پیاده سازی مشابه از LSH می‌تواند روی یک کلاستر محاسباتی اجرا شود، هرچند همین نگرانی‌ها در مواجهه با مشکلات مهندسی و حافه اشغال شده توسط Hashها مطرح

می‌شود برای مثال، میلیاردها patch در تعداد کمی از پایگاه‌ها قرار گرفته باشند. در الگوریتم ما، ما از ایده‌ای مشابه LSH با نگاشت patchها در مجموعه بلوک‌هایی با ابعاد پایین استفاده می‌کنیم اما با انجام اینکار تنها برای مجموعه کاری فعال از مسئله حافظه اجتناب می‌شود.

**توصیف گره‌ها و کلمات بصری:** مقالات متعددی که روی تشخیص یا رندر گیری بر اساس تصویر کار می‌کنند نیاز به شناسایی منظره‌ها یا اشیاء با ظاهر مشابه دارند. برای توصیف مناظر، GIST یک توصیف گر معمولی است که طبقه بندی‌های کلی مناظر (کوه‌ها، درختان، خیابان‌ها و غیره) را با استفاده از یک توصیف گر طیف رنگ فشرده شناسایی می‌کند که ابعاد ادراکی را تشخیص می‌دهد. نمودار بعدی برای بهبود تشخیص منظره، دستیابی به نرخ‌های تشخیص وضعیت نشان داده شده بود. برای توصیف اشیا، اغلب ویژگی‌های SIFT با تراکم بالا محاسبه شدند یا با داشتن داده‌های خلوت، نقطه‌های مطلوب استفاده شده اند. برای تشخیص منظره یا شی، اغلب کلمات بصری، و بسته‌ای از مدل کلمات ممکن است استفاده شوند که ویژگی‌های سطح پایین نمودارها همچون توصیف گره‌های SIFT یا patchها را محاسبه می‌کنند.

**تشخیص شی/منظره:** پژوهش‌های متعددی تشخیص در مجموعه‌های تصویر بزرگ را بررسی کردند. آنها شامل تسخیص مقیاس پذیر تعداد زیادی از تصاویر با استفاده از دیکشنری‌های مقداردهی شده از کلمات بصری یا درخت‌هایی از کلمات بصری می‌شوند. در پژوهش آینده، همانند این پژوهش، فشردگی ساختار داده برای کارایی خیلی اهمیت دارد، بنابراین آنها patchهای تصویر فشرده را به صورت یک یا دو عدد صحیح با مسیری در میان یک درخت از پیش محاسبه شده نمایش می‌دهند. هرچند پژوهش ما از ویژگی‌های محاسبه شده متراکم استفاده می‌کند با این وجود داده‌ها باید در ساختار داده ما ذخیره شوند. استفاده از یک تصاویر رنگی ۳۲ در ۳۲ با ۸۰ میلیون تصویر و الگوریتم‌های هم ترازی جهانی ساده عملکرد تشخیص بالایی در مقایسه با ناحیه‌ای از تصویر در شناساگرهای مبتنی بر طبقه بندی نشان داده اند. چون عملکرد بالا تنها با تعداد زیادی از تصاویر به دست می‌آید، این ما را برای مقیاس پذیر کردن الگوریتم خود ترغیب می‌کند. نهایتا، با ارتقای الگوریتم، نشان داده شده است که توصیف گره‌های SIFT با نمونه گیری متراکم و یک الگوی طبقه بندی نزدیک ترین همسایه ساده می‌توانند عملکرد ناحیه خوبی برای طبقه بندی تصویر ارائه کنند.

**تکمیل، گسترش و ویرایش تصویر:** نشان داده شد که اشیاء نامطلوب بزرگ می‌توانند از عکس با استفاده از میلیون‌ها تصویر و توصیف گره‌های GIST حذف شوند اگر یک تطابق "سطح منظره" هم

۵۰

تراز شده در هرجایی از مجموعه داده وجود داشته باشد. با استفاده از توصیف گره‌های جهانی ساده، حذف یک شی روی یک CPU یک ساعت طول می‌کشد یا در صورتی که با ۱۵ عدد CPU موازی سازی شده باشد، بسته به تعامل، ۵ دقیقه طول می‌کشد. زمانی که یک شی کوچکتر است یا به طرز نادرستی هم تراز شده باشد با استفاده درست از الگوریتم ما تطابق‌های محلی بیشتری به دست می‌آید. بازیابی تصاویری که به اشتباه ظاهر شدند با تطابق با یک مجموعه تصویر بزرگ با استفاده از کلمات بصری انجام می‌شود. به طور مشابه، رندرگیری گرافیک کامپیوتری می‌تواند در واقعیت گیرایی با استفاده از تعداد زیادی از عکس‌های واقعی گسترش یابد. سیستم Sketch2Photo به یک کاربر اجازه می‌دهد یک پیش‌طرح را به یک منظره عکاسی شده با بارگیری هزاران تصویر کاندید و با استفاده از ابتکاری برای اتصال انها به هم تبدیل کند. این ۲۰ دقیقه یا بیشتر زمان می‌برد تا نتیجه نهایی را تولید کند. برای پژوهش آینده، ممکن است پیاده سازی الگوریتم خود را با زمان تعامل سریعتر کاربر در ترسیم به کار ببندیم.

**رندرگیری مبتنی بر تصویر از مجموعه‌های تصویر بزرگ:** گردشگری تصویر [105] نشان داد که مجموعه‌ای از تصاویر پیرامون یک راهنمای گردشگری به طور خودکار می‌تواند به یک مدل سه بعدی خلوت مبدل گردد. بعدا، بازسازی و تطابق مشابهی روی مقیاس بزرگی انجام می‌شود، بالای یک میلیون بازسازی خودکار تصویر از روم انجام شده است. پژوهش‌های نامحدود روی تصویر نشان می‌دهد که با جابجایی‌های یکپارچه دوربین مجازی، جابجایی به چپ/راست، جلو یا چرخش، می‌توان ۶ میلیون تصویر ساخت. در اینجا روی رندرگیری مبتنی بر تصویر کار تمرکز نمی کنیم اما تناظرهای کاندیدا می‌توانند از ساختار PatchWeb برای چنین بازسازی‌هایی گردآوری شوند.

**افزایش تفکیک پذیری تصویر:** وضوح یک تصویر با تفکیک پذیری پایین می‌تواند با مجموعه‌ای آزمایشی از تصاویر نمونه با تفکیک پذیری پایین و بخش‌های با تفکیک پذیری بالا از آنها و یافتن patchهای مشابه با تفکیک پذیری پایین آنها افزایش یابد. در پژوهش قبل لازم بود که مجموعه آزمایشی از همان طبقه از تصویر باشد و تنها شامل ۶ عکس با تفکیک پذیری پایین می‌شد زیرا در غیر اینصورت جستجوی نزدیک ترین همسایه متراکم کند می‌شد. امیدواریم این کار را بتوان با میلیون‌ها تصویر فرا تفکیک پذیر انجام داد و در صورت نیاز طبقه بندی نصویر را خودکار تشخیص دهیم. اخیرا، نشان داده شد که فراتفکیک پذیری تصویر می‌تواند در یک عکس با تطابق مقیاس‌ها انجام شود. با این وجود، تصور جزئیات تفکیک پذیری بالای محتمل می‌تواند با یک مجموعه تصویر مناسب انجام پذیرد،

۵۱

افزایش تفکیک پذیری می‌تواند ۱۰ تا ۱۰۰ برابر بیان شود که در آن الگوریتم‌های فرا تفکیک پذیری سنتی شکست خواهند خورد.

نهایتا، با استفاده از یک تصویر تنها، یا یک مجموعه از تصاویر، مقایسه تصویر می‌تواند با یافتن عناصر تکراری که مشابه به نظر می‌رسند و جابجایی آنها با یکی از عناصر تکراری انجام پذیرد. امیدواریم که برای اینها و سایر برنامه‌های کابردی، وب‌های ما بتوانند در شناسایی نواحی تکراری حتی در مجموعه‌های تصویر بزرگ یاری کنند.

### ۴.۳.۳ الگوریتم PatchWeb

در این بخش، الگوریتم تطابق را توجیه خواهیم کرد (در شکل ۳. ۴ نشان داده شد). به یاد آورید که مسئله یک تصویر درخواستی A و یک مجموعه تصویر $B_1, B_2, ..., B_n$ برای پاسخگویی به درخواست‌های هر patch در A می‌دهد که patch نزدیک ترین همسایه در مجموعه B است. این کار را ابتدا با ساخت یک وب انجام می‌دهیم: یک گراف از نزدیک ترین هم‌سایه‌ها در میان مجموعه تصاویر B از پیش محاسبه می‌کنیم. سپس تصویر A را در مقابل وب، با محاسبه نزدیک ترین همسایه‌هایی که تصویر A را به مجموعه B نگاشت می‌دهند، درخواست می‌کنیم. برای کارایی، همه ی تخمین‌های تطابق را انجام می‌دهیم و یادآوری می‌کنیم که بدون از دست رفتن هیچ توصیف گر نمونه گیری شده متراکم می‌تواند به جای patchها استفاده شود. ما تابع فاصله کلی patch را به صورت D می‌نویسیم: بهترین تناظر برای یک پیکسل دلخواه آن است که این فاصله را کمینه می‌کند. برای نمونه بسته به برنامه، D ممکن است معیار $L^2$ بین patchها یا معیار $L^2$ بین ویژگی‌های SIFT مقایسه کند.

این الگوریتم از ایده‌های مشابهی همچون الگوریتم PatchMatch استفاده می‌کند که تطابق‌های نزدیک ترین همسایه متراکم در میان دو تصویر را محاسبه می‌کند. الگوریتم PatchMatch با ابتدا مقداردهی اولیه با تناظرهای تصادفی و سپس بهبود تناظرها با ارائه تناظرهای مخاطبین جدید، و سپس گرفتن تناظرهای مخاطب فعلی و جدید، سپس تکرار تا همگرایی کار می‌کند. در اینجا نیز رویکرد مشابهی می‌گیریم. تصاویر متعددی را تا جایی که بتوانیم در حافظه قرار می‌دهیم، تناظر خوبی میان تکه‌های این تصاویر می‌یابیم، سپس برخی از این تناظرها را روی دیسک ذخیره می‌کنیم و برخی تصاویر جدید که هنوز تناظر خوبی ندارند را در حافظه بارگذاری می‌کنیم. در سطح خیلی بالا، الگوریتم

ما می‌تواند در یافتن خودکار کلاسترهایی از خوشه‌های مشابه با یک فرایند تصادفی تکراری استفاده شود.

ما با هر تصویر $B_i$ در مجموعه یک حوزه نزدیک ترین عمسایه (NNF) کار کنیم که برای هر مختصات پیکسل، بهترین تطابق patch در باقی مانده مجموعه تصویر ذخیره می‌شود. به خصوص، NNF به صورت $f_i: \mathbb{R}^2 \mapsto \mathbb{R}^3$ تعریف شده است پس $f_i(x,y) = (x', y', j)$، که در آن $(x,y)$ مختصات در تصویر $B_i$ است و $(x', y')$ نزدیک ترین همسایه در دیگر تصویر $B_j$ است. ما همچنین فاصله تکه $D_{patch}$ وابسته به این تناظر را ذخیره می‌کنیم. به یاد داشته باشید که $D_{patch}$ باید ذخیره شود زیرا در مجموعه‌های تصویر بزرگ، تصویر هدف $B_j$ ممکن است در حافظه نباشد. با دسته کردن مختصات هدف و فاصله تکه در یک عدد صحیح ٦٤ بیتی، می‌توانیم NNF را به صورت فشرده به صورت یک تصویر ٦٤ بیتی تک کاناله ذخیره کنیم. به هر کدام از مختصات x و y، ١٢ بیت، ١٦ بیت برای نمایه تصویر و ٢٤ بیت برای فاصله patch اختصاص می‌دهیم، بیشینه تفکیک پذیری ٤٠٩٦ پیکسلی و بیشینه اندازه مجموعه داده تصاویر ٦٥ کیلو اختصاص می‌دهیم و در آینده تعداد بیت‌ها را برای آدرس دهی تصاویر بیشتر افزایش می‌دهیم. هر NNF را در کنار تصویر $B_i$ آن روی دیسک ذخیره می‌کنیم و نیز NNF را برای کمینه سازی فضای ذخیره سازی و الزامات IO با ZIP فشرده می‌کنیم.

### ٤.٣.٤  ساخت وب با آرمیدگی [22]

در ابتدا وب B را با هر تصویر در بر گیرنده یک NNF نگاشت شونده به (١-,١-,١-)، با فاصله patch بی نهایت مقداردهی اولیه می‌کنیم. بنابراین مقدار بی نهایت، با هر تناظر پیشنهادی جایگزین می‌شود.

بعد، تناظرهای در وب را با آرمیدگی بهبود می‌دهیم. مکررا یک مجموعه کاری از تصاویر مجموعه‌ای از تصاویر B را نمونه برداری می‌کنیم. اندازه مجموعه کاری را حداالامکان طولانی می‌گیریم به طوری که

---

Relaxation .٢٢

مجموعه کاری هنوز در حافظه اصلی جای شود و همه ی تصاویر در مجموعه کاری از دیسک بارگیری شده اند زیرا NNFهای وابسته هستند.

همه ی مختصات $(x, y)$ هر تصویر $B_i$ در مجموعه کاری را برای تعداد تکرار $m$، یا به ترتیب اسکن یا عکس آن حلقه اسکن میکنیم. در موقعیت پیکسل کنونی یک تناظر $f_i(x,y)$ ذخیره شده است. مجموعه‌ای از مختصات تناظر $Z$ را پیشنهاد می‌کنیم که با استفاده از پنج عملگر مختلف ساخته شده است. به خصوص، این یک مجموعه است، بنابراین ما تکراری‌ها را ذخیره نمی کنیم و نیز هم موقعیت مرجع $(x, y, i)$ و هم موقعیت پایانه $(x', y', j)$ را ذخیره می‌کنیم. با مشاهدات تجربی، دریافتیم مجموعه پیش رو از عملگرها نسبت به هر زیر مجموعه دیگری: انتشار، جستجوی تصادفی، enrichment، binning و نمونه گیری یکنواخت، کاراتر است. چهار عملگر اول را در بخش‌های اول نمایش دادیم. عملگر آخر، نمونه گیری یکنواخت، به طور ساده یک patch را به صورت یکنواخت تصادفا از مجموعه کاری، به استثنای تصویر فعلی بر می‌دارد.

پس از اجرای این عملگرها، مجموعه‌ای از تناظرهای بهبود یافته ی کاندید داریم، هر‌یک از اینها در پیکسل فعلی مرجع و در یک تصویر دیگر هدف دارد. ما فاصله مرتبط با هر تناظر $(x, y, i) \to (x', y', j)$ را محاسبه می‌کنیم: اگر کمتر از فاصله فعلی ذخیره شده در NNF در موقعیت $(x, y)$ است، آنگاه NNF را یا تناظر بهتری به روز می‌کنیم. به منظور دستیابی به مجموعه بهتری از تناظرها، برای هر تناظر، تناظر مکمل $(x, y, i) \to (x', y', j)$ را تست می‌کنیم و اگر کمتر از فاصله ذخیره شده در NNF در موقعیت هدف $(x, y, i)$ باشد آن را نیز بروز می‌کنیم. این عملیات Enrichment معکوس است. سرانجام، پس از تعدادی تکرار روی مجموعه کاری فعلی، مجموعه کاری جدید را به روز می‌کنیم.

### ٥.٣.٤   انتخاب مجموعه کاری

گزینه‌های متعددی برای ساخت یک مجموعه کاری وجود دارد. ساده ترین استراتژی، نمونه گیری یکنواخت از مجموعه‌ای از تصاویر است. یک استراتژی پیشرفته تر، نگهداری کسر ثابتی از مجموعه کاری اصلی، نمونه گیری یک کسر ثابت جدید از تصاویر جدید به طور یکنواخت و نمونه گیری تصاویر باقی مانده براساس عملگر enrichment به کار برده شده در دو مجموعه اول از تصاویر انتخابی است. آزمایشات مقدماتی نشان می‌دهند که استراتژی دوم سریع تر همگرا می شود. فرض کنید که این

بخاطر کلاستر تصاویر در گروه‌های دارنده patch مرتبط است. برای کشف بیشتر در این مورد می‌توان بر روی پژوهش‌های آتی سرمایه گذاری کرد.

### ٤,٣,٦   موازی سازی

الگوریتم با هسته‌های متعدد یا گره‌ها در یک کلاستر موازی سازی می‌شود. گره‌های محاسباتی متعدد حتی می‌توانند NNF مشابه را مستقلا در مجموعه‌های کاری مختلف، در یک زمان بهبود ببخشند. تنها همگام سازی که نیاز به انجام آن داریم ذخیره NNF است کپی روی دیسک باید قفل شود و با NNF در حافظه با انتخاب نزدیک ترین همسایه با فاصله کمتر ترکیب شود.

### ٤,٣,٧   صف بندی یک وب با آرمیدگی

در این قسمت ما یک وب می‌سازیم. می‌توانیم دقیقا همان الگوریتم آرمیدگی را برای درخواست وب استفاده کنیم. فرض کنید ما یک تصویر A داریم که در مقابل یک وب از تصاویر $B_1, B_2, ..., B_n$ را تقاضا می‌کنیم. به سادگی تصویر A را به مجموعه تصاویر خود می‌افزاییم و یک وب با ترکیب آنها می‌سازیم و برخی محدودیت‌ها نیز ایجاد می‌کنیم: اول، برای کارایی، تصویر A باید همیشه در هر مجموعه کاری، حتی روی ماشین‌های چند پردازنده انتخاب شده باشد یعنی NNFهای متعددی وابسته به تصویر A وجود خواهند داشت.

به سادگی این NNFها را با یافتن نزدیک ترین همسایه درمیان همه‌ی پردازنده‌ها پس از انجام آرمیدگی کاهش می‌دهیم. دوم، برای اجتناب از خرابی وب موجود برای مجموعه تصویر B، NNFها را برای تصاویر B در حالت فقط خواندنی بارگذاری می‌کنیم. سوم، مجدد در مورد کارایی، کمکی در یافتن

پیوندهای بهتری میان تصاویر B نمی‌کند. بنابراین به جای ایجاد حلقه در سراسر مجموعه کاری به ترتیب اسکن و اسکن معکوس، به سادگی حلقه‌ای را در تصویر A رسیدن به همگرایی تکرار می‌کنیم.

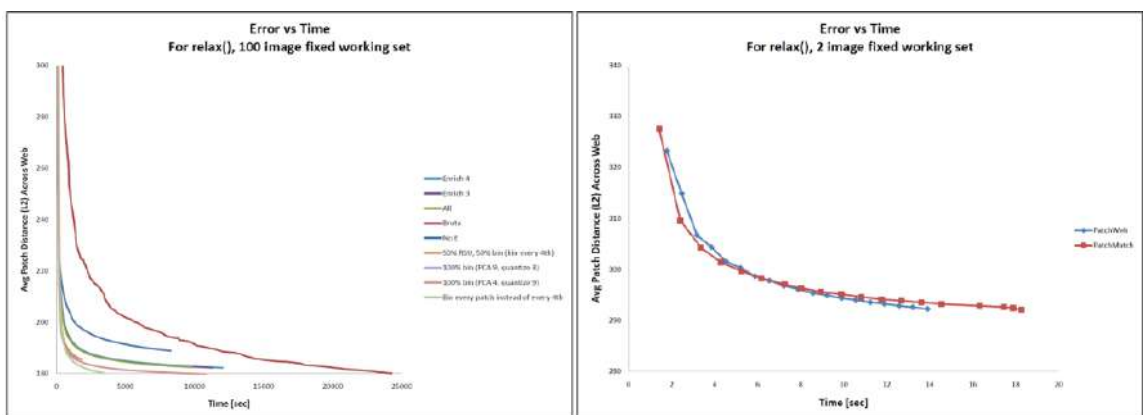

(الف) $n = 2$      (ب) $n = 100$

شکل ۴. ۴ – عملکرد ساخت یک وب. مقایسه با PatchMatch در همه جفت تصاویر. سمت چپ تصاویر $n = 2$ است الگوریتم و PatchMatch عملکرد مشابهی دارند. سمت راست تصاویر $n = 100$ است، بهترین نسخه از الگوریتم پایین‌ترین خط است درحالی که بالاترین خط PatchMatch است. برای یک فاصله patch دلخواه، PatchMatch را با یک فاکتور ۸-۱۰ با موفقیت اجرا می‌کنیم.

### ۸.۳.۴   تحلیل همگرایی

ما تعداد تکرارهای مورد نیاز برای همگرایی را تشریح کردیم. در کل، تعداد تکرارهای مورد نیاز برای ساخت وب و درخواست وب می‌تواند متفاوت باشد و بسته به برنامه کاربردی دارد.

برای ایجاد وب، در شکل ۴. ۴ عملکرد PatchWeb را در مقابل با الگوریتم strawman مقایسه کردیم: PatchMatch را روی همه‌ی جفت‌های ممکن از تصاویر اجرا کردیم و بهترین نزدیک‌ترین هم‌سایه سراسری را یافتیم. این یک مقایسه معقولانه است زیرا PatchMatch عملکردی ناحیه‌ای در یافتن تناظرهای میان دو تصویر دارد. عملکرد الگوریتم با افزایش تعداد تصاویر n افزایش می‌یابد زیرا الگوریتم PatchMatch اندازه $O(n^2)$ زمان می‌گیرد، در حالی که PatchWeb زمانی نزدیک به خطی می‌گیرد به هر دوره با عوامل تجربی همچون اندازه مجموعه کاری مشخص می‌شود. در یک بررسی الگوریتم را با PatchMatch برای مورد $n = 2$ مقایسه کردیم و یافتیم عملکرد الگوریتم با

۵۶

PatchMatch قابل قیاس است. این یک حالت خاص است، چون از عملیات مشابهی، و نه enrichment یا نه binning برای کمک بیشتر برای تصاویر بیشتر استفاده می‌کنیم.

عملکرد الگوریتم به طور تجربه مشخص شده است. خوب است زیرا به ما اجازه می‌دهد تا الگوریتمی انعطاف پذیر داشته باشیم که به خوبی مقیاس پذیر است. اما آیا ما می‌توانیم عملکرد الگوریتم را محدود کنیم؟ یک ورودی ترکیبی را در نظر بگیرید که n تصویر دارد که دو تا از آنها یکسان هستند، آنها را تصاویر ۱ و ۲ می‌نامیم. یافتن هرکدام از آنها چقدر زمان می‌برد؟ فرض کنید اندازه مجموعه کاری m است. بخاطر binning، اگر تصاویر در یک مجموعه کاری یکسان انتخاب شده باشند، با احتمال تقریبا ۱ می‌توانند یکدیگر را بیابند. فرض کنید برای سادگی، سیاست مجموعه کاری نمونه گیری تصادفی در انتخاب مجموعه کاری است. احتمال اینکه دو تصویر ما در یک مجموعه کاری باشند برابر است با:

$$p = \frac{\#Subsets\ of\ containing\ 1,2}{\#subsets} = \frac{\binom{n-2}{m-2}}{\binom{n}{m}} = \frac{m(m-1)}{n(n-1)} \approx \frac{m^2}{n^2} \qquad (4.1)$$

این یک توزیع هندسی است بنابراین متوسط اتخاب $m^2/n^2$ از مجموعه‌های کاری قبل از اینکه تصاویر همگدیگر را بیابند نیاز است. هر مجموعه کاری به زمان $O(m)$ برای پردازش نیاز دارد بنابراین روی یک پردازنده تنها، زمان کلی $n^2/m$ صرف می شود. بنابراین اگر بتوانیم اندازه مجموعه کاری $m$ را افزایش دهیم $n$ افزایش می‌یابد. چون حافظه خلوت داریم، الگوریتم زمان کلی $O(n)$ صرف می‌کند اما همین که حافظه پر می‌شود الگوریتم $O(n^2)$ زمان می‌گیرد. هرچند، در برنامه‌های متعدد، لازم نیست تا تطبیق دقیقی بیابیم، تنها یک تطبیق معقول با اندکی خطا کفایت می‌کند. همچنین در آینده، با کشف مسائل بیشتری در انتخاب مجموعه کاری، ممکن است قادر به بهبود عملکرد باشیم.

### ۴.۳.۹   برنامه‌های کاربردی

در حال حاضر برنامه‌ای برای PatchWeb به غیر از آزمایشات مقدماتی با تکامل تصویر که تنها بخشی از آنها موفق بودند، پیاده سازی نشده است. دو چالشی که در تکامل تصویر با آنها مواجه شدیم آن است که (۱) الگوریتم‌های ترکیبی تمایل به افتادن در بهینه‌های محلی دارد، زیرا تقریبا هر patch ممکن در جایی در مجموعه داده نمایش داده شده با شد و (۲) تکه‌های ترکیب شده از منظره، معنایی درست، یا ویژگی‌های روشنایی درستی ندارند. برای برنامه‌های کاربردی، در پژوهش آینده ممکن است

۵۷

فرا تفکیک پذیری، حذف نویز، طبقه بندی تصویر با توصیف گره‌های متراکم، گسترش تصاویر شخصی، نقاشی با اعداد، یا فشرده سازی تصاویر یا فیلم مطرح شود.

### ۴.۳.۱۰ خلاصه

به طور خلاصه، در این فصل ما انواع موازی الگوریتم PatchMatch و نیز PatchWeb را نمایش دادیم که به نگرانی‌های مقیاس پذیری در مجموعه‌های تصویر بزرگ اشاره کردیم. ما دو عملگر جستجوی اضافی معرفی کردیم: enrichment و bunning، که به فرایند جستجو، به ویژه در مجموعه‌های تصویر بزرگ کمک می‌کنند. همچنین برنامه‌های کاربردی بالقوه از PatchWeb پیشنهاد کردیم.

# فصل پنجم

# برنامه‌های کاربردی و ویرایش تصویر

## ۵. ۱    مقدمه

هرچه عکاسی کامپیوتری و دیجیتالی گسترش می‌یابد، محققان روش‌هایی برای ویرایش سطح بالای فیلم و تصاویر دیجیتالی برای رسیدن به اهداف مطلوب توسعه می‌دهند. برای مثال، الگوریتم اخیر برای بازسازی تصویر، با تصاویر اجازه ی تغییر اندازه به نسبت ابعاد جدید می‌دهد؛ کامپیوتر به طور خودکار به تشابه خوبی از محتوای تصویر اصلی اما با ابعاد جدید تولید می‌کند. سایر الگوریتم‌های تکامل تصویر به کاربر اجازه می‌دهند به سادگی بخش ناخواسته‌ای از تصویر را پاک کنند و کامپیوتر به طور خودکار ناحیه را با تطابق‌هایی از باقی مانده تصویر پر می‌کند. الگوریتم‌های بُرزنی تصویر، تسخیر ناحیه‌هایی از تصویر و حرکت انها به اطراف را ممکن می سازند؛ کامپیوتر به طور خودکار باقی مانده تصویر را برای تشابه گیری با تصویر اصلی با توجه به نواحی حرکت یافته ترکیب می‌کند.

در همه ی این سناریوها، تعامل کاربر به دلایل متعددی اهیمت دارد: اول، این الگوریتم‌ها گاهی به دخالت کاربر برای رسیدن به نتایج بهتر نیاز دارند. الگوریتم‌های بازسازی تصویر، برای مثال، گاهی نظارت‌های کاربری برای تعیین یک یا چند ناحیه که باید بدون تغییر بمانند را نیاز دارند. با این وجود، بهترین الگوریتم‌های تکامل، ابزارهایی برای هدایت نتیجه با اشاره‌هایی به کامپیوتر پیشنهاد می‌کنند. این روش‌ها کنترل‌های اینچنینی برای بهینه سازی مجموعه اهدافی که برای کاربر و نه کامپیوتر شناخته شده اند فراهم می‌آورند. دوم، کاربر اغلب نمی تواند این اهداف را به طور قیاسی تفصیل کند. این فرایندهای هنری ایجاد تصویر مطلوب، استفاده از آزمون و خطا را می‌طلبند زیرا کاربر در جستجوی نتیجه‌ای با توجه به معیارهای شخصی در ملاحظه تصویر است.

نقش تعامل در فرایند هنری بر دو ویژگی در قالب ویرایش تصویر ایده آل دلالت دارد: (۱) ابزارها باید انعطاف پذیری در انجام انواع وسیعی از عملیات ویرایش یکپارچه برای کابران در کشف ایده‌های آنها فراهم آورد و (۲) عملکرد این ابزارها باید با اندازه کافی سریع باشد که کاربر به سرعت نتایج میانی را در فرایند آزمون و خطا ببیند. اغلب رویکرهای ویرایش سطح بالا تنها یکی از این معیارها را برآورده

۵۹

می‌کنند. این روش‌ها براساس patchهای نمونه برداری شده متراکم کوچک در مقیاس‌های چندگانه هستند و قادر به ترکیب هردو ساختار تصویر الگویی و پیچیده هستند که از لحاظ کیفی مشابه تصویر ورودی هستند [1]. بخاطر توانایی آنها در نگهداری ساختارها، این طبقه بندی از تکنیک‌ها را ویرایش تصویر ساختاری می‌نامیم. متاسفانه، تاکنون این روش‌ها در معیار دوم شکست خورده اند – آنها در کاربردهای تعاملی به جز در تصاویر کوچک کند هستند.

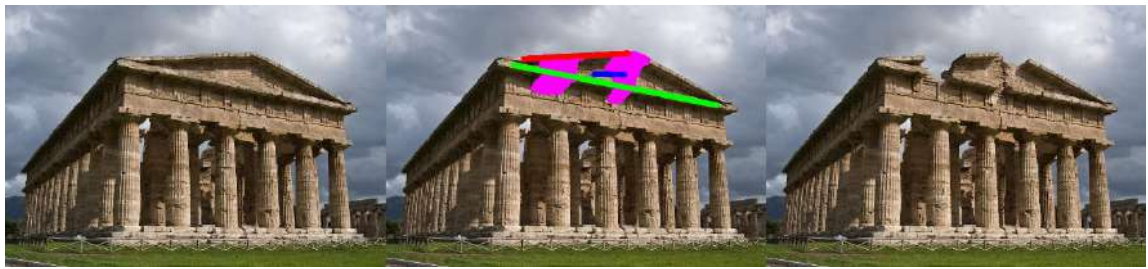

| (ج) حفره ترمیم شده | (ب) حفره‌ها و محدودیت‌ها | (الف) تصویر اصلی |

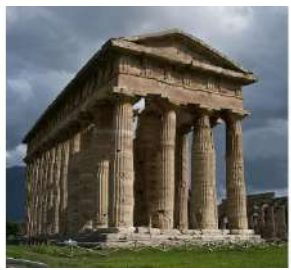 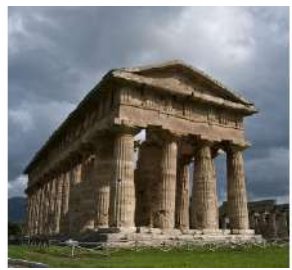 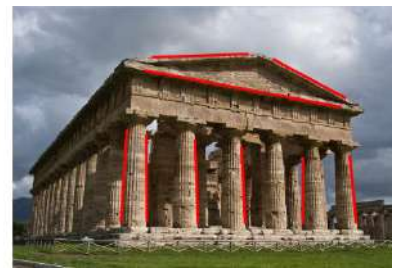

| (و) بُرزنی برای اصلاح تصویر | (هـ) بازسازی با استفاده از محدودیت‌ها | (د) محدودیت‌های خطی کاربر |

شکل ۵٫۱ – ویرایش ساختاری تصویر، (الف) تصویر اصلی، (ب) یک حفره علامت زده شده و از محدودیت خطی (قرمز/سبز/آبی) برای پیوستگی خط سقف استفاده می‌کنیم. (ج) حفره ترمیم شده در تصویر. (د) محدودیت‌های خطی توسط کاربر برای بازسازی. (هـ) بازسازی با استفاده از محدودیت‌ها به طور خودکار دو ستون را حذف می‌کند. (و) کاربر سقف بالایی را با بُرزنی انتقال می‌دهد.

## ۵. ۲    پژوهش‌های مرتبط

روش‌های نمونه گیری مبتنی بر تکه، ابزار معروفی در ترکیب فیلم و تصویرو تحلیل می‌شوند. برنامه‌ها شامل ترکیب الگو، تکمیل فیلم و تصویر، خلاصه سازی، ویرایش و بازسازی و ترکیب تصویر، اختلاط

تصویر، ترکیب نمای جدید، حذف نویز و غیره. ما سپس برخی از این برنامه‌ها را بررسی خواهیم کرد و درجه تعامل آنها را تشریح می‌کنیم.

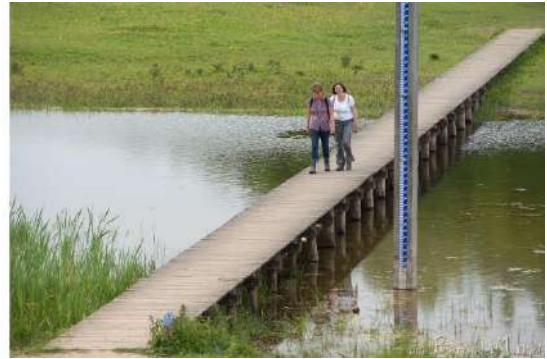
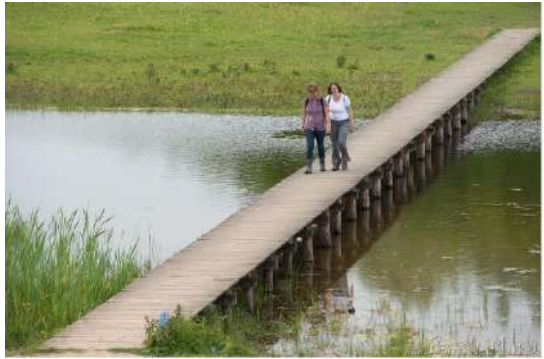

(ب) تصویر خروجی                                    (الف) تصویر ورودی

شکل ۵. ۱ – نمونه‌ای از کاربرد تکمیل تصویر. (الف) تصویر ورودی اصلی؛ (ب) بخشی از تصویر توسط کاربر انتخاب شده است و در تصویر نهایی حذف شده است. نتیجه مطلوب خروجی به علت بافت‌های تکراری تصویر ورودی است.

**تکمیل و ترکیب الگو:** Efros و Leung [37] یک روش ترکیب الگوی غیر پارامتری ساده معرفی کردند که مدل قبلی را بر اساس نمونه گیری patchهایی از یک نمونه الگو و چسباندن آنها در تصویر ترکیب شده اجرا می‌کند. بهبودهای بیشتر، رویکردهای جستجو و نمونه گیری را برای نگهداری ساختاری بهتر تغییر می‌دهند. ترتیب پر کردن حریصانه این الگوریتم‌ها گاهی ناسازگاری‌هایی را در زمان تکمیل کردن حفره‌های بزرگ با ساختارهای پیچیده ایجاد می‌کند [3, 29, 35, 36, 65, 69]. اما Wexler و همکاران [119] مسئله تکامل را به عنوان یک بهینه سراسری برای رسیدن به تکامل‌های سازگار سراسری بیشتر در نواحی ازد ست رفته بزرگ فرموله سازی کردند. این الگوریتم بهینه سازی چند مقیاسی تکرار شونده، مکررا در جستجوی patchهای نزدیک ترین همسایه برای همه ی پیکسل‌ها به طور موازی است. گرچه پیاده سازی اصلی کند بود (یک دقیقه برای تصاویر کوچکتر از یک مگاپیکسل)، الگوریتم بررسی شده در اینجا، این تکنیک را برای تصاویر بزرگتر در نرخ‌های تعاملی عملی ساخت. رویکردهای مبتنی بر patch در حال حاضر در ترکیب الگو رایج شده اند. در این زمینه،

Lefebvre و Hoppe [67] از الگوهای بروز رسانی موازی مرتبط و حتی پیاده سازی‌های مبتنی بر GPU بلادرنگ استفاده کردند.

**کنترل و تعامل پذیری**: یکی از مزایای الگوهای نمونه گیری patch آن است که آنها تعامل زیادی با مقیاس خوب دارند. برای مثال، در ترکیب الگو، روش Ashikhmin [3] در سراسر فرایند با مقداردهی اولیه پیکسل‌های خروجی با رنگهای مطلوب کنترل را به کاربر می‌سپارد. قالب تحلیل تصویر Hertzmann و همکاران [52] از تصاویر مجاور به عنوان "لایه‌های‌هادی"، در ایجاد انواع تاثیرات شامل فرا تفکیک پذیری، انتقال الگو، فیلترهای هنری، و الگو با اعداد استفاده می‌کند. در حوزه تکمیل تصویر، نتایج پر کردن با تفسیر ساختارهایی که در درون و بیرون ناحیه از دست رفته قرار دارند نشان داده شدند. Fang و Hart [39] ابزاری برای تغییر شکل منحنی‌های ویژگی تصویر در هنگام نگهداری الگوها رونمایی کردند که تنظیمات بهتری نسبت به ابزارهای ویرایشی، و نه در نرخ‌های تعاملی فراهم می‌کند. Pavic و همکاران [87] یک سیستم تکمیل تعاملی مبتنی بر patchهای بزرگ و چسباندن آنها رونمایی کردند. گرچه سیستم آنها تعاملا هر تکه انفرادی را می‌چسباند اما کاربر باید به طور دستی روی هر ناحیه تکامل کلیک کند پس فرایند کلی می‌تواند هنوز خسته کننده باشد.

**باز سازی تصویر**: روش‌های متعدد باز سازی تصویر، انحراف یا برش با استفاده از برخی معیارهای برجسته برای اجتناب از بدشکل شدن نواحی مهم تصویر را به کار می‌برند. حکاکی درز [23] [5, 94] از یک رویکرد حریصانه ساده برای اولویت بندی درزها در یک تصویر استفاده می‌کند که می‌تواند به درستی آنها را در بازسازی حذف نماید. گرچه حکاکی درز سریع است، اما ساختارها را به خوبی نگهداری نمی کند و تنها کنترل محدودی روی نتایج ارائه می‌کند. Simakov و همکاران [98] فریم بندی مسئله بازسازی تصویر و عکس به صورت بیشینه سازی شباهت‌های میان patchهای کوچک در تصاویر خروجی و اصلی پیشنهاد کردند و یک تابع هدف مشابه و الگوریتم بهینه سازی مستقلا توسط Wei و همکاران [118] به عنوان روشی برای ایجاد الگوهای خلاصه برای ترکیب سریع تر ارائه شده بود. متا سفانه، رویکرد Simakov و همکاران در مقایسه با حکاکی درز به شدت کند است. برنامه‌های بُرزنی تصویر و بازسازی محدود شده ما تابع هدف و الگوریتم تکرارشونده مشابهی همپون Simakov و همکاران را با استفاده از الگوریتم نزدیک ترین همسایه برای دستیابی به سرعت‌های تعاملی به کار

---

[23]. Seam carving

۶۲

گرفته اند. در همه ی این رویکردهای پیشین، روش اصولی کنترل کاربر قادر به تعریف و حفاظت نواحی مهم از اعوجاج ا ست. در مقابل، سیستم معرفی شده محدودیت‌های کاربر محوره خاص را در فرآیند بازسازی برای حفاظت صریح از خطوط در برابر ترکیب یا شکستن، با محدود کردن نواحی تعریف شده توسط کاربر به تبدیلات مشخص همچون مقیاس پذیری یکنواخت و غیر یکنواخت و ثابت کردن خطوط یا اشیا به موقعیت خروجی مشخص، یکپارچه می‌کند.

**بُرزنی تصویر** یک نظم آرایی محتوا در تصویر براساس ورودی کاربر بدون دقت چندانی است. بُرزدن همزمان توسط Simakov و همکاران [98] و Cho و همکاران [25] رونمایی شده بود که از patchهای تصویر بزرگتر و Belief Propagation در فرموله سازی MRF استفاده کردند. بُرزدن نیازمند کمینه سازی تابع خطای سراسری است. در مقابل در همه ی پژوهش‌های قبلی، روش بُرزدن ما کاملا تعاملی ا ست. چون این وظیفه ممکن ا ست د شوار و محدود شده با شد، این الگوریتم‌ها همواره نتیجه مورد انتظار را تولید نمی کنند. بنابراین تعامل پذیری اهمیت دارد چون به کاربر اجازه می‌دهد برخی ساختارهای مهم را در مقابل بُرزدن حفاظت کند و با سرعت بهترین نتیجه را در میان جایگزین‌ها انتخاب کند.

## ۵. ۳   ابزارهای ویرایش

در این بخش، برخی ابزارهای ویرایش تعاملی جدید که الگوریتم ایجاد می‌کند را تشریح می‌کنیم. اول، باید رویکرد ترکیب تشابه دو طرفه را دوباره مرور کنیم. این روش براساس یک اندازه گیری فاصله دوطرفه میان جفت‌های تصویر- تصویر (ورودی) مرجع S و یک تصویر هدف (خروجی) T است. اندازی گیری شامل دو عبارت می‌شود: (۱) عبارت تمامیت تضمین می‌کند که تصویر خروجی حدالامکان شامل همان اطلاعات بصری ورودی باشد و بنابراین خلاصه مفیدی است. (۲) عبارت وابستگی تضمین می‌کند که خروجی به ورودی وابسته است و ساختارهای بصری جدید(مصنوع) در این فرمول هزینه ایجاد می‌کنند [1]. به طور ر سمی، اندازه گیری فا صله به  سادگی به  صورت مجموع

میانگین فاصله همه‌ی patchها در S با مشابه ترین patchها به خود (نزدیک ترین همسایه)در T تعریف شده است و برعکس:

$$d_{BDS}(S,T) = \overbrace{\frac{1}{N_S}\sum_{s\subset S}\min_{t\subset T}D(s,t)}^{d_{complete}(S,T)} + \overbrace{\frac{1}{N_T}\sum_{t\subset T}\min_{s\subset S}D(t,s)}^{d_{cohere}(S,T)} \qquad (5.1)$$

که در آن فاصله $D$، $SSD$ (مجموع مجذور اختلافات) مقادیر پیکسل patch در فضای رنگ $L^*a^*b^*$ است. برای بازسازی تصویر، ما به دنبال حلی برای تصویر $T$ هستیم که $d_{BDS}$ را تحت محدودیت‌هایی از ابعاد خروجی مطلوب کمینه می‌کند. یک تخمین اولیه برای تصویر خروجی، $T_0$ در نظر بگیرید این فاصله تعاملا با یک الگوریتم شبیه EM کمینه شده است. در گام $E$ از هر تکرار $i$، مقادیر NN از $S$ و $T_i$ محاسبه شده اند و "انتخاب patch" برای تلفیق رنگ‌های پیکسل هر همسایه مشترک انجام شده است. در گام $M$، همه‌ی "انتخاب"های رنگ برای تولید یک تصویر جدید $T_{i+1}$ میانگین گرفته شده اند. برای اجتناب از گیر افتادن در کمینه محلی، یک فرایند "مقیاس گذاری تدریجی" به کار گرفته شده است: $T$ به یک نسخه با تراکم پذیری پایین از $S$ مقداردهی اولیه می‌شود و به تدریج با یک عامل کوچک، یا تعدادی تکرار EM پس از هر مقیاس‌گذاری تا رسیدن به ابعاد نهایی، تغییر اندازه می‌یابد. سپس، هردوی $T$ و $S$ برای تفکیک پذیری خوب تر به تدریج با تکرارهای EM بیشتر تا رسیدن به تفکیک پذیری خوب نهایی، نمونه برداری می‌شوند.

علاوه بر بازسازی تصویر، Simakov و همکاران [98] نشان دادند که این روش می‌تواند برای سایر وظایف ترکیب اختلاط همچون، بُرزنی، برش خودکار تصویر و تشابه در ویدئو استفاده شود. به علاوه، اگر یک ناحیه از دست رفته (یک "حفره") در یک تصویر به صورت $T$ و بقیه تصویر به صورت $S$ تعریف کنیم و تمامیت را حذف کنیم، با الگوریتم تکمیل ویدئو و تصویر Wexler [119] و همکاران، دقیقا به هدف می‌رسیم. نگاشت‌های وزنی مهم می‌توانند در هم ورودی (مثلا برای تاکید بر یک ناحیه مهم) و هم در خروجی (مثلا برای هدایت فرایند تکمیل در داخل مرزهای حفره) استفاده شوند.

الگوریتم NNF تصادفی داده شده در بخش 3. 4 هیچ محدودیت صریحی روی انحرافات به جز در کمینه سازی فاصله patch قرار نمی دهد. هرچند، با تغییر جستجو به روش‌های مختلف، می‌توانیم محدودیت‌های محلی روی آن انحرافات برای ارائه کنترل بیشتر به کاربر در فرایند ترکیب معرفی کنیم.

برای بازسازی تصویر، می‌توانیم به سادگی ناحیه‌های پُراهمیت را برای شناسایی نواحی که هرگز نباید تغییر یابند یا غیر یکنواخت مقیاس گذاری شوند همانند پژوهش قبلی، پیاده سازی کنیم. هرچند، می‌توانی صریحا محدودیت‌هایی تعریف کنیم که با روش‌های قبلی پشتیبانی به خوبی پشتیبانی نشده اند همچون، خطوط مستقیمی که باید راست باقی بمانند یا ا شائی که باید به موقعیت دیگری حرکت کنند در حالی که تصویر بازسازی شده است. الگوریتم می‌تواند اشیا را نسخه برداری (کپی و چسباندن) کند یا چیز دیگری تعریف کند که باید حفره حاصل را جابجا کند. اشیا می‌توانند به طور یکنواخت یا غیر یکنواخت (مانند برای "ر شد" درختان یا ساختمان‌ها به صورت عمودی) در محتوای ویرایش تصویر ساختاری تغییر مقیاس یابند. همه ی اینها به طور طبیعی با علامت گذاری چند ضلعی‌ها و خطوط در تصویر انجام شده اند. یک جعبه مرزی ساده برای اشیا اغلب کفایت می‌کند.

برخی از این ابزارهای ویرایش تصویر جدید هستند و بقیه قبلا در بخش‌های محدودی استفاده شدند. هرچند، ما باور داریم که مجموعه این ابزارها، در ارتباط با بازخورد تعاملی سیستم – قابلیت‌های ویرایش تصویر قدرتمند جدید و تجریه تعامل کاربری منحصر به فردی ایجاد می‌کنند.

### ۵.۳.۱    محدودیت‌های فضای جستجو

تکمیل تصویر نواحی از د ست رفته بزرگ یک کار چالش برانگیز ا ست. حتی اغلب روش‌های مبتنی بر بهینه سازی سراسری پیشرفته می‌توانند ناسازگاری‌هایی در جایی که محتوای ساختاریافته نشان داده نشده ایجاد کنند (مانند یک خط راست که از میان ناحیه از دست رفته می‌گذرد) به علاوه، در برخی موارد، مرزهای ناحیه ازدست رفته، اندکی یا هیچ محدودیتی برای یک تکمیل تصویر محتمل فراهم می‌کند. Sun و همکاران [106] به رویکرد انتشار ساختاری هدایت شده ارائه کردند که در آن، کاربر منحنی‌هایی در بالای یال‌ها رسم می‌کند که از خارج از حفره شروع می‌شوند و جایی که باید از میان آنها بگذرد را تعریف می‌کنند. این روش ساختار را به خوبی در منحنی انتشار می‌دهد اما فرایند کلی هنوز

کند است (اغلب برای یک تصویر ۰٫۵ مگاپیکسلی در حدود چند دقیقه) و گاهی مداخله دستی اضافی می‌طلبد.

در پژوهش ما، ما همان رویکرد تعامل کاربر را با اجازه دادن به کاربر برای رسم منحنی‌هایی در پیرامون ناحیه از دست رفته استفاده کردیم. منحنی‌ها می‌توانند برچسب‌های متفاوتی (با استفاده از رنگ‌های متفاوت نشان داده شد) برای مشخص کردن انتشار ساختارهای مختلف داشته باشد.

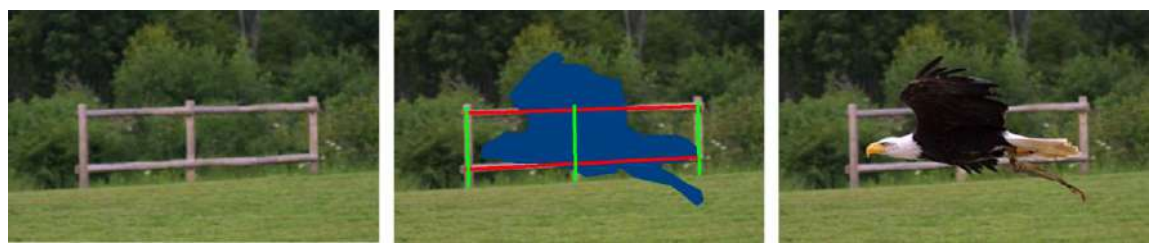

(الف) تصویر ورودی        (ب) حفره و محدودیت‌های کاربر        (ج) نتیجه تکمیل تصویر

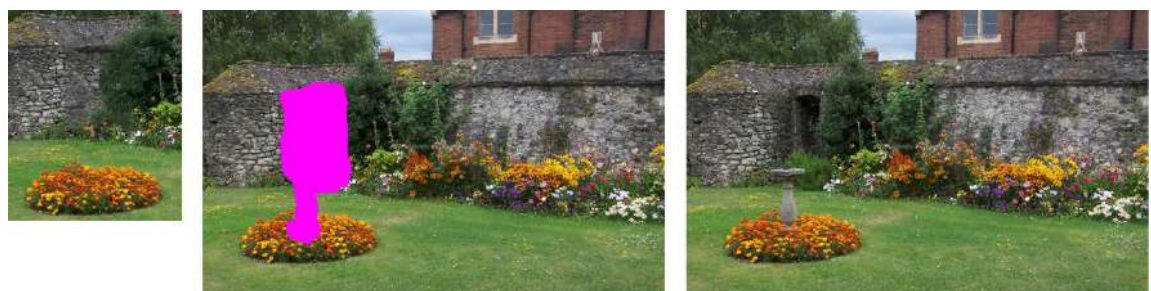

(د) تصویر ورودی        (هـ) حفره        (و) تکمیل تصویر

شکل ۵. ۳ – مثال‌هایی از تکمیل تصویر هدایت شده. (الف) پرنده از ورودی حذف شده است. (ب) کاربر ناحیه تکمیل را علامت‌گذاری می‌کند و محدودیت‌های جستجو را برچسب می‌زند. (ج) تولید خروجی در چند ثانیه. (د) گلها از ورودی حذف شده اند. (هـ) کاربر محدوده را مشخص می‌کند. (و) تولید خروجی

هرچند برخلاف Sun و همکاران، که فرایندهای تکمیل مجزایی را برای منحنی‌ها و نواحی الگو شده به کار می‌برند، سیستم ما هر دو را در قالب یکپارچه یکسانی ترکیب می‌کند. این با محدود کردن فضای جستجو برای پیکسل‌های برچسب زده شده درون ناحیه حفره تا نواحی بیرون حفره با همان برچسب‌ها انجام شده است. (متناقضاً، آوردن این محدودیت‌های اضافی به ویژگی‌های همگرایی ایجاد شده با محدود کردن فضای جستجو هم راستایی دارند). شکل ۵. ۱ تاثیر این محدودیت‌های فضای جستجو

برای تکمیل تصویر را نشان می‌دهد. علاوه بر ساختارهای یال و منحنی، ابزار یکسانی می‌تواند برای شناسایی محتوای مشخصی در برخی بخش‌های حفره استفاده شود. این نوع از تعامل شبیه به رویکرد "بافت با اعداد" [52] در هنگام به کارگیری تکمیل تصویر است. ما مثال‌هایی از این موارد را در شکل‌های ۵. ۱ و ۵. ۳ نشان داده‌ایم.

### ۲.۳.۵  محدودیت‌های تغییر شکل

تعدادی از روش‌های بازسازی اخیر به کاربر اجازه ی علامت گذاری نواحی مهم از نظر معنایی برای استفاده از سایر سرنخ‌های شناخته شده به صورت خودکار می‌دهد (مثلا، یال‌ها، چهره‌ها، نواحی برجسته). یک سرنخ مهم که در روش‌های قبلی چشم پوشی شد خطوط و اشیای با یال‌های مستقیم هستند که در تصاویر منظره‌های ساخته شده به دست بشر (تصاویر درون خانه، ساختمان‌ها، جاده‌ها) و در منظره‌های طبیعی (تنه ی درخت و خطوط افق) رایج هستند. بنابراین نگه داشتن چنین خطوطی برای تولید خروجی‌های محتمل اهمیت دارد. هرچند علامت‌گذاری یک خط به عنوان یک ناحیه مهم در تکنیک‌های موجود معمولا خط را در تصویر خروجی نمایان می‌سازد اما تضمین نمی کند که خط خمیده یا شکسته نشود (شکل ۵. ۱۰ را ببینید). علاوه بر آن ما اهمیت نمی دهیم که آیا خط در خروجی کوتاه تر یا بلند تر می‌شود یا چقدر راست باقی می‌ماند. مثال‌های اضافی از محدودیت‌های خط راست در شکل‌های ۵-٤ و ۵-۵ نشان داده شده است.

**محدودیت‌های مدل:** برای اینکار، ما فرموله سازی BDS از یک بهینه سازی آزاد برای یک مسئله بهینه سازی محدود شده را با محدود کردن بعد موقعیت نزدیک ترین همسایه ممکن در خروجی زیرمجموعه مشخصی از نقاط ورودی گسترش می‌دهیم. با پیروی از معادله (۵. ۱)، فرض کنید به همه ی پیکسل‌های متعلق به ناحیه (یا خط) $L_k$ در تصویر ورودی $S$ اشاره می‌کند و فرض کنید $\{q_j\} \in T | NN(q_j) \in L_k$ به همه ی نقاط $q_j$ در خروجی $T$ که نزدیک ترین همسایه در ناحیه $L_k$

قرار دارد اشاره می‌کند (نشانه گذاری‌های شکل ٥-٦ را ببینید). سپس حاصل معادله هدف (١. ٥) می‌شود:

$$argmin\ d_{BDS} \quad s.t. \quad \mathcal{M}_k\left(p_i, NN(p_i), q_j, NN(q_j)\right) = 0 \qquad (٢. ٥)$$

که در آن ما مدل $k$ به صورت $\mathcal{M}_k\ (k \in 1 ... K)$ را برای تصدیق داریم. معنی $\mathcal{M}_k\ () = 0$ به صورت مقابل است: در مورد خطوط، تصدیق یک مدل یعنی که ضرب نقطه‌ای خط سه برداری ($l$ ، در مختصات همگن) و همه‌ی نقاط تصویر خروجی باید صفر با شند برای مثال: $NN(p_i)^T l = 0$. در مورد نواحی، در اینجا به تبدیلات همگن دو بعدی ($H$) محدود می‌شویم و بنابراین تصدیق مدل یعنی که فاصله‌ی همه‌ی نقاط و نقاط $NN$ متناظر در تصویر خروجی باید صفر باشد، برای مثال: $H_{kP_i} - NN(p_i) = 0$.

اکنون این سوال باقی می‌ماند که بهتر است چطور این محدودیت‌ها را روی راه حل اعمال کنیم. ما شاهده کردیم که در طی فرایند (مقیاس گذاری تدریجی) خطوط و نواحی تنها با وجود کمبود فضا بد شکل می‌شدند. این به ما فرصتی می‌دهد تا این بد شکلی‌ها را با تنظیمات کوچکی پس از هر تکرار EM تصحیح کنیم.

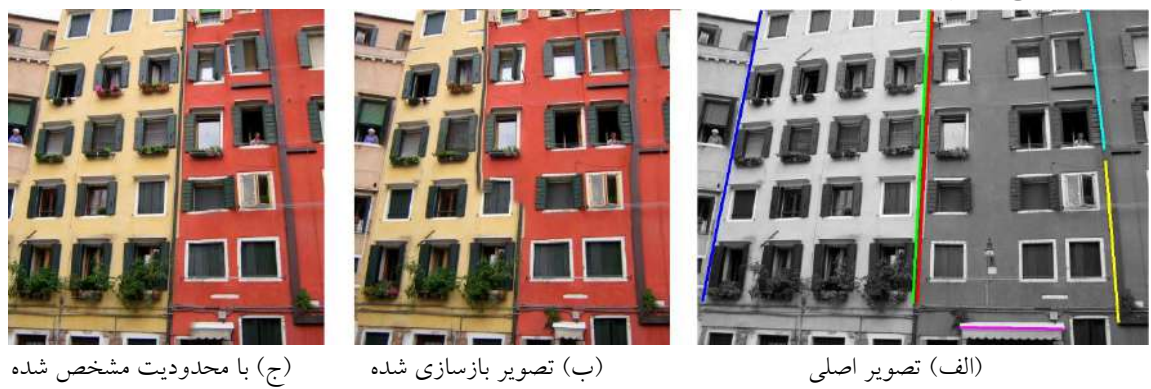

شکل ٥. ٤ – محدودیت‌های مدل. (الف) تصویر اصلی (ب) بدون محدودیت، با ایجاد مرزهای شکسته بازسازی شده است. (ج) زمانیکه محدودیت‌های نشان داده شده به صورت خطوط رنگی اضافه شده اند، مرزها راست باقی مانده اند.

بنابراین، به منظور تصدیق محدودیت‌های بیشتر، ما یک الگوی تصحیح تعاملی با استفاده از روش تخمین قوی RANSAK [41] پس از هر تکرار به کار می‌بریم. فرض می‌کنیم که در هر تکرار اغلب

موقعیت‌های $NN(p_i)$ و نیز $q_j$ تقریبا مدل مطلوب را تصدیق می‌کنند و این مدل را با دور ریختن نقاط غیر مقید انجام می‌دهیم. مدل تخمینی برای افکندن نقاط خروجی در روی $\widehat{NN(p_i)}$ و $\widehat{q_j}$ و تصحیح اساسی حوزه‌های NN استفاده شده است. برای نواحی، نقاط غیر مقید صحیح هستند اما ما نتایج بهتری برای خطوط با خارج کردن نقاط غیر مقید از فرایند انتخاب به دست می‌آید.

ما مدلهای زیر را برای بازسازی محدود شده و کاربردهای بُرزنی سودمند یافتیم: خطوط آزاد، با نقاط خروجی محدود شده به خطی راست با شیب و انتقال نامحدود (شکل ۵. ۵ را ببینید)؛ خطوط با شیب ثابت، با شیبی یکسان با ورودی، اما انتقال آزاد (شکل ۵. ۴ را ببینید)؛ خطوط با موقعیت ثابت، با شیب ثابت و انتقال توسط کاربر، برای آنکه هیچ مدلی برای تخمین وجود ندارد اما نقاط هنوز روی خطی که طول آن تغییر می‌کند قرار گرفته اند (شکل ۵. ۱۱را ببینید که خط آب با یک خط محدود به پایین کشیده شده بود)؛ انتقال نواحی، با انتقال آزاد اما مقیاس ثابت (شکل ۵. ۱۰ را ببینید که ماشین و پل به عنوان ناحیه انتقال علامت زده شدند)؛ و نواحی تغییر مقیاس یافته، با مقیاس یکنواخت تعریف شده توسط کاربر و انتقال آزاد (شکل ۵. ۵ را ببینید).

**محدودیت‌های سخت:** محدودیت‌های مدل تشریح شده در بخش قبلی معمولا در نگهداری خطوط و نواحی استفاده شده اند. هرچند، در موارد دشوار، با عوامل مقیاسی بزرگ – یا هنگامی که

محدودیت‌هایی میان محدودیت‌های مختلف وجود دارند – آنها نمی‌توانند به طور خودکار به خوبی تصدیق شوند.

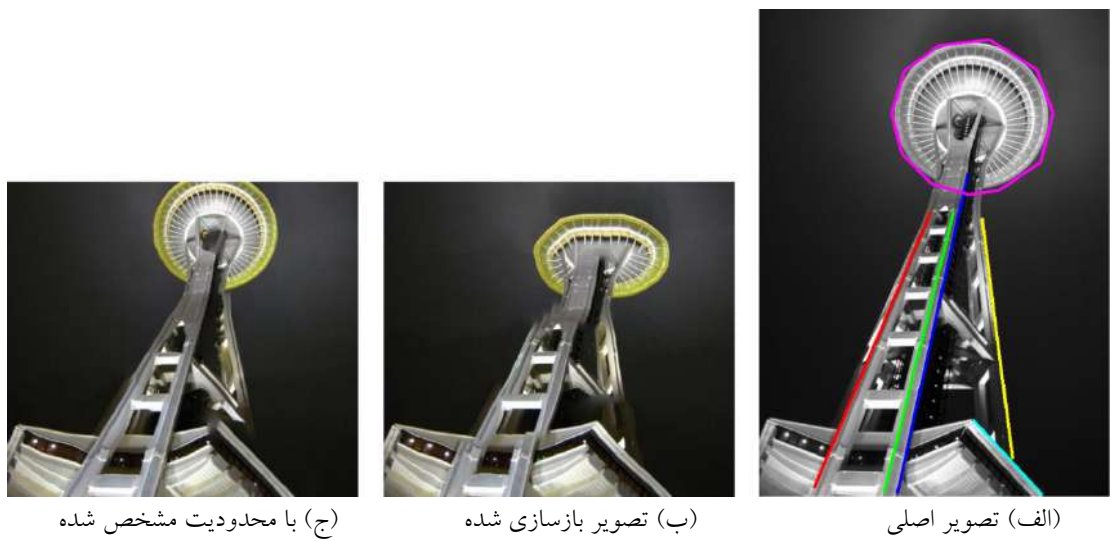

(ج) با محدودیت مشخص شده     (ب) تصویر بازسازی شده     (الف) تصویر اصلی

شکل ۵. ۵ – محدودیت‌های مقیاس پذیر یکنواخت و خطوط آزاد. (الف) تصویر اصلی (ب) بدون هیچ محدودیتی بازسازی شده است. (ج) محدودیت‌ها با خطوط رنگی، خطوط راستی را ایجاد می‌کنند و دایره برای پُرکردن فضای محدود شده کاهش مقیاس یافته است.

در موارد دیگر، همچون در بُرزدن تصویر، کاربر ممکن است بخواهد موقعیتی از یک ناحیه در خروجی به عنوان محدودیت سخت در بهینه سازی تعریف کند. این می‌تواند در قالب ما، با ثابت کردن حوزه‌های NN نقاط ناحیه معادل برای انحرافات مطلوب براساس محدودیت‌های سخت انجام شود.

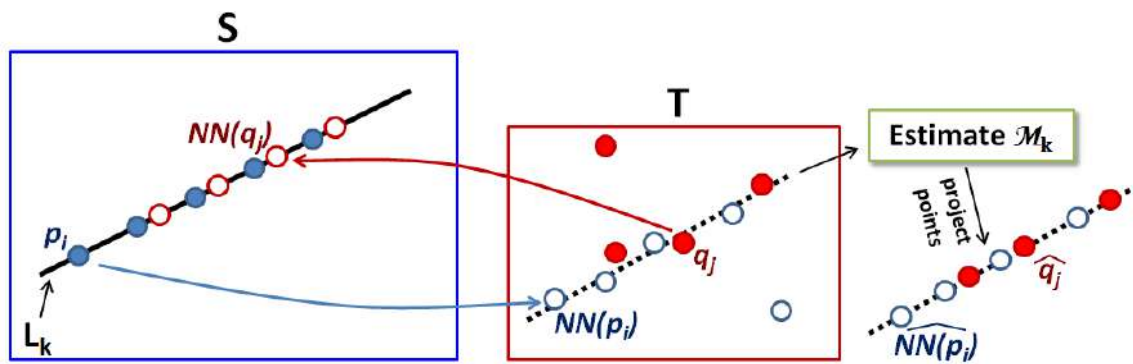

شکل ۵. ۶ – محدودیت‌های مدل. $L_k$ یک خط راست در تصویر مرجع S است. برای نقطه $p_i$ روی $L_k$، نزدیکترین همسایه در T، $NN(p_i)$ است. یک نقطه $q_i$ در T نزدیکترین همسایه $NN(q_j)$ دارد که روی $L_k$ قرار می‌گیرد. همه‌ی نقاط به شکل $NN(p_i)$ و $q_i$ گردآوری شده است و بهترین خط $M_k$ در T محاسبه شده است. سپس نقاط را در خط تخمینی قرار داده‌ایم.

پس از هر تکرار EM ما به سادگی انحرافات را برای موقعیت خروجی تصحیح کردیم، به طوری که سایر نواحی پیرامون اشیا به طور تدریجی خود شان را با این نواحی محدود شده هم تراز کردند. برای یک شی که از موقعیت خروجی اصلی خود جابجا می‌شود، سه گزینه ابتکاری برای کاربر در تشخیص مقدار اولیه محتوای حفره قبل از شروع بهینه سازی ارائه می‌کنیم: *تعویض*، که در آن سیستم به سادگی پیکسل‌های میان موقعیت‌های قبلی و جدید را تعویض می‌کند؛ *درون یابی*، که در آن سیستم با آرامی حفره‌های مرزها را درون یابی می‌کند (همانند پژوهش Wexler و همکاران [119])؛ *نسخه برداری*، که در آن سیستم به سادگی شی را در موقعیت اصلی آن نسخه برداری می‌کند. برای انتقالات کوچک، همه ی اینها خروجی‌های مشابهی تولید می‌کنند اما برای اشیا بزرگ و حرکت بزرگ، این گزینه‌ها منجر به نتایج کاملا متفاوتی می‌شوند (شکل ۵. ۱۱ را ببینید).

**مقیاس گذاری ساختاری محلی**: یک ابزار نشان داده شده در شکل ۵. ۸ به کاربر اجازه می‌دهد یک شی را علامت بزند و آن را در هنگام نگهداری ساختاری و الگویش تغییر مقیاس دهد (برخلاف مقیاس

گذاری منظم). ما این کار را با مقیاس پذیری تدریجی شــی و اجرای اندکی تکرار EM پس از هر تغییر مقیاس در تفکیک پذیری سخت، همانند فرایند بازسازی سراسری انجام می‌دهیم.

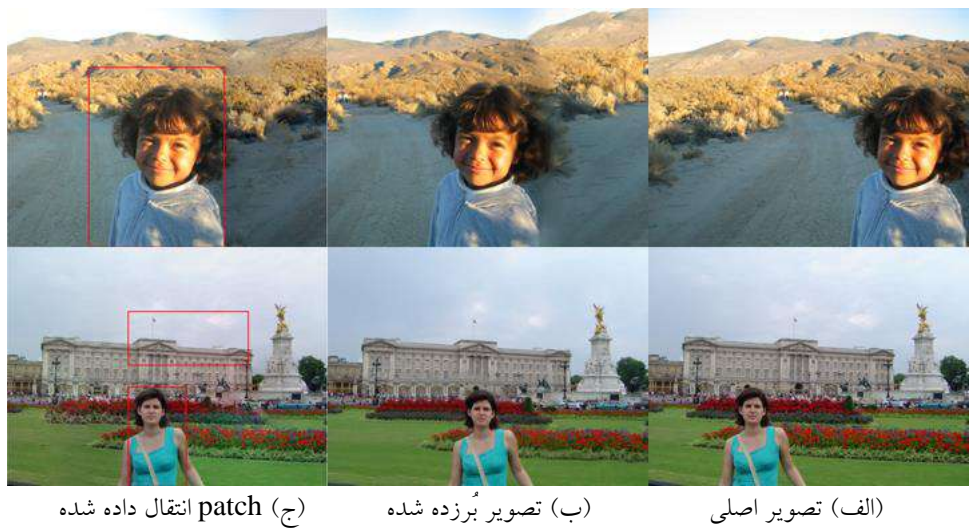

(الف) تصویر اصلی     (ب) تصویر بُرزده شده     (ج) patch انتقال داده شده

شکل ۵. ۷ – مثال‌هایی از بُر زدن. تصاویر ورودی نشان داده شده اند. کاربر یک ناحیه سخت برای حرکت مشخص می‌کند و الگوریتم پس زمینه را به روشی سازگاز تکمیل می‌کند. اینکار نتایج انتقال patch است که اجرای آن چند دقیقه طول می‌کشد ولی الگوریتم آن را به صورت تعاملی اجرا می‌کند.

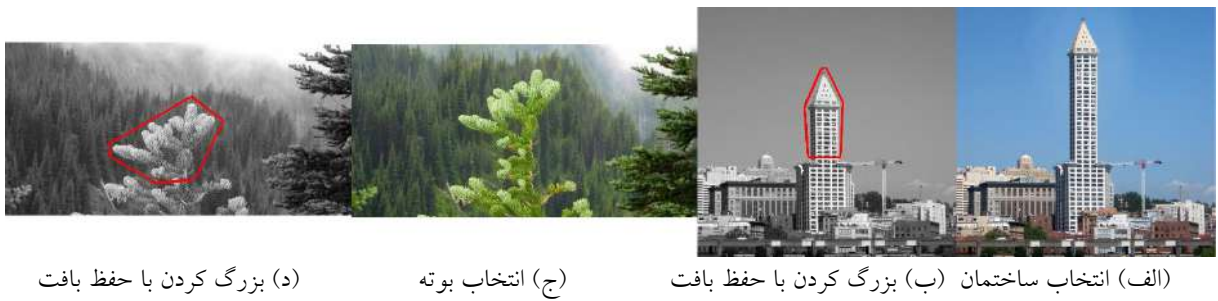

(الف) انتخاب ساختمان   (ب) بزرگ کردن با حفظ بافت     (ج) انتخاب بوته     (د) بزرگ کردن با حفظ بافت

شکل ۵. ۸ – مثال‌هایی از ابزار مقیاس گذاری محلی. در هردو مثال، کاربر یک چند ضلعی مرجع مشخص می‌کند و سپس یک مقیاس گذاری غیر یکپارچه به چند ضلعی، با حفظ الگو اعمال می‌کند.

۷۲

### ۵.۳.۳ بررسی پیاده سازی

تغییرات کوچک جهت گیری و مقیاس در برخی مدل‌های دگردیسی (مثلا خطوط آزاد و نواحی تغییر مقیاس یافته) می‌تواند با تنظیم ساده موقعیت‌های patch‌های موجود انجام شود. برای تغییرات مقیاس و زاویه بزرگ که ممکن است نیازمند عوامل بازسازی پیچیده باشد، ممکن است patch‌ها را چرخش/تغییر مقیاس نیز دهیم. در هر یک از موارد بالا، ما از یک نسخه وزنی از معادله (۵. ۱) استفاده می‌کنیم و وزن patch‌هایی که در خطوط و نواحی مهم قرار دارند تا ۲۰٪ افزایش می‌دهیم. همچنین یادآوری می‌کنیم که سطوح خوب‌تر از روی هم انباشتن، تخمین‌های اولیه بهتری هم دارند و بنابراین، مسئله جستجو

ساده تر است و تکرارهای EM کمتری نیاز است، و در سطوح خوب تر از تعدادی از تکرارهای EM استفاده می‌کنیم که خطی بودن را کاهش می‌دهد.

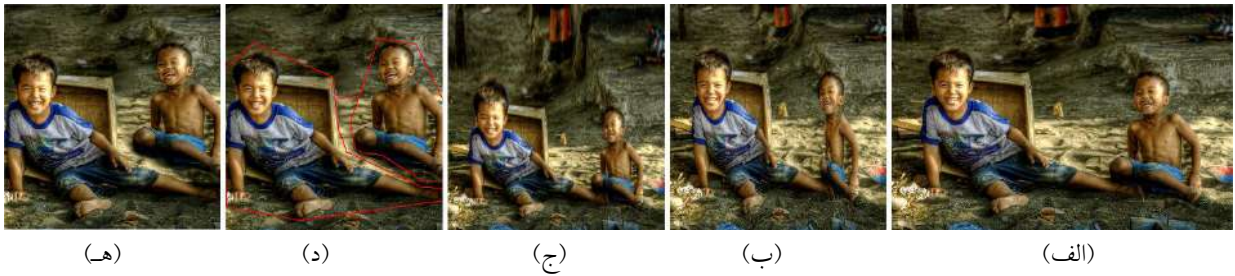

(هـ) (د) (ج) (ب) (الف)

شکل ۵. ۹ – بازسازی. (الف) تصویر ورودی، (ب) Rubinstein و همکاران، (ج) Wang و همکاران، (د) محدودیت‌های – کاربر برای الگوریتم PatchMatch، (هـ) نتیجه الگوریتم PatchMatch

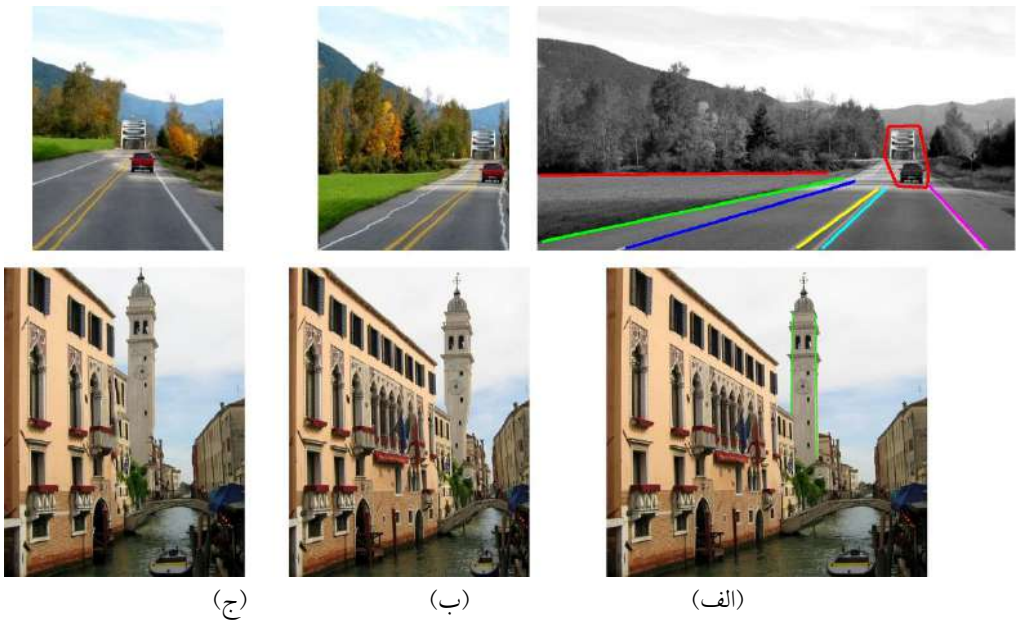

(ج) (ب) (الف)

شکل ۵. ۱۰ – بازسازی. (الف) تصویر ورودی با محدودیت‌ها، (ب) Rubinstein و همکاران، (ج) نتیجه الگوریتم بازسازی

برای سطوح آخر لایه‌ها روی هم رفته، تطابق سراسری کمتر ضرورت دارد بنابراین می‌فهمیم که کاهش شعاع جستجوی تصادفی به $w = 1$ چندان کیفیت را تحت تاثیر قرار نمی دهد.

5. 4    نتایج، تشریح و پژوهش آینده

همانطور که تشریح کردیم، قالب نزدیک ترین همسایه ما می‌تواند برای بازسازی، تکمیل حفره، و بُرزدن محتوای تصویر استفاده شود. ما عملکرد و کیفیت سایر روشهای رقیب در این حوزه را مقایسه کردیم.

شکل‌های ۵. ۹ و ۵. ۱۰ تفاوت‌های میان نتایج روش ما و نتایج Rubinstein و همکاران و Wang و همکاران را نشان می‌دهند. در شکل ۵. ۹ هردو روش باز سازی موجود شکل یک کودک را در عکس تغییر می‌دهند با این حال، سیستم ما به ما اجازه می‌دهد به سادگی تصویر یک کودک را بُر بزنیم بنابراین دستیابی به پس زمینه و فضای بیشتر برای بازسازی به طور خودکار به یک روش محتمل بازسازی شده است. در شکل ۵. ۱۰، می‌بینیم که "حکاکی حفره" اعوجاج‌های هندسی غیر قابل اجتنابی در خطوط راست ایجاد می‌کند و عناصر تکرار شونده را مقایسه می‌کند. در مقابل، روش ما به ما اجازه می‌دهد خطوط منظره را نگه داریم و عناصر تکرار شونده را حذف کنیم.

با ایجاد نواحی ماسک، کاربران می‌توانند به صورت تعاملی حفره‌ها را پر کنند. برای مثال، در شکل ۵. ۱، ما از سیستم خود برای انجام برخی اعمال روی تصویر معبد استفاده کرده‌ایم.

ابزارهای بُرزنی ما ممکن است برای تغییر سریع ابعاد معماری و طرحها همانطور که در شکل ۵. ۱۱ نشان داده شد استفاده شوند. ساختمان‌های قابل قبول از نظر بصری –گرچه فانتزی است– می‌توانند به سادگی با الگوریتم ساخته شوند زیرا معماری معمولا از الگوهای تکرار شونده استفاده می‌کند. بُرزدن می‌تواند همچنین برای حرکت افراد یا اشیا ارگانیک روی پس زمینه‌ها همانطور که در شکل ۵. ۷ نشان داده شد استفاده شود. هرچند، در برخی موارد محدودیت‌های خطی برای ثابت نگه داشتن هر عنصر خطی در پس زمینه که با پیش زمینه اشتراک دارد همانطور که در شکل ۵. ۹ که در آن محدودیت خطی برای ایجاد سایه روی ماسه استفاده شده است نشان داده شد مورد نیاز هستند.

با همه‌ی رویکردهای ترکیب تصویر، الگوریتم ما برخی شکست‌های موردی هم دارد. برجسته ترین آنها، برای ورودی‌های آسیب دیده همانند مورد آزمایش ترکیب در بخش ۳. ۸. ۱ است که ویژگی‌های همگرایی ضعیفی داریم. به علاوه، ویرایش‌های پیچیده یک تصویر گاهی می‌توانند نتیجه "سایه اندازی"

۷۵

یا "ماتی" ایجاد کند که در آن الگوریتم به سادگی نمی تواند یک حوزه کمینه محلی بزرگ را از آ سیب حفظ کند.

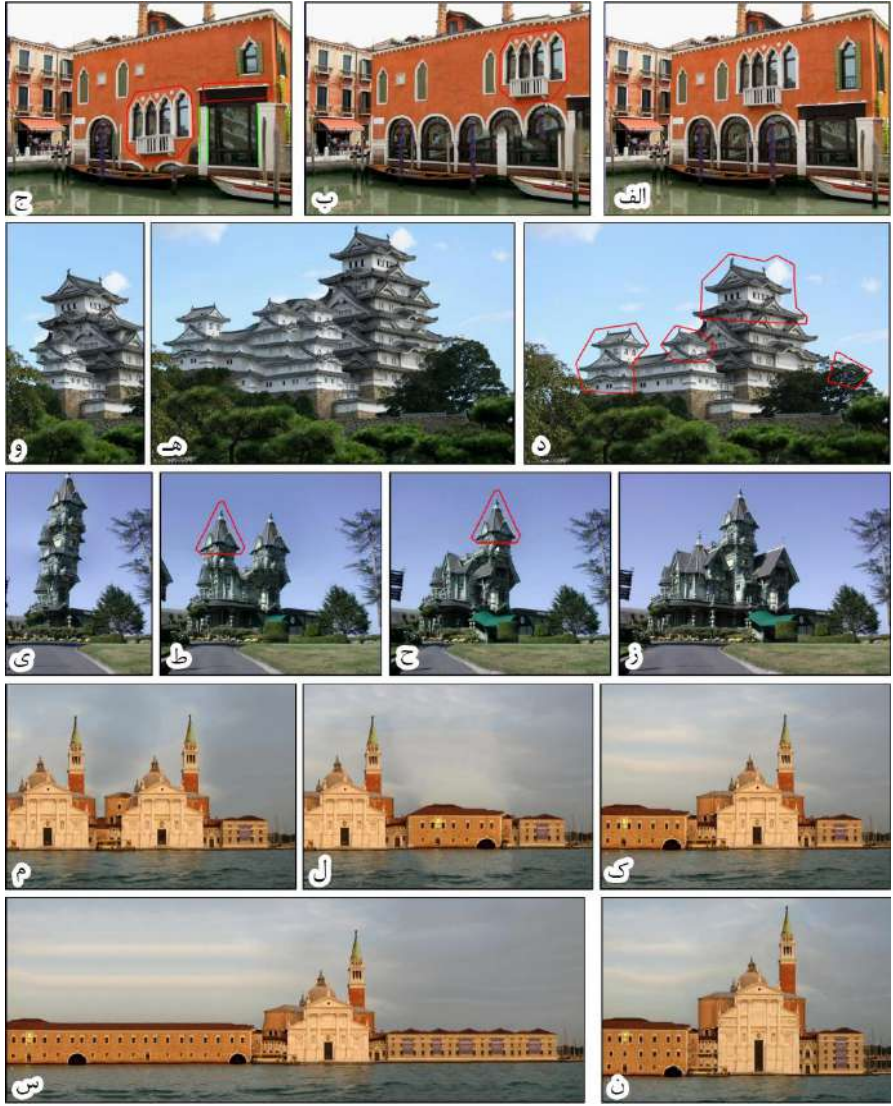

شکل ۵. ۱۱ – تغییر معماری با بُر زدن موقعیت پنجره برای یک (الف) ورودی تصویر به (ج) راست یا (د) به پایین ترین طبقه از ساختمان حرکت کرده است. (د-و) محدودیت‌های سخت روی یک ساختمان برای ایجاد اصلاحات مختلف استفاده شدند. (ز) خانه دیگر و (ح-ی) ترکیباتی از بازسازی و بُرزدن، انواع مختلفی از مدل سازی مجدد ایجاد میکنند. (م-ک) ساختمان‌های (ک) تصویر ورودی (ل) به طور افقی (م) نسخه برداری شده اند، (د) به طور عمودی کشیده شده اند و (س) گسترش یافته اند.

هرچند، اشـاره کردیم که سـرعت الگوریتم، تعریف محدودیت‌های اضـافی را ممکن می‌سـازد یا به سـادگی الگوریتم را با ورودی تصـادفی برای دسـتیابی به راه حل‌های مختلف از نو اجرا می‌کند. گرچه

چنین آزمون‌های تکراری می‌توانند هزینه سنگینی با الگوریتمی آهسته تر ایجاد کنند در آزمایشات ما، از چنین اکتشافات فضایی از دستکاری تصویر خلاقانه لذت بردیم.

در میان و سعت هیجان انگیزی از پژوهش‌های آینده در این حوزه، ما برخی از مهمترین‌ها را در اینجا برجسته می‌کنیم:

**توسعه الگوریتم تطابق**: ممکن است کسی از استراتژی k-وابستگی [109] با الگوریتم بهره ببرد. در کل، همانطور که در فـ صل ۲ تـ شریح کردیم، دریافتیم که الگوریتم نمونه برداری تـ صادفی بهینه تابعی از ورودی‌ها است. با کشف این سبک و سنگین‌ها و بررسی‌های بیشتر پیاده سازی‌های GPU، دستیابی به سرعت بیشتر با رونمایی برنامه‌های جدید پردازش فیلم و تصویر بلادرنگ ممکن می‌شود.

**سایر برنامه‌های کاربردی**: گرچه ما در این فصل روی دستکاری تصویر خلاقانه تمرکز داریم به پیاده سازی تکمیل، بازسازی و بُرزدن فیلم هم چشم داریم. Fitzgibbon و همکاران [42] نتایج جدیدی برای ترکیب منظره جدید با استفاده از روابط نمونه برداری patch و روابط هندسی به دست آوردند. جستجوی نزدیک ترین همسایه patch برای راه حل‌های مبتنی بر یادگیری سودمند نشان داده شده بود. ما همچنین فرض کردیم که با کار در حوزه مناسب، به کار بردن تکنیک‌های گفته شده در انجام باز سازی، پرکردن حفره و بُرزدن در هند سه‌ سه بعدی یا حرکت چهار بعدی با دنباله‌های شبیه سازی حجمی می‌تواند ممکن شود. نهایتا، گرچه بهینه سازی که انجام شده است هیچ رابطه همسایگی صریحی برای جلوگیری از انحرافات غیر پیوسته ندارد، ما باور داریم آن ممکن است سرعت آن به عنوان کامپوننتی در سیستم‌های تصویری نیازمند سرعت، تخمین‌های متراکم تناظر تصویر همچون ردیابی شی سودمند باشد.

خلاصـه، ما الگوریتم تطابق سـریع جدید خود را برای انواع گسـترده‌ای از رویکردهای بهینه سـازی مبتنی بر patch به کار بسـتیم، به طوری که ترکیب تصـویر می‌تواند اکنون در رابط تعاملی بلادرنگ به کار رود. به علاوه، سـادگی الگوریتم، تعریف متنوعی از کنترل‌های معنایی سـطح بالا با کاربرانی که می‌توانند ذاتا بهینه سازی را با محدود سازی فرآیند جستجوی نزدیک ترین همسایه جهت دهند را ممکن می‌سـازد. ما باور داریم نمایه‌هایی از انواع کنترل‌های تعاملی ایجاد کردیم که با اسـتفاده از این تکنیک امکان پذیر هستند.

# فصل ششم
# پرده‌های فیلم

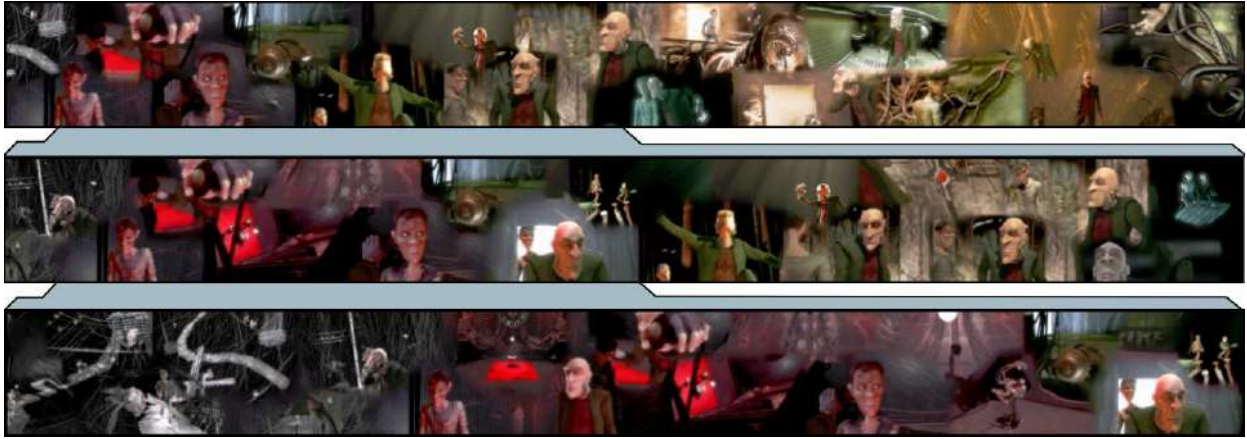

شکل ۶.۱ – یک پرده چند مقیاسی یک فیلم ورودی به صورت یک تصویر مختصر قابل بزرگنمایی و یکپارچه را نمایش می‌دهد که می‌تواند با جابجایی در فیلم استفاده شود. بصری سازی، حاشیه مشخص میان فریم‌ها را با ارائه پیوستگی مکانی و نیز بزرگنمایی‌های پیوسته با تفکیک پذیری‌های زمانی خوب تر، حذف می‌کند. این شکل سه سطح مقیاس گسسته یک فیلم را نمایش می‌دهد. خطوط میان هر سطح مقیاس، ابعاد متناظر میان مقیاس‌ها را نشان می‌دهد.

## ۶.۱  مقدمه

در سال‌های اخیر، تعداد فیلم‌های موجود در شکل دیجیتالی به تدریج افزایش یافت. هرچند، با توجه به حجم وسیعی از داده‌های درگیر، برای یک کاربر سخت است که منظره‌ای را در یک ویدئوی تنها بیابد یا سریعا محتوای یک ویدئو را با جستجو در یک نما درک کند. برای مثال، یک فیلم معمولا شامل ۲۰۰۰۰۰ فریم می‌شود و ممکن است حتی ۲۰ برابر تصاویر خام قبل از ویرایش آن به فیلم نهایی وجود داشته باشند. برای کمک به مردم در جستجو، درک و جابجایی فیلم، هم نرم افزار تجاری و هم سیستم‌های تحقیقاتی کشف کردند چطور فیلم به شکل تصاویر استاتیک خلاصه کنند.

در حوزه تجاری، ابزارهای ناوبری فیلم شامل آنچه در منوی فیلم‌ها و خطوط زمانی که دنباله ای از تصاویر را شامل می شوند همانطور که در نرم افزارهای ویرایش فیلم یافت می شوند، هستند. در مورد

خطوط زمانی، یک پیش نمایش برای هر تصویر از فیلم ممکن است نشان داده شود یا دنباله‌ای از تصاویر ممکن است در وقفه‌های زمانی با تناظر خطی با زمان‌های هر منظره نمایش داده شود. هر کدام از این نمایش‌ها موجب یک سبک و سنگینی می‌شوند: منوها به سادگی قابل درک هستند اما ممکن است کامل نباشند، جزئیات زمانی مهم را در نظر نگیرند. خطوط زمانی ویرایش فیلم کامل تر هستند زیرا شامل جزئیات دقیق تری می‌شوند – هر تصویر به صورت یک بلاک مجزا نشان داده شده است.- و آنها بزرگنمایی با جزئیات زمانی را فراهم می‌آورند اما معمولا انتقال ادراک در میان سطوح بزرگ نمایی کاستی‌هایی دارد و می‌تواند از نظر بصری در هنگام تماشا در سطح بزرگ نمایی بالا، گیج کننده باشد.

در این فصل، یک رویکرد جدید برای بصری سازی خطوط زمان و جستجو در میان مناظر یک فیلم ارائه می‌کنیم که قابلیت درک یک منوی تکامل یافته و ناوبری خط زمان ویرایش فیلم را فراهم می سازد. پژوهش اخیر در خلاصه سازی فیلم بیشتر روی انتخاب فریم‌های مهم –"فریم‌های کلیدی" نامیده می‌شد- و تمرکز داشت. هرچند، چنین سیستم‌هایی در کل یک یا چند بعد مهم از تجربیات کاربری در جستجو در میان فیلم را در نظر نمی گیرند. اول، استفاده از حاشیه مشخص در میان فریم‌های مربعی حاصلی از این است که فیلم چطور به صورت افقی نمایش داده می شود و یک نمایش بصری بهینه از فیلم اهمیت ندارد. حاشیه‌های مشخص در میان فریم‌ها، محتوای تصویر برجسته و گیج کننده اضافی با ایجاد نمایش‌های گسسته سخت در تجزیه بصری تعریف می‌کنند. در مقابل، رویکرد ما از تکنیک‌هایی الهام گرفته است که هنرمندان در طول تاریخ برای نمایش حرکات به صورت پیوسته بدون حاشیه‌های مشخص در سبک‌های هنری مانند پرده‌های هنری قرون وسطایی و ایرانی، حاشیه نویسی هنری و علائم رنگی برای هنر فیلمبرداری (شکل ۶. ۲) استفاده شده است. دوم، برای برنامه‌هایی مانند ویرایش فیلم، نمایش مختصر آنلاین به صورت بلادرنگ باید ممکن باشد. سوم، باور داریم که کاربران باید قادر به بزرگنمایی در خلاصه فیلم برای نمایش جزئیات زمانی با مقیاس خوب باشند. همانطور که بیان کردیم نا پیوستگی‌های شدید در طرح مکانی ایراد دار هستند، ما همچنین باور داریم که نا پیوستگی‌های بزرگنمایی مکانی نیز نامطلوب هستند، چون می‌توانند حواس کاربر را پرت کنند. تحریک انتقالات میان

مقیاس‌ها می‌تواند به کاربر در تلفیق محتوا و جهت‌گیری انتقال کمک کند. در حالی که برخی سیستم‌های قبلی برخی اهداف گفته شده را عملی می‌سازند، هیچ سیستمی همه‌ی آنها را عملی نمی کند.

یک قالب یکپارچه برای ایجاد خلاصه‌های یکپارچه از فیلم ارائه می‌کنیم که به کاربر اجازه می‌دهد به طور پیوسته در جزئیات زمانی بیشتری بزرگنمایی کند [2]. این خلاصه‌ها را پرده‌های چند مقیاسی می‌نامیم. نمونه‌ای از پرداه در شکل ۶. ۱ نشان داده شده است.

دستاوردها عبارتند از: مجموعه‌ای از معیارها برای ارزیابی کیفیت یک خلاصه بهینه؛ خلاصه‌های یکپارچه آمیخته شده در عرض‌شان با حفاظت از ناحیه دلخواه همراه آن؛ یک نمایش چند مقیاسی با بزرگنمایی پیوسته در میان مقیاس‌های زمانی؛ و تفسیر [24] پرده‌ها به صورت بلادرنگ یا پیش محاسبات محدود.

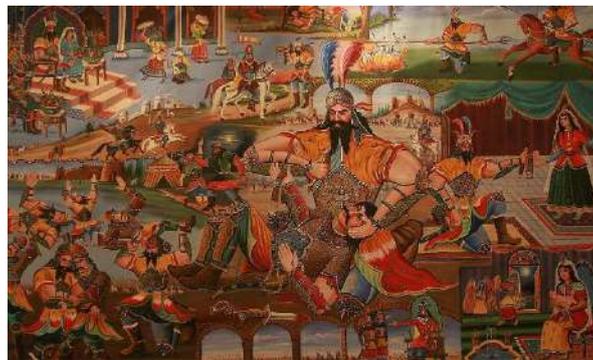

شکل ۶. ۲ – الگوریتم از تکنیک‌های هنری نمایش حرکات بدون حاشیه‌های مشخص همچون پرده‌های فیلم (شبیه نگارگری قدیمی قهوه‌خانه‌ای) و علائم رنگی برای هنر فیلم برداری الهام گرفته است.

۶٫۱٫۱    معیارها

برای ساخت پرده‌های فیلم چند مقیاسی، چهار کیفیت مهم پیشنهاد می‌کنیم که در خلاصه سازی بصری فیلم به ترتیب اولویت سخت مطلوب هستند.

---

[24]. Rendering

ابتدا، باید با نمایش تنها موجودیت‌های بصری که در فیلم مرجع ظاهر می‌شوند، *منسجم* با شد. این معیار تمایل ما به حذف مرزهای فریم را تقویت می‌کند زیرا چنین ساختارهایی در محتوای مرجع وجود ندارند.

دوم، باید نمایش دهنده رویدادها به به *ترتیب زمان وقوع* که با ترتیب زمانی آنها در فیلم متناظر می‌شود باشد. در این پژوهش همانند سایر مقالات، از ترتیب چپ به راست استفاده کردیم اما باید بالا به پایین و راست به چپ هم ممکن باشد. گرچه طرح افقی همه‌ی شکل‌های پنجره را پر نمی‌کند ناوبری با نگاشت طبیعی "چپ" و "راست" به "قبل" و "بعد" موثر واقع می‌شود.

سوم، باید در فضای مقیاس، با انتقالات بصری آرام از مقیاس‌های زمانی سخت به نرم، پیوسته باشد.

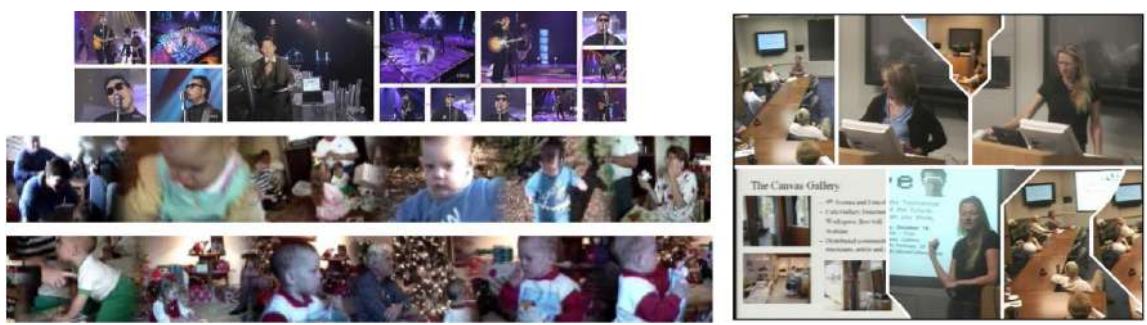

شکل ۳.۶.۳ – نمونه‌ای از چینش‌ها برای ایجاد پرده‌های فیلم در پژوهش‌های قبلی با استفاده از فریم‌های کلیدی و دیاگرام ورونوی، طرح مستطیلی و اختلاط تصاویر با حاشیه نرم

چهارم، شامل اغلب محتوای بصری منحصر به فرد در فیلم مرجع حدالامکان با همه پیکسل‌های موجود و *کامل* باشد. از آنجایی که پرده‌ها پیکسل‌های خیلی کمی نسبت به تصویر مرجع دارند، محتوای منحصر به فرد تر را در سراسر محتوای تکراری براساس اهمیت اولویت بندی می‌کنیم.

مشاهده می‌کنیم که دو تا این نیازمندی‌های پرده‌ها، وابستگی و تکمیل، قبلا با درک تشابه دوطرفه (BDS) فرموله‌سازی شدند و با موفقیت به مسائل مرتبط با اختلاط تصویر و فیلم اعمال شدند. به علاوه، روش‌های اخیر برای تخمین تشابه دوطرفه چنین الگوریتم‌های ترکیب تصویر تجربی برای کاربرد تعاملی کیفیت بالا ایجاد کردند که در فصل ۵ نشان داده شد. بنابراین، سیستم براساس الگوریتم ترکیبی Simakov و همکاران ساخته شد و الگوریتم PatchMatch توصیف شده در فصل ۳ به عنوان پایه و

۸۱

اساس آن، با محدودیت‌های اضافی برای اعمال به معیار دیگری از رویدادها به ترتیب زمان وقوع و پیوستگی مقیاس-مکان است.

## ۶. ۲  پرده‌های تک مقیاسی

برای سادگی، ما با در نظر گرفتن یک شی در یک مقیاس بزرگنمایی تنها شروع می‌کنیم. تابع هدف برای پرده تک مقیاسی بهینه از پژوهش Simakov و همکاران [98] الهام گرفته است، که یک تابع هدف کمینه برای بازسازی تصویر تعریف کرد. هدف شامل دو خطای مکمل $d_{complete}$ و $d_{cohere}$ می‌شود که تضمین کند هر patch در تصویر مرجع در تصویر هدف هم یافت شود و برعکس. هدف کامل به صورت زیر است:

$$d_{BDS}(S,T) = \overbrace{\frac{1}{N_S}\sum_{s \subset S} \min_{t \subset T} D(s,t)}^{d_{complete}(S,T)} + \overbrace{\frac{1}{N_T}\sum_{t \subset T} \min_{s \subset S} D(t,s)}^{d_{cohere}(S,T)} \quad (۶.۱)$$

که این معادله همچون معادله (۵.۱) برای ویرایش تصویر در نظر گرفته شده است. در اینجا $S$ مرجع یا تصویر اصلی است، $T$ تصویر هدف است، patchهای تصویر مستطیلی کوچک $s$ و $t$ از تصاویر مرجع و هدف نمونه‌برداری شده‌اند و تعداد patchهای مرجع و هدف $N_S$ و $N_T$ هستند. patchها با اندازه ثابت هستند و از بخش‌های ۷ در ۷ در تصاویر استفاده می‌کنیم. فاصله $D(S,T)$، فاصله فضای رنگی میان این patchهای مربع است و از فاصله $L^2$ در فضای RGB استفاده می‌کنیم. در هنگام بازسازی، تصویر مرجع $S$ ابعاد متفاوتی نسبت به تصویر بازسازی شده $T$ خواهد داشت. Simakov و همکاران [98] همچنین از تابع انرژی مشابهی برای فیلم‌های کوتاه شده با فرض اینکه $S$ و $T$ نشان دهنده حجم تصاویر در ابعاد مختلف و $s$ و $t$ نشان دهنده patch (جعبه)های کوچک فضا-زمان باشند، هستند [2].

رویکرد پرده تک مقیاسی از تعریف مشابهی از بهدینگی در معادله ۶.۱، با تغییرات مقابل در عبارات تعریف شده استفاده می‌کند: ما با یک فیلم ورودی $S$ با سه بعد، برای تولید یک تصویر خلاصه (پرده) $T$ با دو بعد آغاز می‌کنیم. ما تکه‌های تصویر $s$ و $t$ را می‌گیریم، بنابراین جمع در $d_{cohere}$ در سراسر همه ی تکه‌های تصویر $p \times p$ در تصویر هدف $T$ گرفته شده است و جمع در $d_{complete}$ در سراسر همه ی patchهای تصویر $p \times p$ در هر فریم از فیلم مرجع $S$ گرفته شده است. همچنین انتظار متحمل شدن حداقل یک ترتیب زمانی بی قاعده در خلاصه نهایی، با پیشرفت زمان تقریبی از چپ به راست

داریم. این با تعریف یک عبارت اضافی برای تابع فاصله $D(S,T)$ درمیان تکه‌های $s$ و $t$ انجام شده است که بعد زمان فیلم $S$ را به محور $x$ از خلاصه $T$ نگاشت می‌دهد.

$$D(S,T) = D_{color}(s,t) + \alpha(\tau_s - \beta x_t)^2 \qquad (۲.۶)$$

که در این $\tau_s$ زمان در فیلم ورودی را نشان می‌دهد، $x_t$ موقعیت افقی در پرده خروجی را نشان می‌دهد، $\alpha$ یک پارامتر کاربر است که کنترل می‌کند زمان چطور باید به صورت خطی از چپ به راست افزایش یابد و $\beta$ یک عامل تناسب فضا-زمان انتخاب شده است بنابراین طرف راست از پرده $T$ با آخرین فریم از فیلم ورودی $S$ منطبق می‌شود به یاد داشته باشید که $D_{color}$ فواصل فضای رنگی میان patchها است که در بالا توصیف شده است.

در اصل، با یک فیلم ورودی دلخواه $S$، بهترین خلاصه $T$ از یک تفکیک پذیری مشخص شده توسط کاربر می‌تواند با بهینه سازی معادله (۶.۱) یافت شود. هرچند، این در کل یک مسئله NP-سخت است، بنابراین، تقریب‌های کارا باید با حل کارای آن تعریف شوند. Simakov و همکاران [98] یک الگوریتم‌هادی تکرار شونده پیشنهاد کردند که با یک تخمین اولیه آغاز می شود و $T$ را برای تولید یک تصویر هدف با خطای کمینه تصحیح می‌کند. این الگوریتم تکرار شونده می‌تواند با الگوریتم PatchMatch توصیف شده در فصل ۳ برای انجام ترکیب تصویر در نرخ‌های تعاملی سرعت گیرد. پیاده سازی‌های ما به این رویکردها وابسته اند اما برنامه همچنان تعاملی خواهد بود. انجام بهینه سازی روی فیلم ورودی $S$ ناکارآمد خواهد بود زیرا در اغلب موارد این فیلم نمی تواند حتی در حافظه اصلی جای گیرد. حتی با حافظه کافی، بهینه سازی برای گام پیش پرداز خیلی کند، خواهد بود.

به همین دلیل، ما بهینه سازی را به دو گام، با شروع از سطحی از سراسر فریم‌ها و سپس پیشرفت به سطح بهتری از patchهای تصویر با فریم‌ها تقسیم می‌کنیم. اول، ما یک زیر مجموعه از فریم‌های کلیدی از فیلم ورودی می‌یابیم که یه تقریب سطح فریم از هدف (۶.۱) را در سراسر مجموعه فریم‌های ورودی، بهینه می سازد. این می‌تواند همچنین به صورت خوشه‌بندی در سراسر مجموعه فریم‌ها نشان داده شود. دوم، با استفاده از تنها این فریم‌های کلیدی به عنوان یک بعد محدود شده $S$ معادله ۶.۱ را در سطحی از patchهای تصویر کوچک بهینه می‌کنیم. چون گام خوشه‌بندی محاسبات د شواری ا ست

۸۳

می‌تواند به صورت بهینه با یک نمونه برداری با نرخ ثابت از فریم‌های ورودی یا حتی با استفاده از انتخاب فریم دستی جابجا شود.

## ۶.۳  پرده‌های چند مقیاسی

متاسفانه، فرموله سازی پرده‌های یک مقیاسی تو صیف شده در بخش قبلی معیارهای گفته شده از یکپارچکی در میان مقیاس‌ها را در خود جای نمی دهد. درواقع، در دو مقیاس مختلف، زیرمجموعه‌ای از فریم‌های کلیدی انتخاب شده در گام اول ممکن است جالب نباشند. و حتی اگر اینگونه باشد، ما هنوز با چالش‌هایی از درون یابی در میان تصاویر با اندازه مختلف و تفاوت فراوانی در محتوا مواجه هستیم.

بنابراین، ما باید اهداف خود در دستکاری پرده‌های چند مقیاسی را تغییر دهیم و گام سومی برای ترکیب پرده‌های میانی در میان مقیاس‌ها اضافه کنیم. در بخش‌های فرعی در زیر هر گام از تولید پرده چند مقیاسی را تشریح می‌کنیم. تولید پرده تک مقیاسی با دو گام اول کاملا تشابه دارد اما بدون محدودیت زیر مجموعه که کلمات کلیدی هر سطح بزرگ نمایی در سطوح بزرگنمایی بهتر حاضر هستند.

### ۶.۳.۱   خوشه‌بندی فریم کلیدی

در فرآیند خوشه‌بندی خودکار، برای یافتن تعداد ثابتی از فریم‌های کلیدی $n_1$ برای سخت ترین سطح (از $n_1 = 24$ ا استفاده می‌کنیم) تلاش می‌کنیم. برای سطوح بهتر $i = 2, 3, ..., d$، که $d$ تعداد سطوح برای دستیابی به بیشینه نرخ نمونه برداری زمانی (مثلا ۲ فریم در ثانیه) است، تعداد فریم‌های کلیدی $n_i$ را به صورت هندسی افزایش می‌دهیم ($n_i = n_{i-1} * \rho$) که در آن در پیاده سازی ما $\rho = 2$ است.)

برای تابع هدف معادله (۶.۱)، خوشه‌بندی خود را به صورت یک پیش محاسبه آفلاین انجام می‌دهیم. فرض کنید برای انتخاب مجموعه‌ای از فریم‌های کلیدی K در سطح بزرگ نمایی اول که تابع هدف را کمینه می‌کند تلاش می‌کنیم. برای انجام این خوشه‌بندی به صورت کارا، ما نیاز به ارزیابی سریع معادله (۶.۱) داریم. فرض کنید $f_1, f_2, ..., f_n$ فریم‌های گسسته از فیلم ورودی را نشان می‌دهند. ما یک

۸٤

ماتریس پیوستگی نامتقارن $n \times n$ با نام $A$ محاسبه می‌کنیم که در آن $A_{ij}$ یک جهت در تابع هدف ما (۶. ۱) است:

$$A_{ij} = \frac{1}{N_{f_i}} \left( \sum_{s \subset f_i} \min_{t \subset f_j} D_{color}(s,t) + \alpha \left( \tau_{f_i} - \tau_{f_j} \right)^2 \right) \quad (۶.۳)$$

$D_{color}$ فاصله رنگی میان $s$ و $t$ را نشان می‌دهد و $\tau_{f_i}$ و $\tau_{f_j}$ زمان مربوط به فریم‌های $i$ و $j$ را نشان می‌دهد. به یاد داشته باشید که فریم‌های کلیدی در K مقادیر زمان مشخصی دارند، بنابراین از آنها برای زمان مربوطه به طور مستقیم به جای زمان نگاشت زمان مکانی $\beta x_t$ در معادله (۶. ۲) استفاده می‌کنیم. به علاوه، جایی که $\left( \tau_{f_i} - \tau_{f_j} \right)^2$ از حد آستانه تجاوز می‌کند همواره اختلافات رنگ patch را نشان می‌دهد و بنابراین می‌توانیم گام پیش پردازش را با حذف محاسبات ماتریس پیوستگی خارج یک دسته قطر مرکزی تسریع کنیم.

با ماتریس پیوستگی $A$ و مجموعه‌ای از فریم‌های کلیدی $K = \{K_1, ..., K_m\}$، معادله (۶. ۱) را به صورت زیر تقریب می‌زنیم:

$$d_{BDS}(K) = \overbrace{\frac{1}{n} \sum_{i=1}^{n} \min_{j=1...m} A_{i,k_j}}^{d_{complete}(K)} + \overbrace{\frac{1}{m} \sum_{i=1}^{m} \min_{j=1...m} A_{k_i,j}}^{d_{cohere}(K)} \quad (۶.۴)$$

عبارت دوم $d_{cohere}(K)$ صفر است، چون عناصر قطری $A$ صفر هستند: هر یک از فریم‌های کلیدی نیز در مجموعه‌ای از فریم‌های ورودی وجود دارند. به یاد داشته باشید که این تنها یک محدوده ورودی برای هدف در ست است همانطور که برای $d_{complete}$ عملگر کمینه در معادله ۶. ۳ تنها در حوزه یک فریم کلیدی به جای مجموعه‌ای از فریم‌های کلیدی گرفته شده است.

در اینجا برای کمینه سازی این هدف الگوریتم خوشه‌بندی k-medoids [12] به کار گرفته شده است. فریم‌های کلیدی در هر سطح با گرفتن فریم‌های کلیدی سطح قبلی و حل مسئله برای بهترین فریم‌های کلیدی قبلی برای افزودن به طرح یافت می‌شوند. در حالی که فریم‌های موجود حذف نشدند یا تغییر نیافتند. بنابراین فریم‌های کلیدی در هر سطح یک زیرمجموعه از فریم‌های سطح بعدی هستند.

این فرآیند انتخاب فریم کلیدی خودکار معمولا در برجسته کردن اطلاعات مهم موثر هستند ولی می‌تواند به صورت بهینه با نمونه گیری فرعی ساده در زمان (مثلا ۲ فریم در ثانیه) جابجا شود یا کاربر

می‌تواند به صـورت دسـتی فریم‌های کلیدی را برای یک نتیجه بهتر انتخاب کند. نمونه گیری فرعی یکنواخت این مزیت را به ما می‌دهد که پرده نهایی تقریبا به صـورت خطی با زمان حرکت کند، که ممکن اسـت برای برنامه‌های ویرایش فیلم مطلوب باشـد که طول رویداد اهمیت دارد. به علاوه، هیچ فرایند مکانی نیاز ندارد. انتخاب دستی مزیت درک سطح بالای انسان از شخصیت‌ها، نمودار، و داستان را به همراه دارد [2].

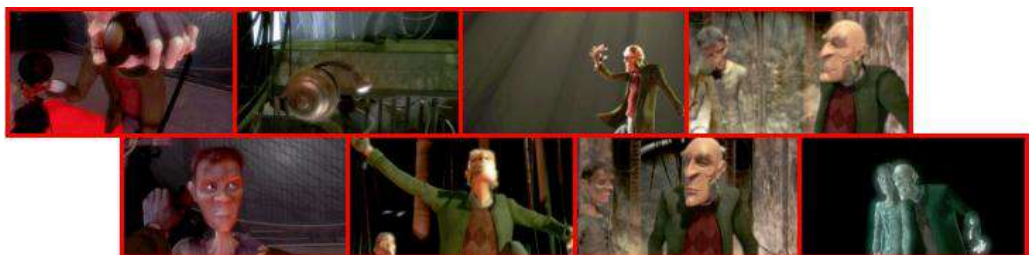

شکل ۶. ٤ - طرح آجری با استفاده از تخمین اولیه برای بهینه سازی

### ۲.۳.۶ پرده‌های مجزا

با فریم‌های کلیدی دلخواه، ما برای پرده بهینه به طور مسـتقل در هر مقیاس با کمینه سـازی هدف در معادله (۶. ۱) با روش Simakov و همکاران [98] مسـئله را حل کردیم که گام‌های سـرازیری در تابع هدف را در بر می‌گیرد [25]. از یک طرح آجری به عنوان تخمین اولیه خود ا ستفاده شده ا ست: فریم‌های کلیدی در دو ردیف مرتب شده اند، در ترتیب افزایش زمانی که در شکل ۶. ٤ نشان داده شد. سپس این تصویر ورودی را با یک عامل ٪۷۵ در جهت y با تقویت ساختارهای تکرار شونده برای ادغام و ترکیب یکپارچه رویداد باز سازی شده ا ست. این فرایند از طریق آزمایشات به د ست آمده ا ست؛ به یاد دا شته باشـید که گرچه از یک طرح ویژه همانند تخمین اولیه اسـتفاده می‌کنیم، اما تنها patchهای فریم‌های کلیدی خودشـان در عبارت $S$ اسـتفاده شـده اند بنابراین نواحی تخمین اولیه که با فریم‌های کلیدی چندگانه اشتراک دارند با فرایند بازسازی ترکیب یا فشرده خواهد شد.

این روش با موفقیت، پرده را با حذف ویژگی‌های ا ضافه ت صویر، خلا صه سازی عنا صر تکراری و ترکیب نواحی تصویر به صورت یکپارچه خلاصه می‌کند. هرچند، می‌تواند گاهی اطلاعات معنایی سطح

---

۲۵. این بار در سطح patch، برخلاف سطح فریم در بخش قبلی

۸٦

بالا همچون چهره‌ها را از بین ببرد، بنابراین، ما یک عامل وزنی برای هر patch تعریف می‌کنیم و وزن patchهایی که با چهره‌ها اشتراک دارند، افزایش می‌دهیم.

برای کارایی، پرده‌ها را در کاشی‌های گسسته از حدود و سعت ۵۰۰ پیکسل محاسبه می‌کنیم. اگرچه این در اصـــل خروجی را تحت تاثیر می‌گذارد، دریافتیم در عمل تاثیر اندکی دارد، چون زمان از تابع فاصله در معادله ۶. ۲ تعاملات میان نواحی پرده که خیلی منفصل هستند را محدود می‌کند. از یکپارچگی در میان کاشی‌ها با اشتراک کاشی‌های خروجی خودداری می‌کنیم و در فرآیند بازسازی یک محدودیت سخت تعریف می‌کنیم که رنگ‌ها در ناحیه اشتراکی باید کاشی‌های محاسبه شده قبلی را تطابق دهند. پیاده سازی این محدودیت‌های سخت مشابه با آنچه در فصل ۵ آورده شد است به استثنای آنچه اینجا پس از هر تکرار از الگوریتم بازسازی آورده شد، می‌باشد. همچنین در اینجا کپی‌های سایه دار از کاشی‌های مشخص را با ترکیب آلفا برای اجتناب از ناپیوستگی‌های بصری شدید ترکیب میکنیم.

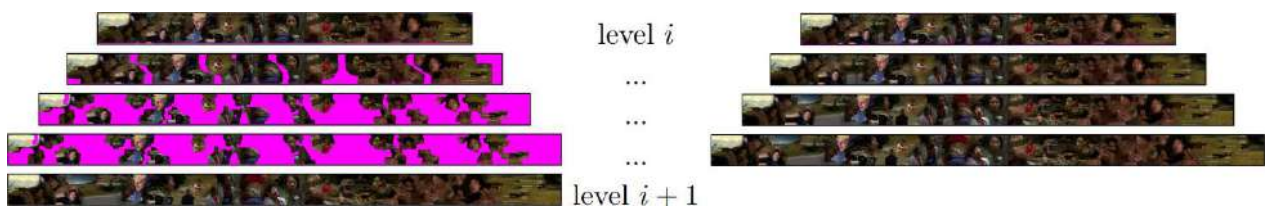

شکل ۶. ۵ - بزرگنمایی پیوسته. چپ: "محدودیت‌های "جزیره" روی دنباله بزرگنمایی محرک در میان یک پرده مجزا در یک مقدار بزرگنمایی کوچکتر (بالا) و مقدار بزرگنمایی بزرگتر(پایین). نواحی نشان داده شده در هر دو تصویر در حرکت با سرعت ثابت خطی محدود شده اند. نواحی قهوه‌ای محدود نشده اند. درون یابی با تصویر پایینی به عنوان تخمین اولیه آغاز می‌شود و تصویر را به اندازه‌های کوچکتر با حفظ جزیره‌ها با حرکت خطی به صورت محدودیت‌های سخت بازسازی می‌کند. جزیره‌ها به صورت تصویری کوچتر با تحمیل همگرایی توسعه می‌یابند.

راست: نتایج بازسازی ابنگونه است که نواحی (قهوه‌ای) محدوده نشده ترکیب شده اند. بزرگنمایی دنباله‌ای، یکپارچگی زمانی را نشان می‌دهد.

در همگام تفسـیر پرده‌ها، در حالت کیفیت بالا، همه‌ی کاشـی‌ها برای همه‌ی سـطوح بزرگنمایی از پیش محاسـبه شـده اند. در این هنگام پرده‌ها به صـورت یک فرآیند تعاملی با تفکیک پذیری پایین تر، برای کاشی‌های بزرگنمایی شده، محاسبه می‌شود. حتی در هنگام اجرا در حالت تعاملی با تفکیک پذیری پایین تر، فرایند بازسـازی حدود ۰٫۵ ثانیه برای اجرای هر کاشـی زمان می‌برد بنابراین، کاشـی‌ها را در

۸۷

حافظه نهان می‌گذاریم و از وقفه در رابط کاربری جلوگیری می‌کنیم و از یک thread پس زمینه برای پیش ذخیره کا شی‌ها در حافظه نهان در یک هم‌سایگی کوچک از موقعیت فعلی کاربر در جهت x و در فضای مقیاس استفاده می‌کنیم.

٦.٣.٣ بزرگنمایی پیوسته

در اینجا، ما پرده‌های گسسته را در مجموعه ثابتی از سطوح گسسته محاسبه کردیم. برای تصدیق نیاز به یکپارچگی در انیمیشن در میان سطوح گسسته، اکنون برای پرکردن حجم فضا-زمان میان هر جفت از پرده‌های گسسته به روش یکپارچگی زمانی تلاش می‌کنیم. فرض کنید یک پرده کوچک در سطح بزرگنمایی $i$ به صورت $A$ و یک پرده بزرگتر در سطح $i+1$ به صورت $B$ داریم؛ هدف تکمیل کردن ناحیه فضا-زمان میان $A$ و $B$ است.

الگوریتم پرکردن این ناحیه به صورت زیر پیش می‌رود: اول، سیستم ما نواحی متناظر در میان دو پرده مجزا $A$ و $B$ را شناسایی می‌کند. این نواحی به صورت خطی در امتداد شکاف میان $A$ و $B$ حرکت می‌کنند. نهایتا، نواحی باقی مانده با ترکیب‌های تشابه دوطرفه پُر می‌شوند.

**یافتن تناظرها**: در بخش ١.٣.٦ تضمین کردیم که فرم‌های داخل $A$ زیر مجموعه‌ای از فریم‌های داخل $B$ هستند. بنابراین می‌توان تناظرهای میان $A$ و $B$، مجددا با استفاده از الگوریتم تناظر سریع فصل ٣ پیدا کرد. این تناظرهای متراکم میان patchهای مربع کوچک با اندازه $p \times p$ را انتشار می‌دهد.

**شناسایی نواحی تناظر یکپارچه**: سپس برای یافتن نواحی متناظر میان دو پرده که اندازه آستانه‌ای دارند تلاش می‌کنیم. ما یک گراف با یال $E$ می‌سازیم و اجزای متصل بزرگ با حداقل اندازه $R$ را انتخاب می‌کنیم. گراف با یال $E$ راس‌هایی شامل مختصات $A$ دارد و یال‌های افقی و عمودی در صورتی نصب می شوند که حوزه تناظر ملایم با شد برای مثال بردار حوزه کمتر از حد آ ستانه تغییر کرده با شد. این نواحی بزرگی تناظر در میان دو تصویر به ما می‌دهد که آنها را "جزایر" می‌نامیم. در ناحیه فضا-زمان ما بین $A$ و $B$ را پر می‌کنیم، ما از نواحی "جزیره" متناظر به عنوان محدودیت‌های سخت ا ستفاده می‌کنیم و آنها به حرکت با سرعت خطی ثابت محدود شده اند. به خاطر داشته با شید که اگر یک فریم کلیدی از ردیف بالایی به پایینی (یا برعکس) تغییر کند ممکن است در طی بزرگنمایی ایراد دار باشد.

بنابراین، ما با اختیار خود گذر دوم از خوشه‌بندی مجموعه‌های پیوسته نیازمند خوشه‌بندی فریم کلیدی را اجرا می‌کنیم.

**بازسازی**: بعد، با محدودیت‌های داده شده، ما بخش‌های محدود نشده از بعد فضا-زمان را که با قهوه‌ای در چپ شکل ۵ .۶ نشان داده شد را پر می‌کنیم. ما در تصویر بزرگتر $B$ شروع می‌کنیم و به تفکیک پذیری‌های کوچکتر تا رسیدن به اندازه $A$ بازسازی می‌کنیم. برای هر تفکیک پذیری، مکررا الگوریتم خلاصه سازی را برای بازسازی تصویر فعلی با تصویر مرجع $S$ در معادله (۱ .۶) به صورت پرده بزرگتر $B$ اجرا می‌کنیم. ما عبارت $d_{complete}$ را از هدف حذف می‌کنیم زیرا بعد ورودی به محض انتقال به مقیاس گسسته بعدی تغییر می‌کند. همچنین فرآیند را با بازسازی تنها در بهترین مقیاس تسریع می‌کنیم.

**شرایط مرزی**: چون تخمین اولیه ما در $B$ شروع می‌شود، شرط مرزی در پایین بعد فضا-زمان اعمال می‌شود. هرچند، ما نیاز به تضمین این داریم که دنباله انیمیشن در پرده کوچکتر $A$ پایان یابد. برای انجام اینکار ما از بعد ورودی $S$ خود به عنوان ترکیبی وزنی از تکه‌های $A$ و $B$ با درون یابی خطی عامل وزن دهی ا استفاده می‌کنیم بنابراین انیمیشن به $A$ در فریم آخر همگرا می‌شود. ما همچنین جزایر را با بزرگ کردن آنها همانطور که $A$ را در بعد فضا-زمان رونمایی کردیم رشد می‌دهیم.

یک مثال از درون یابی فضا-زمان یکپارچه در شکل ۵ .۶ سمت راست نشان داده شد. همانند بخش قبلی، این شکل با کاشی‌های اشتراکی به صورت آنلاین محاسبه شده است.

### ۴ .۶ پیاده سازی و نتایج

نتایج نشان دهنده مقیاسهای پرده در شکل‌های ۱ .۶ و ۶ .۶ نشان داده شدند. فضا و زمان استفاده شده توسط الگوریتم ارائه شده براساس سطح تقریب‌های ایجاد شده متفاوت اند. پیش پردازش‌های خوشه‌بندی برای یک فیلم کوتاه (۱۰ دقیقه با نمونه برداری ۲ فریم در ثانیه) حدود ۷۵ دقیقه طول می‌کشد: ۱۵ دقیقه برای خوشه‌بندی واقعی و ۶۰ دقیقه برای محاسبه ماتریس پیوستگی مربعی. برای خوشه‌بندی فریم‌ها در فیلم‌های طولانی، محاسبه ماتریس پیوستگی زمان اجرای زیادی صرف می‌کند. به

عنوان یک تقریب، تمام ماتریس پیوستگی نیاز نیست محاسبه شواد و تنها دسته‌ای در فاصله آستانه قطر اصلی همانطور که در بخش ۶.۳.۱ بیان شد نیاز به محاسبه دارد [2].

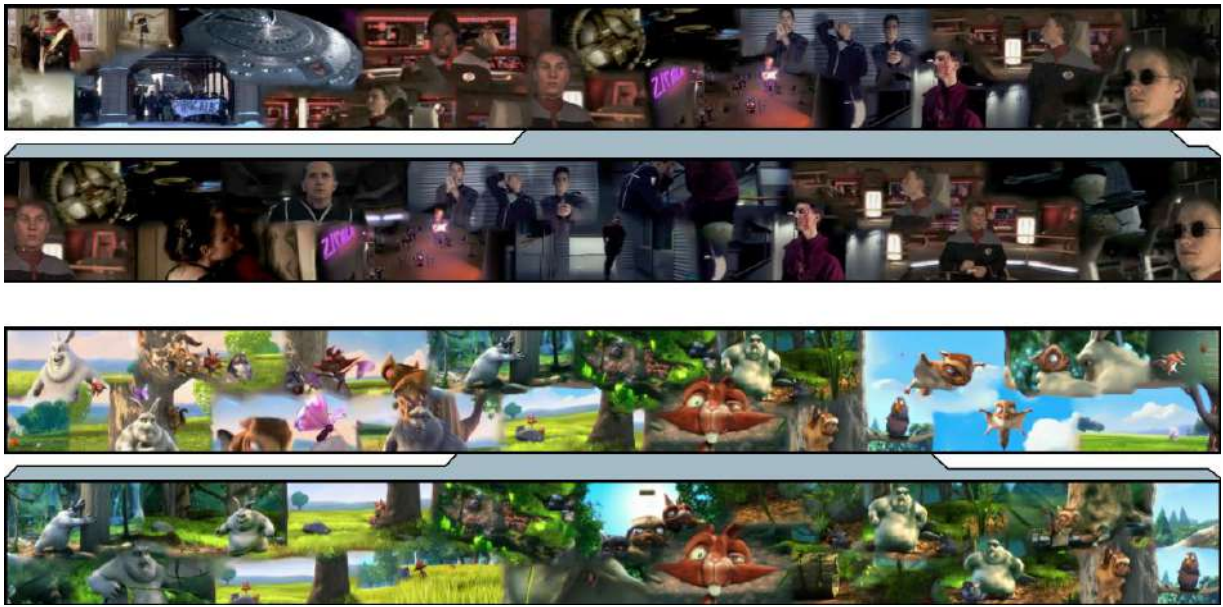

شکل ۶.۶ – پرده‌های مجزا برای دو فیلم مختلف. خطوط میان هر سطح بزرگنمایی نواحی متناظر در هر مقیاس را نشان می‌دهد.

با این وجود، حتی با این تسریع، محاسبه ماتریس پیوستگی برای یک فیلم ۷۵ دقیقه‌ای ۶ ساعت طول می‌کشد. به یاد داشته باشید با که فرآیندهای خوشه‌بندی بهینه هستند و می‌توانند صرف نظر شوند یا می‌توانند با فرآیند کارآتری جایگزین شوند.

به محض اینکه فریم‌های کلیدی انتخاب شـدند، پرده‌های کیفیت بالای با تفکیک پذیری بالا می‌توانند به صورت آفلاین محاسبه شوند. هر کاشی گسسته (به وسعت ۵۰۰ پیکسل) حدود ۱۰ ثانیه برای محاسبه زمان می‌برد. درون یابی بزرگنمایی پیوسته از کاشی‌های با اندازه یکسان استفاده می‌کند، و حدود ۱۰ ثانیه برای محاسبه حرکت بزرگنمایی میان مقیاس‌ها زمان می‌برد. متناوبا، رابط می‌تواند در نرخ‌های تعاملی در کیفیت و تفکیک پذیری‌های پایین‌تر اجرا شود. پیاده سازی تعامل ما کاشی‌های با وسعت ۵۰۰ پیکسل از هر پرده گسسته ۰٫۵ ثانیه‌ای را محاسبه می‌کند و هر کاشی بزرگنمایی می‌تواند در حدود ۲ ثانیه محاسبه شود. همانطور که قبلا تشریح شد، برای اجتناب از وقفه در رابط کاربری، ما کاشی‌ها را

۹۰

در همسایگی ناحیه فضای مقیاس فعلی کاربر از پیش بارگذاری می‌کنیم. بنابراین کاربر می‌تواند در پیرامون فضای مقیاس به صورت تعاملی حرکت کند.

## 6. 5    تشریح و پژوهش آینده

در هر گام از الگوریتم ارائه شده، ما تلاش می‌کنیم همه‌ی چهار معیار توصیف شده در بخش 6. 1. 1 را نگه داریم. در انتخاب فریم کلیدی، تحت محدودیت‌هایی تمام فریم‌ها پذیرفته یا رد می‌شوند. در حالی که ترتیب وقوع و یکپارچگی میان مقیاس‌ها با استفاده از محدودیت‌های زیر مجموعه به دست آمده اند. در تولید پرده ثابت، سیستم بهینه سازی‌هایی برای تشابه دو طرف با به کار بردن فاصله زمانی اضافی در ترتیب وقوع را استفاده می‌کند. در اینجا پیوستگی مقیاس-مکان به صورت صریح انجام نمی شود بلکه صریحا از محدودیت‌های انتخاب فریم کلیدی مشتق شده است. نهایتا، هنگام ایجاد حرکات بزرگنمایی، محدودیت پیوستگی با ایجاد و درون یابی "جزایر" انجام شده است، گردش یکپارچه رای پر کردن فضای باقی مانده بهینه شده است و ترتیب وقوع رویدادها صریحا از پرده‌های ثابت مشتق شده است.

ممکن است کارآمد به نظر برسد که قالب ارائه شده هیچ فرآیند تشخیص بریدگی صریحی همانند بسیاری از تکنیک‌های تحلیل فیلم رایج دیگر ندارد. هرچند، در فیلم‌های ویرایش شده، برش‌ها دیده نمی شوند بنابراین ما هیچ اهمیت مکانی در موقعیت برش قائل نمی شویم: تنها تشابه میان فریم‌ها اهمیت دارد. برای مثال، اگر یک منظره شامل سری سریعی از صحنه‌های جابجا شونده بین دو شخصیت باشد، خلاصه سازی آن با استفاده از فریم‌های کلیدی اندکی برای هر شخصیت به جای یک فریم کلیدی برای هر صحنه محتمل است.

**انتخاب فریم کلیدی:** الگوریتم انتخاب فریم کلیدی، یک راه حل تئوری ساده ارائه می‌کند که همان تابع انرژی را به صورت ترکیب صحنه مجزا بهینه می‌کند. هرچند، سیستم به استفاده از روش انتخاب فریم کلیدی وابسته نیست و درواقع سایر روش‌های انتخاب فریم کلیدی ممکن است برای برنامه‌های خاصی سودمند باشند. برای مثال، انتخاب فریم کلیدی یکنواخت دو مزیت سرعت –پیش محاسبه‌ای نیاز نیست- و یک نگاشت خطی از مختصات x پرده ارائه می‌کند. انتخاب فریم کلیدی دستی مزایای

ادراک انسانی و زیبایی شناختی در انتخاب فریم‌های کلیدی حاوی اطلاعات مفید و با زیبایی بصری ارائه می‌کند.

**موارد شکست:** گرچه الگوریتم ترکیب تشابه دوطرفه، محتوای بصری تکراری را حذف و ترکیب می‌کند، اغلب اتفاق می‌افتد که دو فریم کلیدی مجاور در یک طرح اولیه، محتوای بصری کاملا متفاوتی دارند. در این موارد، الگوریتم با ایجاد یک خط تیره و تار در امتداد مرز شکست می‌خورد. این روش از تشخیص چهره به عنوان یک پیش پردازش استفاده می‌کند: هنگامی که شکست می‌خورد می‌توانیم برخی چهره‌های لکه دار را در پرده نهایی ببینیم. اما هزینه‌ی مثبت‌های کاذب پایین است بنابراین می‌توانیم یک بخش پیچیده از منحنی ROC را برای تخمین نواحی چهره ا ستفاده کنیم. تنها نتیجه آن ا ست که برهی نواحی غیر چهره ممکن است در فشرده سازی شکست بخورند. با این وجود، تشخیص گر چهره ممکن است در کار روی چهره‌ها شکست بخورد (مثلا ربات‌ها، چهره‌های ریش دار، چهره‌های کارتونی). تشخیصی چهره تنها گزینه برای حفظ ساختار نیست؛ این روش شناسه‌های پیچیده‌ای از سایر اشیا و نواحی مهم در فیلم را پشتیبانی می‌کند که می‌توانند یا به صورت دستی یا تکنیک‌های بینایی کامپیوتری) مثلا تشخیصی گره‌های ناحیه برجسته و تشخیص گره‌های شی) تامین شوند.

توسعه دیگر در ایجاد پرده‌های محرک خواهد بود که در آن، ناحیه زیرین ماوس شروع به "نمایش" یک دنباله کوتاه یا دنباله‌های متعدد در پرده به منظور بصری سازی محتوای پویا از آن منظره می‌کند. بهبودهای دیگری در تولید پرده می‌توانند با افزودن یک عبارت سازگاری مرزی برای فریم‌های کلیدی مجاور برای تقویت پرده‌های پیوسته تر به د ست آیند. در این پژوهش ما یک بهینه سازی ارائه کردیم که در کمینه سازی یک تابع انرژی با تعریف گام‌های متعدد از تقریبات ارائه شود. در آینده به کشف امکان ساخت یک پرده با کمینه سازی مستقیم یک تابع هدف یکپارچه به جای گام‌های جایگزین متعدد با تقریب‌های متفاوت علاقمند هستیم.

خلاصه، ما تکنیک خلاصه سازی فیلم جدیدی ارائه کردیم که پیوستگی مکانی و بصری سازی‌های قبل بزرگنمایی فیلم را ایجاد می‌کند. ما یک تابع هدف برای چنین خلاصه‌هایی ارائه کردیم و این هدف را با ا ستفاده از یک سری گام با محا سبه گام‌های بلادرنگ بهینه می‌کنیم. امیدواریم که این تکنیک برای ناوبری فیلم، ویرایش فیلم و نگهداری نتایج جستجوی فیلم مفید واقع شود.

# فصل هفتم
# کاربردهای بینایی کامپیوتری

## ۷. ۱   مقدمه

محاسبه تناظرهای میان نواحی تصویر مسئله اصلی در مسائل بینایی کامپیوتری است. از مسائل سنتی مانند ردیابی الگو و ردیابی جریان حرکتی، تا پردازش تصویر سطح پایین مانند ترکیب الگو و رنگ آمیزی، تا وظایف تحلیل تصویر سطح بالا مثل تشخیص شی، طبقه بندی و تقسیم بندی تصویر. جستجوهای تناظر می‌توانند به صورت محلی، که جستجو در پنجره مکانی محدود شده‌ای انجام می‌شود یا سرا سری، که همه جایگزین‌های ممکن در نظر گرفته می شوند، با شند. تناظرها می‌توانند همچنین به صورت خلوت، با تشخیص تنها در یک زیر مجموعه از نقاط کلیدی مهم یا متراکم، با تشخصی در هر پیکسل یا در یک شبکه متراکم در ورودی طبقه بندی شوند.

برای کارایی، الگوریتم‌های رایج متعددی تنها از تناظرهای محلی یا خلوت استفاده می‌کنند. جستجوی محلی می‌تواند تنها جایگزین‌های کوچک را شناسایی کند بنابراین پالایش‌های با تفکیک پذیری‌های چندگانه اغلب استفاده شده است، اما حرکت‌های بزرگ اشیا کوچک می‌تواند از دست رود.

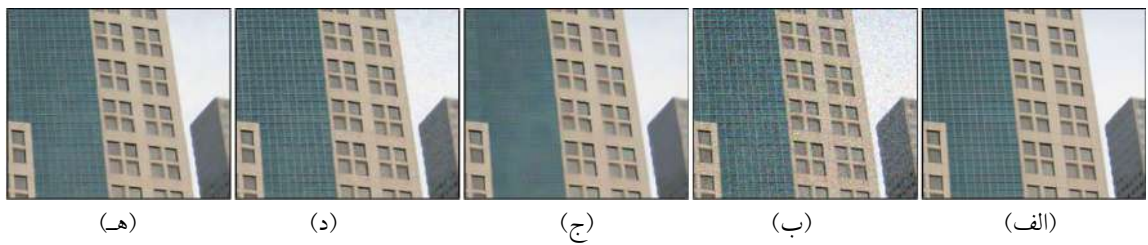

(هـ)   (د)   (ج)   (ب)   (الف)

شکل ۷. ۱ – حذف نویز با PatchMatch سراسری. (الف) تصویر اصلی (ب) خراب شده با نویز گووسی. (ج) Buades و همکاران. (د) با میانگین گیری از patchهای مشابه، در یک پنجره محلی کوچک نویز حذف شده است. (هـ) استفاده از PatchMatch برای جستجوی غیرمحلی، بهبود ویژگی‌های تکرار شونده را در بر دارد، اما نواحی نویزی باقی می‌مانند.

تناظرهای نقاط اصلی خلوت معمولا برای همترازی، بازسازی‌های سه بعدی و شناسایی و تشخصی شی استفاده شده اند. این روش‌ها روی مناظر الگودار با تفکیک پذیری بالا خوب کار می‌کنند اما در

سایر موارد کارایی کمی دارند. روش‌های پیشرفته تر [20, 99] با تطابق‌های خلوت شروع می شوند و سپس آنها را انتشار می‌دهند. بنابراین، چنین روش‌هایی می‌توانند از آرمیدگی فرضیات خلوت و محلی بهره ببرند. به علاوه، برنامه‌های تحلیل متعدد و برنامه‌های ترکیب [98, 43, 37, 29] به طور یکپارچه‌ای به تناظرهای سراسری متراکم برای عملکرد مناسب نیاز دارند.

الگوریتم تطابق PatchMatch کلی، تناظرهای سراسری متراکم را سریع تر از رویکردهای قبلی می‌یابد همچون کاهش ابعاد (مثلا PCA) ترکیب شده با ساختارهای درختی مانند درخت‌های kd، درخت‌های VP و TSVQ. الگوریتم ارائه شده یک نزدیک ترین هم سایه تقریبی در یک تصویر برای هر مستطیلی کوچک (مثلا ۷ در ۷) در تصویر دیگر با استفاده از یک استراتژی تپه نوردی تصادفی می‌یابد.

مسائل بینایی، سناریوهای تطابق چالش برانگیز را در بر می‌گیرند بنابراین ما نیاز به تمام توان الگوریتم تطابق کلی داریم. برای مثال، برای مسائلی مانند تشخیصی شی، حذف نویز و تشخیص تقارن، ما در تشخیص تطابق‌های داوطلب متعدد برای هر patch تقاضه شده تلاش می‌کنیم. بنابراین ما الگوریتم نزدیک ترین همسایه‌های k را نیاز داریم. همچنین، برای مسائلی مانند فرا تفکیک پذیری، تشخیص شی، طبقه بندی تصویر و ردیابی T ممکن است ورودی‌ها در مقیاس‌ها و چرخش‌های متفاوتی باشند بنابراین انواع الگوریتم تطابق مختلفی نیاز داریم که در میان این ابعاد جستجو می‌کنند. سوم، برای مسائلی مانند تشخیص شی، patchها برای تغییر در ظاهر و هندسه سه بسیار بزرگ هستند بنابراین ممکن است رد این موارد به تطابق با انواع توصیف گرهای تصویر نیاز داشته باشیم.

در این فصل، کاربردهایی از الگوریتم جدید برای مسائل بینایی کامپیوتر زیر ارائه می‌کنیم: حذف نویز محلی، تشخیص کپی، تشخیص تقارن، SIFT و تشخیص شی. باور داریم که الگوریتم PatchMatch کلی می‌تواند به عنوان یک کامپوننت کلی در نوعی از روش‌های بینایی کامپیوتری موجود و آینده به کار گرفته شود و قابلیت آن برای حذف نویز تصویر، یافتن جعل در تصاویر، تشخیص تقارن و تشخیص شی را بررسی می‌کنیم.

## ۷.۲   حذف نویز میانگین غیر محلی

برای حذف نویز تصویر، Buades و همکاران [21] نشان دادند که نتایج کیفیت بالا می‌توانند با حذف نویز میانگین غیر محلی به دست آیند. یافتن patchهای مشابه در تصویر و سپس میانگین گیری

۹٤

آنها. پژوهش‌های زیر مجموعه نشان دادند که این روش مبتنی بر patch می‌تواند برای دستیابی به نتایج جدید با انجام گام‌های فیلترینگ اضافی توسعه یابند. هنگامی که Buades و همکاران برای تکه‌های مشابه در پنجره جستجوی محدودی جستجو می‌کردند، Brox و همکاران [19] نشان دادند که یک روش مبتنی بر درخت می‌تواند برای دستیابی به کیفیتی بهتر برای برخی ورودی‌ها استفاده شود. هرچند آنها فاصله را برای patchهای دور افزایش می‌دهند بنابراین جستجو هنوز محدود به برخی نواحی محلی است.

الگوریتم kNN که در فصل ۳ توصیف شد می‌تواند برای یافتن patchهای مشابه در یک تصویر استفاده شود بنابراین می‌تواند به عنوان یک کامپوننت در این الگوریتم‌های حذف نویز نیز استفاده شود [4]. بنابراین روش ساده‌ای با استفاده از آنچه گفته شد و الگوریتم kNN برای این کار پیاده‌سازی شده است. این روش با آزمون هر patch مرجع از یک تصویر با انجام یک جستجوی محلی در سراسر همه patch ی‌ها در یک فا صله ثابت r از تکه مرجع با محا سبه یک فا صله $d$ ی با وزن $L^2$ در میان patch مرجع و هدف و محاسبه یک میانگین وزنی برای رنگ پیکسل مرکزی با تابع وزنی $f(d)$ کار می‌کند.

برای اســتفاده از الگوریتم kNN ما در این قالب حذف نویز، می‌توانیم به ســادگی تعدادی از همسایه‌های k را انتخاب کنیم و برای هر patch مرجع، از kNN آن در سراسر تصویر به صورت لیستی از patchهای هدف استفاده کنیم.

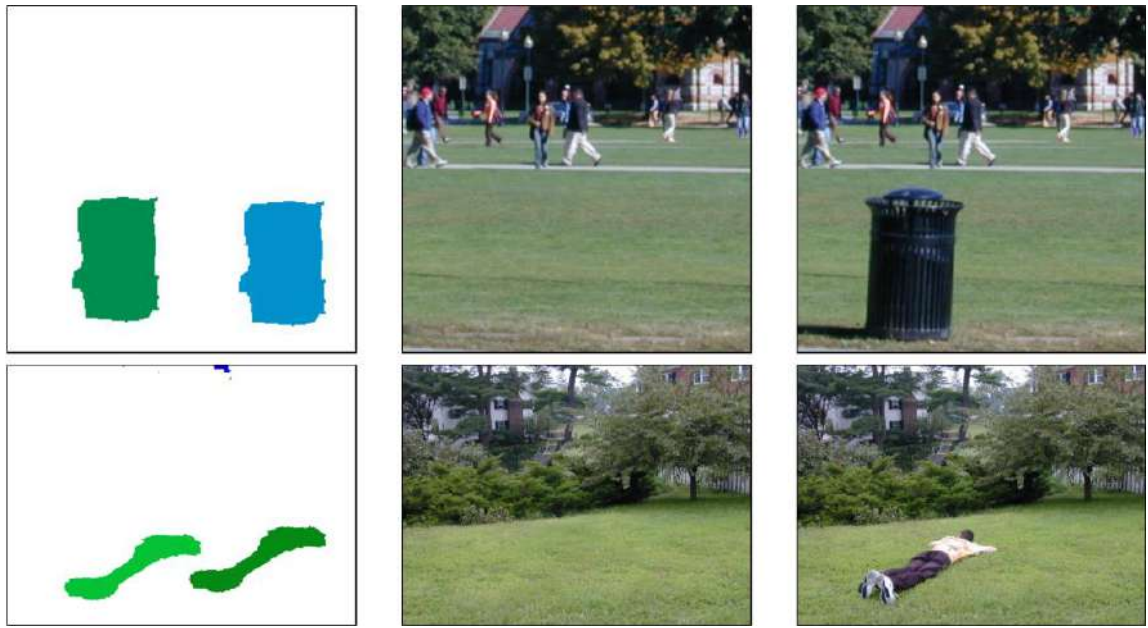

(الف) تصاویر اصلی   (ب) تصاویر جعل شده   (ج) شناسایی بخش‌های جعل شده

شکل ۷. ۲ – تشخیص نواحی تصویر جعل شده با روش کپی کردن بخش‌های دیگر. (الف) تصاویر اصلی که تغییر نکرده است. (ب) تصاویر جعل شده. (ج) نواحی کپی شده با الگوریتم kNN و کامپوننت‌های مرتبط با آن.

## ۷. ۳   تشخیص کپی

یک تکنیک برای جعل تصاویر دیجیتالی، حذف یک ناحیه از یک تصویر با کپی برداری ناحیه‌ای دیگر است.

۹۶

روش‌های تشخیص چنین جعل‌هایی اخیرا ارائه شده است [11, 89]. این روش‌ها، تجزیه تصویر به patchهای مربع یا اشکال نامنظم، استفاده از PCA یا DCT برای رد بخش‌های فرعی در تصویر با توجه به نویز یا فشرده سازی و مرتب کردن بلاک‌های حاصل برای تشخیص تکرار را پیشنهاد می‌کنند.

ما می‌توانیم الگوریتم kNN خود را برای تشخیص نواحی کپی شده استفاده کنیم. به جای مرتب کردن همه‌ی بلاک‌ها در یک لیست مرتب، می‌توانیم برای هر تکه، k-NN آن را به عنوان کاندیدهای کپی شده در نظر بگیریم و نواحی کپی شده را با تشخیص "جزایر" متصل از patchهایی که همه نزدیک ترین همسایه‌های مشابهی دارند شناسایی کنیم.

به خصوص، ما یک گراف می سازیم و کامپوننت‌های متصل از گراف را برای شناسایی نواحی کپی شده ا ستخراج می‌کنیم. رئوس گراف مجموعه‌ای از همه مختصات پیکسل $(x, y)$ از تصویر هستند. برای هر مختصات $(x, y)$، یک یال افقی یا عمودی در گراف می‌سازیم اگر kNNها به ترتیب با هم‌سایه در $(x+1, y)$ یا $(x, y+1)$ مشابه با شند، دو مجموعه $A$ و $B$ از kNN را مشابه می‌نامیم اگر برای هر جفت از نزدیک ترین همسایه‌ها $(ax, ay) \in A$ و $(bx, by) \in B$ ، نزدیک ترین همسایه‌ها در یک فاصله آستانه $T$ از یکدیگر باشند و هردو یک فاصله patch کمتر از یک آستانه فاصله بیشینه داشته باشند.

نهایتا، کامپوننت‌های متصل در گراف را تشخیص می‌دهیم و هر کامپوننت را با یک ناحیه در بالای کمینه اندازه ناحیه کپی شده $C$ در نظر می‌گیریم تا ناحیه کپی شده باشد.

مثال‌هایی از پیاده سازی تشخیص کپی در شکل ۷. ۲ نشان داده شدند. به یاد داشته باشید که نواحی کپی شده به درستی شناسایی شده اند. هرچند، ناحیه کپی دقیقا مثل ناحیه حذف شده نیست زیرا نمونه اولیه ما برای نویز، نتیجه فشرده سازی، یا آراستن بزرگ نیست.

## ٤ .٧  تشخیص تقارن

تشخیص ویژگی‌های متقارن در تصاویر اخیرا مورد توجه بوده است. یک بررسی تکنیک‌ها برای یافتن تقارن‌های چرخشی و انعکا سی تو سط Park و همکاران [86] انجام شده ا ست. روش‌هایی نیز برای یافتن تقارن‌های انتقالی در شکل شبکه‌های منظم توسعه داده شده اند.

چون تطابق‌های الگوریتم kNN ما، ویژگی‌های غیر محلی را تکرار کردند می‌توانند به عنوان یک کامپوننت در الگوریتم‌های تشخیص تقارن استفاده شوند [4]. تقارن‌ها با نقاط خلوت دلخواه تشخیص

۹۷

داده شــده اند. مانند تشــخیص گرههای مرز یا SIFT یا نقاط یال. در مقابل روشهای خلوت، الگوریتم ارائه شده میتواند توصیف گرهای نمونه گیری شده متراکم همانند توصیف گرهای SIFT و patch را تطبیق دهد و تقارنها میتوانند با آزمون حوزه تناظر تراکم ایجاد شده، یافت شوند. الگوریتم ما ممکن است قادر به یافتن کامپوننتهای متقارن حتی در موردی که هیچ نقطه دلخواهی وجود ندارد باشد.

برای نمایش اینکه چگونه روش میتواند برای تشخیص تقارن استفاده شود، یک الگوریتم ساده برای تقارنهای انتقالی در عناصر جابجا شده روی یک شبکه ارائه میکنیم. ابتدا، ما الگوریتم kNN را اجرا میکنیم. توصیف گر برای الگوریتم تکههای ۷ در ۷ است. سپس فاصله patch را با استفاده از $L^2$ میان پیکسلهای متناظر پس از تصحیح برای تغییرات محدود در شفاف سازی با نرمال سازی میانه و انحراف استاندارد برای رسیدن به تساوی محاسبه میکنیم.

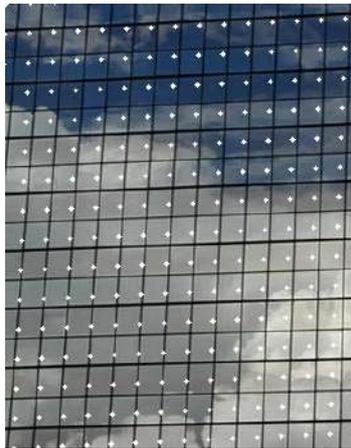

شکل ۷. ۳ – نتایج تشخیص تقارن با استفاده از یک شبکه منظم (با نقاط سفید نشان داده شد).

همچنین نزدیک ترین همســایههای $K = 16$ را یافته و ســپس از RANSAC برای یافتن بردارهای پایه $v_1$ و $v_2$ از شبکه استفاده میکنیم. مختصات جایی که فاصله میان همه ی kNNها کوچک است در درون ناحیه در نظر میگیریم. نتایج تشخیص تقارن در شکل ۷-۳ نشان داده شده است.

## ۷. ۵    تشخیص شی

روشهای تشــخیص شــی شــامل الگوهای بد شــکل [54]، boosted cascades [114]، تطابق ویژگیهای خلوت همچون SIFT [73] و غیره میشــود. الگوریتم ارائه شــده میتواند با ویژگیهای

۹۸

نمونه‌گیری متراکم، شامل patchهای نادرست، patchهای چرخش یافته یا تغییر مقیاس یافته، یا توصیف گرهایی مانند SIFT تطابق یابد. این تطابق‌ها سراسری هستند پس این تناظرها می‌توانند حتی در زمانی که تغییر مقیاس زیادی می‌یابد پیدا شوند [4].

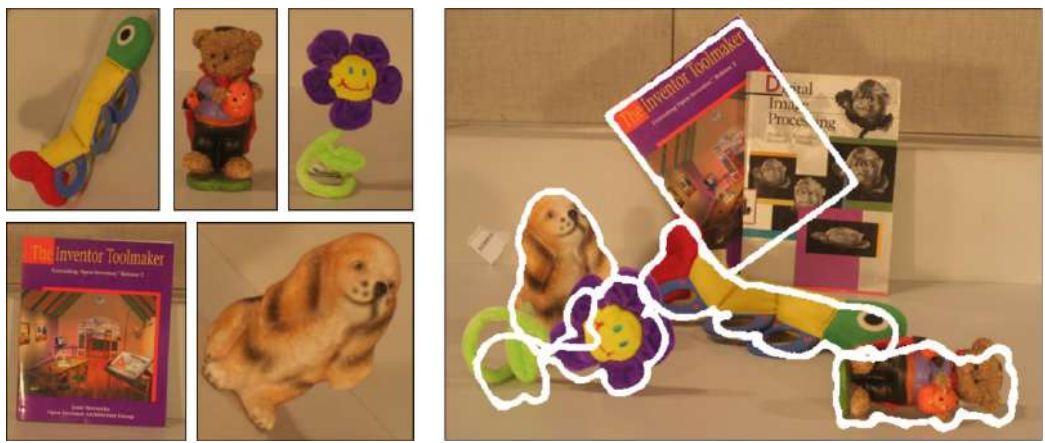

شکل ۷. ٤ – تشخیص اشیا. الگوهای چپ، با تصویر، راست، تطابق دارند. patchهای مربعی، جستجو در تمام چرخش‌ها و مقیاس‌ها همانطور که در بخش ۳. ۵ توصیف شده تطابق یافته‌اند. یک انتقال مشابه با تناظرهای حاصل با RANSAC تطابق یافته است.

در شکل ۷. ٤، مثالی از تشخیص شی نشان داده شده است. الگو را به patchهایی که اشتراک دارند، شکسته‌ایم. این patchها را در مقابل تصویر هدف، با جستجو در چرخش‌ها و محدوده مقیاس‌ها مانند بخش ۳. ۵ جستجو کرده‌ایم. یک تبدیل مشابه از الگو در هدف با استفاده از RANSAC یافت شده است و فاصله تکه را با استفاده از $L^2$، پس از تصحیح روشنایی، همانطور که در تشخیص تقارن نیز انجام دادیم محاسبه کرده‌ایم. نتیجه این است که اشیا تحت انسداد جزئی را در مقیاس‌ها و چرخش‌های مختلف می‌توانیم بیابیم.

برای ثبات بیشتر در تغییرات ظاهر و روشنایی، یک مدل محلی پیشرفته مورد نیاز است. برای مثال، فرض کنید تصاویری از دو شی مشابه با ظاهری مختلف داریم.

ما ممکن است در انتشار برچسب‌ها از یک تصویر به دیگری برای همه اشیا مشابه و پس زمینه تلاش کنیم. پژوهش SIFT [71] نشان می‌دهد که این می‌تواند با کمک تناظر ویژگی‌های SIFT در یک شبکه متراکم با یک ردیابی جریان حرکتی ترکیب شود. حوزه حاصل با کمک یک رویکرد سخت به نرم و

بهینه‌سازی سراسری یافت می‌شود. مانند اغلب روش‌های ردیابی جریان حرکتی، جریان SIFT، محلیت و ملایمت جریان را در نظر می‌گیرد و سپس در همترازی تحت جابجایی‌های بزرگ شکست، ممکن است شکست بخورد. همان‌طور که در شکل ۵.۷ نشان داده شد، می‌توانیم به درستی برچسب‌ها را هنگام حرکت اشیا به میزان زیادی تبدیل کنیم. این کار را با نمونه گیری متراکم توصیف گره‌های SIFT و سپس تطابق این‌ها همان‌طور که در بخش ۶.۳ نشان داده شد، انجام داده‌ایم.

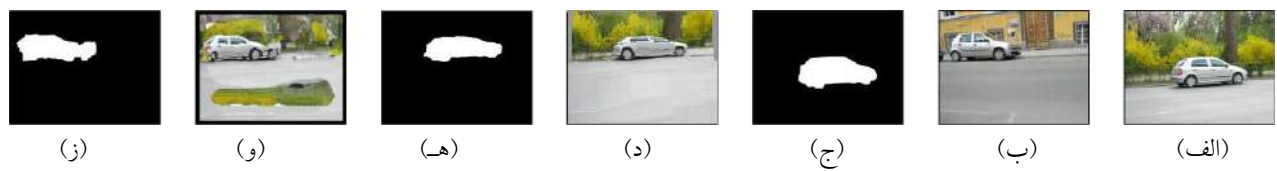

شکل ۵.۷ - تغییر برچسب با (الف) توصیف گره‌های SIFT (الف) ماشین اول؛ (ب) ماشین دوم، و (ج) هر دو ماشین یکسان برچسب زده شده اند؛ (د) برای تطابق با ماشین دیگری با استفاده از SIFT و نیز ماسک برچسب (هـ). (و) با روش ارائه شده و برچسب جابجا شده در (ز). جریان در کل ملایم تر است اما می‌تواند حرکات بزرگ دلخواه را نیز بگرداند.

۱۰۰

# فصل هشتم
# جمع بندی

این پایان نامه، در مورد پیشرفت‌هایی که در تطابق تصویر و ویدئو و کاربردهای مرتبط داشتیم بحث کرد. موارد زیر بررسی شدند:

**الگوریتم‌های تطابق الهام گرفته از تصاویر طبیعی**: یک آرگومان آماری ساختیم که ابتکاراتی که الگوریتم تطابق از آنها استفاده می‌کند را تقویت می‌کند. یک الگوریتم تطابق ، PatchMatch، و کلی سازی‌های آن را معرفی کردیم که از طریق انتقالات دلخواه، چرخش‌ها، توصیف گره‌ها و یافتن نزدیک ترین k همسایه کار میکند. بسته به اینکه کدام نوع از الگوریتم آزمون شده است، بالای یک یا دو مزیت نسبت به تکنیک‌های قبلی وجود دارد. همچنین انواع موازی و یک توسعه ، PatchWeb، برای تطابق در مجموعه‌های تصویر بزرگ معرفی کردیم.

**ویرایش تصویر**: ما نشان دادیم که الگوریتم تطابق سریع می‌تواند کاربردهای تعاملی دیگر در تکمیل تصویر، بازسازی، و بُرزدن داشته باشد. همچنین محدودیت‌های جدیدی برای کمک به هدایت کاربر در فرایند ترکیب معرفی کردیم.

**پرده‌های فیلم**: یک روش جدید از خلاصه سازی ویدئو معرفی کردیم. این پرده‌ها هم قابلیت مکانی، بدوم حاشیه مشخصی میان فریم‌ها، و هم زمانی، که در آن حرکت بزرگنمایی ملایم برای سطوح جزئیات دقیق فیلم را ممکن می‌سازد، ایجاد میکنند. بهینه سازی‌های انجام گرفته برای ایجاد پرده‌ها از الگوریتم تطابق سریع، از تفسیر بلادرنگ پرده‌ها استفاده می‌کند.

**کاربردهای بینایی**: با استفاده از الگوریتم تطابق کلی، تعدادی از کاربردهای بینایی را بررسی کردیم که شامل حذف نویز تصویر، یافتن جعل در تصاویر، تشخیص متقارن، SIFT و تشخیص شی می‌شود.

حوزه‌های متعددی برای پژوهش آینده وجود دارند. در فصل ۲، ذکر کردیم که آرگومان آماری می‌تواند برای ایجاد الگوریتم‌های تطابق کارآمد درآینده استفاده شود به خصوص برای حوزه‌های جدیدی مانند ویدئو که در آن الگوهای جستجوی بهینه را تغییر دهیم یا در شرایط متغییر ، الگوهای جستجوی متفاوتی فراهم کنیم. در بخش ۴. ۲. ۱ ، پیشنهاد کردیم که enrichment می‌تواند در مواردی

۱۰۱

که یک تصویر با خودش تطابق ندارد استفاده شود. در بخش ۵. ۴، پیشنهاد کردیم که معیارهای همگرایی یا پارامترهای مختلف می‌توانند، به خصوص با الگوریتم‌های GPU بهتر، برای ایجاد برنامه‌هایی در بینایی بلادرنگ یا پردازش ویدئو استفاده شوند. نهایتا، در بخش ۶. ۷، برای کاربرد پرده‌های ویدئو، پیشنهاد کردیم که موارد شکست می‌توانند بهبود یابند یا پرده‌ها می‌توانند متحرک باشند.

اکنون توسعه‌های ممکن سطح بالاتری برای پژوهش آینده پیشنهاد می‌کنیم.

ابتدا، الگوریتم خودش می‌تواند گسترش یابد، این کار می‌تواند به طور جزئی، با به کارگیری الگوریتم برای سیگنال‌های صوتی، رشته‌های ژن، یا تطابق مرزی در تصاویر یا سیگنال‌های سه بعدی مانند داده‌های حجیم با نمایش‌های سطحی توری شکل انجام شود. هرچند، نمایش اینکه در حوزه یک مسئله خاص، الگوریتم باید با استفاده از تکنیک‌های جستجوی مبتنی بر اولویت برای آن حوزه تناسب یابد. همچنین، چون توصیف گره‌های بینایی کامپیوتری متعددی براساس نقاط خلوت دلخواه نمونه گیری شدند، می‌توان نوع دیگری از الگوریتم ایجاد کرد که به جای توصیف گره‌های متراکم روی حالت خلوت کار کند. یا جتی به طور کلی تر، ممکن است ایجاد یک الگوریتم، که هیچ تعبیه مکانی خاصی را در نظر نمی گیرد با استفاده از ایده‌های enrichment و binning بیان شده در بخش ۴. ۲ امکان پذیر باشد.

به عنوان توسعه دوم، برای پژوهش آینده، الگوریتم تطابق می‌تواند بیشتر بهینه شود. در یک سطح بالاتر، ذکر کردیم که حداقل چهار اپراتور تطابق ممکن وجود دارند: انتشار، جستجوی تصادفی، enrichment و binning. شاید سایر اپراتورها نیز بتوانند توسعه یابند. به علاوه، تطابق می‌تواند در تفکیک پذیری‌های متعدد تطابق یابد: ناحیه‌هایی که همگرا شدند می‌توانند شناسایی شوند و دیگر تکرار روی آنها انجام نشود. تعیین اینکه کدام ترکیب از این تکنیک‌ها و تحت تنظیم کدام پارامترها خیلی موثر تر است جالب خواهد بود. این تفاوت‌ها به برنامه کاربردی و حتی ورودی خاص استفاده شده بستگی دارند. به علاوه، ترکیب این تکنیک‌ها می‌تواند به صورتی خلاقانه و با پیش محاسبات مختلقی انجام شوند.

نهایتا، ما تعدادی از کاربردهای آینده ممکن را پیشنهاد می‌کنیم که به صورت بالقوه از تکنیک‌های مبتنی بر patch استفاده می‌کنند: اختلاط تصویر، ترکیب هندسه سه بعدی، پر کردن حفره ویدئو، ترکیب الگوری ویدئو، طبقه بندی تصویر و ویدئو، ردیابی بلادرنگ، فرا تفکیک پذیری، نقاشی با اعداد، تحلیل تصویر، و فشرده سازی عکس و ویدئو. ما همچنین می‌توانیم برنامه‌های کاربردی موجود خود را بهبود دهیم. اغلب الگوریتم‌های ما – مانند بُرزدن – خروجی‌های هم خوب و هم بد بسته به محدودیت‌های

۱۰۲

کاربر تولید می‌کند. یک ایده کلی پس از آن، تولید تصاویر خروجی متعدد و انتخاب خودکار بهترین تصویر از این مجموعه است. متناوبا، چنین محدودیت‌هایی ممکن است حداقل در گام‌های اولیه به صورت الگوریتمی ارائه شوند.

بخاطر تنوع گسترده‌ای از توسعه‌ها و کاربردهای الگوریتم ارائه شده، باور داریم الگوریتم ارائه شده می‌تواند کلی سازی شود و برای حوزه‌های مختلف گسترش یابد.

# مراجع

# واژه نامه

| | |
|---|---|
| Image | تصویر |
| Matching | تطابق |
| Randomized | تصادفی |
| Correspondence | تناظر |
| Region | ناحیه |
| Thechnique | شیوه |
| Nearest neighbors | نزدیکترین همسایه |
| Magnitude | اندازه |
| Approximate | تقریبی |
| Convergence | همگرایی |
| Translate | انتقال |
| Rotate | چرخاندن |
| Scale | مقیاس کردن |
| Computer graphics | گرافیک کامپیوتری |
| Computer vision | بینایی کامپیوتری |
| Application | برنامه/کاربرد |
| Denoising | حذف نویز |
| Forgery | تقلب |
| Overlapping | اشتراک |
| Textural patterns | الگوهای بافتی |
| Method | روش |
| Photograph | عکس |
| Feature | ویژگی |
| Resolution | وضوح/تفکیک پذیری |
| Patch-based | بر پایه تکه |
| Computational photography | عکاسی محاسباتی |
| Implementation | پیاده سازی |
| Generalized | کلی |
| Scalability | مقیاس پذیری |
| Tapestries | پرده‌های فیلم |
| Cluster architectures | معماری خوشه‌ای |
| Edit | ویرایش |
| Applications | برنامه‌های کاربردی/ کاربردها |
| Stretching | کشش |
| Distortions | اعوجاج |
| Portion | بخش |
| Multiscale | چند مقیاسی |
| Spatial domain | بعد مکانی |
| Border | مرز |
| Means | متوسط |
| Non-local | غیر محلی |
| Difference | تفاضل |
| Constraint | ثابت |
| Coherence | وابستگی |
| Same | یکسان |
| Propagation | انتشار |
| Correlation | همبستگی |
| Ground truth | محل‌های واقعی |
| Coordinate | مختصات |
| Wide | عریض |
| Dataset | مجموعه داده |
| Similar | مشابه |
| Dissimilar | غیر مشابه |
| Baseline | پایه |
| Suboptimal | حد مطلوب |
| Peak | اوج |
| Descriptor | توصیف گر |
| Reasonable | معقول |
| Common | متداول |
| Nonparametric | غیر پارامتری |
| Nearest-neighbor field | نزدیکترین همسایه درست |
| Brute force | جامع (فراگیر) |
| Dimensionality reduction | کاهش ابعاد |
| Dimension | بعد |
| Space | فضا |
| Incremental | افزایشی |
| Offset | انحراف |
| Sparse | خلوت |
| Transformation | تبدیل |
| Invariant | نامتغیر |
| Interest points | نقاط مورد علاقه |
| Texture | بافت |
| Recomposition | بازترکیب |
| Precomputation | پیش محاسبه |
| Enhancing images | افزایش تصاویر |
| Morphing | شکل گیری |
| Motion estimation | تخمین حرکت |
| Sequence | توالی |
| Map | نگاشت |
| Halting | توقف |
| Reconstruct | بازسازی |
| Heap | پشته |
| Candidate | داوطلب |
| Eigenvectors | بردارهای ویژه |
| Inference | استنتاجی |
| Tile | تکه |
| Visual words | کلمات بصری |
| Image reshuffling | بُرزنی تصویر |
| Image completion | تکمیل تصویر |
| Image retargeting | بازسازی تصویر |
| Global | سراسری |
| Hole | حفره |
| Project | افکندن |
| Swap | تعویض |
| Clone | نسخه برداری |
| Discrete | مجزا |
| Image segmentation | تقسیم بندی تصویر |